\documentclass[margin,12pt]{article}
\usepackage[catalan,english]{babel}
\usepackage{lmodern}
\usepackage[T1]{fontenc}
\usepackage{latexsym}
\usepackage{slashbox}
\usepackage{amsfonts}
\usepackage{amsmath}
\usepackage{amssymb}
\usepackage[mathscr]{eucal}
\usepackage{a4wide}
\usepackage{graphicx}
\usepackage[usenames]{color}
\usepackage{xcolor}
\usepackage{bbold}
\usepackage{bbm}
\usepackage{authblk}
\numberwithin{equation}{section}
\usepackage{multirow}
\RequirePackage{hyperref}
\hypersetup{
	linktocpage,
    colorlinks,
    citecolor=blue,
    filecolor=black,
    linkcolor=blue,
    urlcolor=blue,
    }

\def\be{\begin{equation}}
\def\ee{\end{equation}}
\def\bea{\begin{eqnarray}}
\def\eea{\end{eqnarray}}
\def\ben{\begin{enumerate}}
\def\een{\end{enumerate}}

\title{\begin{flushright}
\small{JLAB-THY-18-2701}\\[4ex]
\end{flushright}
\bf{$\eta^{\prime}\to\eta\pi\pi$ decays in unitarized resonance chiral theory}}
\author[a]{Sergi Gonz\`{a}lez-Sol\'{i}s\thanks{sgonzalez@itp.ac.cn}}
\author[b,c]{Emilie Passemar\thanks{epassema@indiana.edu}}
\affil[a]{{\it{CAS Key Laboratory of Theoretical Physics, Institute of Theoretical Physics, Chinese Academy of Sciences, Beijing 100190, China}}}
\affil[b]{{\it{Department of Physics, Indiana University, Bloomington, IN 47405, USA\newline
Center for Exploration of Energy and Matter, Indiana University, Bloomington, IN 47408, USA}}}
\affil[c]{{\it{Theory Center, Thomas Jefferson National Accelerator Facility, Newport News, VA 23606, USA}}}

\begin{document}\maketitle

\begin{abstract}

We study the hadronic $\eta^{\prime}\to\eta\pi\pi$ decays within the framework of $U(3)_{L}\otimes U(3)_{R}$ Chiral Perturbation Theory including resonance states and the complete one-loop corrections. 
The amplitude is projected in partial waves and unitarized by means of the $N/D$ method resumming both the important $S$-and $D$-wave $\pi\pi$ and the subleading $S$-wave $\pi\eta$ final-state interactions.
The participating scalar multiplet mass and coupling strengths are determined from fits to the Dalitz plot experimental data recently released by the A2 collaboration.
As a byproduct of our analysis, the associated Dalitz-plot slope parameters are found to be $a=-0.072(7)_{\rm{stat}}(8)_{\rm{syst}}\,, b=-0.052(1)_{\rm{stat}}(2)_{\rm{syst}}\,, d=-0.051(8)_{\rm{stat}}(6)_{\rm{syst}}$, which lie in the ballpark of the current experimental and theoretical determinations.

\end{abstract}

\section{Introduction}\label{section1}

Different to Quantum Electrodynamics, a perturbative expansion in terms of the strong coupling cannot be applied to describe QCD processes at low-energies because the coupling strength becomes very large and therefore invalidates such an expansion.
A well-known and particularly successful approach to overcome this limitation is Chiral Perturbation Theory (ChPT) \cite{Gasser:1984gg}, the low-energy effective field theory of QCD.
ChPT is described in terms of eight pseudo-Goldstone bosons associated to the spontaneous chiral symmetry breaking $SU(3)_{L}\otimes SU(3)_{R}\to SU(3)_{V}$ exhibited by QCD i.e. three pions $\pi^{\pm,0}$, four kaons $K^{\pm},K^{0}$ and $\bar{K}^{0}$, and the $\eta$. The theory is constructed by performing a double perturbative expansion, in momenta, $p^{2}$, and quark masses, $m_{q}$, over $\Lambda_{\chi}\sim m_{\rho}\sim1$ GeV\footnote{It corresponds to the scale where ChPT breaks down.}. 
ChPT has been successfully applied for describing numerous processes involving pions and kaons but much less for the $\eta$. 
Actually, the $\eta$ entering ChPT is not the physical one but rather a part of it corresponding to the octet\footnote{Accordingly one should not call this state $\eta$ but rather $\eta_{8}$.}.
In reality, the $\eta$ meson has a second component, coming from the pseudoscalar singlet $\eta_{1}$, which is not systematically included in ChPT due to the emergence of an anomaly. Indeed the $U(1)_{A}$ symmetry is broken (even in the massless case) by the quantum dynamics of QCD itself, preventing the $\eta_{1}$ to be the ninth Goldstone boson.
This makes the $\eta^{\prime}$ too heavy to be included as the ninth pseudo-Goldstone boson.
However, in the limit of the number of colours becoming large, the "large-$N_{C}$ limit", the axial anomaly vanishes and 
the $\eta_{1}$ can be integrated to the Goldstone bosons~\cite{Kaiser:2000gs,HerreraSiklody:1996pm,HerreraSiklody:1997kd}.
In this limit, the (inverse) number of colors $1/N_{C}$ is included in the power counting scheme as $\delta\equiv\{(p/\Lambda_{\chi})^{2},m_{q}/\Lambda_{\chi},1/N_{C}\}$ leading to a combined triple expansion in $\delta\sim p^{2}/\Lambda_{\chi}^{2}\sim m_{q}/\Lambda_{\chi}\sim1/N_{C}$. 
Moreover, the $SU(3)_{L}\otimes SU(3)_{R}$ symmetry is enlarged to $U(3)_{L}\otimes U(3)_{R}$ and, the pseudoscalar octet and singlet states $\eta_{8}$ and $\eta_{1}$ mix allowing for a reasonable dynamical description of the physical $\eta$ and $\eta^{\prime}$ mesons.
 
While the convergence of the $SU(3)$ ChPT perturbative expansion is restricted to low-energies i.e.\,when the energy available in the process is below the mass of the first resonance (i.e. the mass of the $\rho(770)$), the $\eta^{\prime}$ meson in $U(3)$ is heavier ($m_{\eta^{\prime}}\sim958$ MeV) than some resonances.
To describe processes involving the $\eta^{\prime}$ meson the ChPT framework has therefore to be enlarged to include explicitly such resonances.
This is the avenue of Resonance Chiral Theory (R$\chi$T) \cite{Ecker:1988te}. In this theory, the interactions of the pseudoscalar mesons are governed by resonance exchanges. 
For these reasons, predicting observables that include $\eta$ and $\eta^{\prime}$ mesons is, 
typically, more difficult than for pions and kaons.

Measuring $\eta$ and $\eta^{\prime}$ observables is also more complicated 
since more decay channels are allowed and most of them contain hadrons in the final state instead of photons as in the $\pi^{0}$ case.
At present, there is a series of ongoing experiments 
measuring the decays of $\eta$ and $\eta^{\prime}$ mesons with a precision never reached before.
The WASA-at-COSY experiment has recently measured the branching ratios of the decays  $\eta\to e^{+}e^{-}\gamma$, $\eta\to\pi^{+}\pi^{-}\gamma$, $\eta\to e^{+}e^{-}e^{+}e^{-}$ and $\eta\to\pi^{+}\pi^{-}e^{+}e^{-}$ \cite{Adlarson:2015zta}. 
The A2 collaboration at MAMI has measured $\eta\to e^{+}e^{-}\gamma$ \cite{Adlarson:2016hpp}, released a new evaluation of the decay rate distribution of the doubly radiative decay $\eta\to\pi^{0}\gamma\gamma$ \cite{Nefkens:2014zlt} as well as very recently a high-statistic measurement of $\eta\to3\pi^{0}$ \cite{Prakhov:2018tou}. 
The BESIII collaboration has reported the first measurements of the decays $\eta^{\prime}\to e^{+}e^{-}\gamma$ \cite{Ablikim:2015wnx}, $\eta^{\prime}\to\pi^{0}\gamma\gamma$ \cite{Ablikim:2016tuo}, $\eta^{\prime}\to\omega e^{+}e^{-}$ \cite{Ablikim:2015eos}, $\eta^{\prime}\to\pi^{+}\pi^{-}\pi^{+}\pi^{-}$ and $\eta^{\prime}\to\pi^{+}\pi^{-}\pi^{0}\pi^{0}$ \cite{Ablikim:2014eoc}, performed a precise study of $\eta^{\prime}\to\pi^{+}\pi^{-}\gamma$ \cite{Ablikim:2017fll}, provided new measurements of $\eta^{\prime}\to3\pi$ \cite{Ablikim:2016frj} and $\eta^{\prime}\to\pi^{+}\pi^{-}e^{+}e^{-}$ \cite{Ablikim:2013wfg}, and set the first upper bound on $\eta^{\prime}\to\pi^{+}\pi^{-}\mu^{+}\mu^{-}$ \cite{Ablikim:2013wfg}.
Searches of the CP-violating $\eta^{(\prime)}\to\pi^{+}\pi^{-}$ decays are pursued at LHCb \cite{Aaij:2016jaa}.
This experimental progress mades us enter in a precision era for the physics of $\eta$ and $\eta^{\prime}$. 
These experimental advances require revisiting the corresponding theoretical analyses in order to understand better the meson dynamics at low energy in the non perturbative regime of QCD.
Recent studies of some of the aforementioned decays include Refs.\,\cite{Colangelo:2016jmc,Guo:2015zqa,Albaladejo:2017hhj,Kolesar:2017xrl} for $\eta\to3\pi$, Ref.\,\cite{Guo:2011ir} for $\eta^{(\prime)}\to4\pi$, Refs.\,\cite{Stollenwerk:2011zz,Kubis:2015sga,Dai:2017tew} for $\eta^{(\prime)}\to\pi^{+}\pi^{-}\gamma$, Ref.\,\cite{Escribano:2012dk} for $\eta^{\prime}\to\pi^{0}\gamma\gamma$ and $\eta^{\prime}\to\eta\gamma\gamma$, and Refs.\,\cite{Hanhart:2013vba,Escribano:2013kba,Roig:2014uja,Escribano:2015yup,Escribano:2015vjz, Kampf:2018wau,Husek:2017vmo} for transition form factors and Dalitz decays.
 
Studying the hadronic $\eta^{\prime}\to\eta\pi\pi$ decay is particularly interesting theoretically since this decay cannot be described within $SU(3)$ ChPT alone for the reasons given above. 
Therefore, it represents an advantageous 
laboratory to test any of the extensions of ChPT such as the Large-$N_{C}$ $U(3)$ ChPT or R$\chi$T.
In $U(3)$ Large-$N_{C}$ ChPT, the lowest-order (LO) contribution to the amplitude is chirally suppressed giving a branching ratio inconsistent with its measured value. 
The LO is a constant leading to a constant Dalitz-plot distribution in disagreement with the measurements.
The next-to-leading order contribution
is 
found to be the dominant one and considered as the first term in the expansion \cite{HerreraSiklody:1999ss,Escribano:2010wt}.
This fact should not be understood as the sign for a poorly convergent expansion since we can anticipate from our study that, 
as shown in Ref.\,\cite{Beisert:2002ad},
higher order terms, i.e.\,loop corrections in the simultaneous triple chiral expansion scheme, are rather small.
On the other hand, this process can be explained by means of the explicit exchange of the scalar resonances $\sigma$ or $f_0 (550),f_{0}(980)$ and $a_{0}(980)$. 
Based on an effective chiral Lagrangian model, the authors of Ref.\,\cite{Fariborz:1999gr} have shown that the $a_{0}(980)$ resonance indeed dominates the decay.
This was confirmed later on by the $U(3)$ chiral unitary analyses of Refs.\,\cite{Beisert:2003zs,Borasoy:2005du}. 
Notwithstanding, the $\sigma$ is essential to determine the Dalitz-plot parameters \cite{Escribano:2010wt}.
See Ref.\,\cite{Isken:2017dkw} for a recent study of this process using a dispersion approach. 

On the experimental side, the $\eta^{\prime}\to\eta\pi\pi$ decay width represents  $\sim65\%$ of the total width; the PDG reported values for the branching ratios of $42.6(7)\%$ in the charged channel and of $22.8(8)\%$ in the neutral one~\cite{Olive:2016xmw}. 
Experimentally the Dalitz plots parameters associated to the decay are usually extracted from the measurements. 
In the isospin limit, these parameters should be the same in both channels. However, large discrepancies have been reported 
between the VES \cite{Dorofeev:2006fb}, GAMS-4$\pi$ \cite{Blik:2009zz} and BESIII \cite{Ablikim:2010kp} and the A2 \cite{Adlarson:2017wlz} and BESIII \cite{Ablikim:2017irx} measurements. 
The current status calls for clarification. 
The theoretical predictions \cite{Escribano:2010wt,Beisert:2002ad,Fariborz:1999gr,Beisert:2003zs,Borasoy:2005du,Isken:2017dkw,Fariborz:2014gsa} exhibit the same level of disagreement 
to the extent that, for example, the 2011 BESIII paper \cite{Ablikim:2010kp} cites some of the existing calculations of the $Y$-variable quadratic term \textquotedblleft$b$\textquotedblright\,concluding that \textquoteleft the dynamical nature of this term needs further clarification\textquoteright.

In this work, we revisit the $\eta^{\prime}\to\eta\pi\pi$ decays taking advantage of the large number of reconstructed events, $\sim1.23\cdot10^{5}$, for the neutral mode recently collected by the A2 collaboration~\cite{Adlarson:2017wlz}.
This study extends the analyses of Refs.\,\cite{Escribano:2010wt,Guo:2011pa} by 
including the complete one-loop corrections within a $U(3)$ ChPT framework and taking into account the $\pi\pi$ and $\pi\eta$ final-state interactions.
These rescattering effects are accounted using the $N/D$ unitarization method. We extract with accurate precision the associated Dalitz-plot parameters from fits to the A2 Dalitz distributions \cite{Adlarson:2017wlz}\footnote{Contrary to A2 \cite{Adlarson:2017wlz}, the recent BESIII Dalitz plot measurements \cite{Ablikim:2017irx}, consisting of $351016$ and $56249$ events for the charged and neutral channels, respectively, are unfortunately not yet publicly available. 
We therefore postpone the analysis of this data for the near future.}. 

This article is structured as follows.
In section \ref{section2}, we define the kinematics of the process, introduce the Dalitz-plot parametrisation and discuss the current status of the associated parameters.
The relevant Lagrangian is given in section \ref{section3}.
The structure of the decay amplitude is addressed in section \ref{section4} while section \ref{section5} is devoted to its unitarization.
In section \ref{section6} we present the results of our fit to the experimental data for the $\pi^{0}\pi^{0}$ mode from A2.
Different fits are performed. They are organized according to their increasing fulfilment of unitarity. In each of these fits, the mass and the couplings of the scalar resonances are determined as well as  the Dalitz plot parameters.
We start our study considering ChPT including resonances and one-loop corrections. 
In a second step, we show the importance of the $\pi\pi$ $S$-and $D$-wave rescattering effects that nicely accommodate the $\pi^{+}\pi^{-}$ cusp effect seen for the first time in $\eta^{\prime}\to\eta\pi^{0}\pi^{0}$ by the A2 collaboration.
We then explore first the individual effect of the $\pi\eta$ final-state interactions, anticipated to be small, 
before presenting our central description of the process including both $\pi\pi$ and $\pi\eta$ rescattering effects. 
Our results are obtained from a fit to the A2 data. 
Our analysis enables us to extract some information about the $I=1$ $\pi\eta$ scattering phase shift within the allowed physical decay region. Moreover we predict the Dalitz-plot parameters and distribution of the $\pi^{+}\pi^{-}$ decay channel that are found to be
in excellent agreement with the BESIII experimental data. 
Finally, our conclusions are presented in section \ref{conclusions}.

\section{Kinematics and Dalitz-plot parametrisation}\label{section2}

\begin{figure}[h!]
\begin{center}
\includegraphics[scale=0.415]{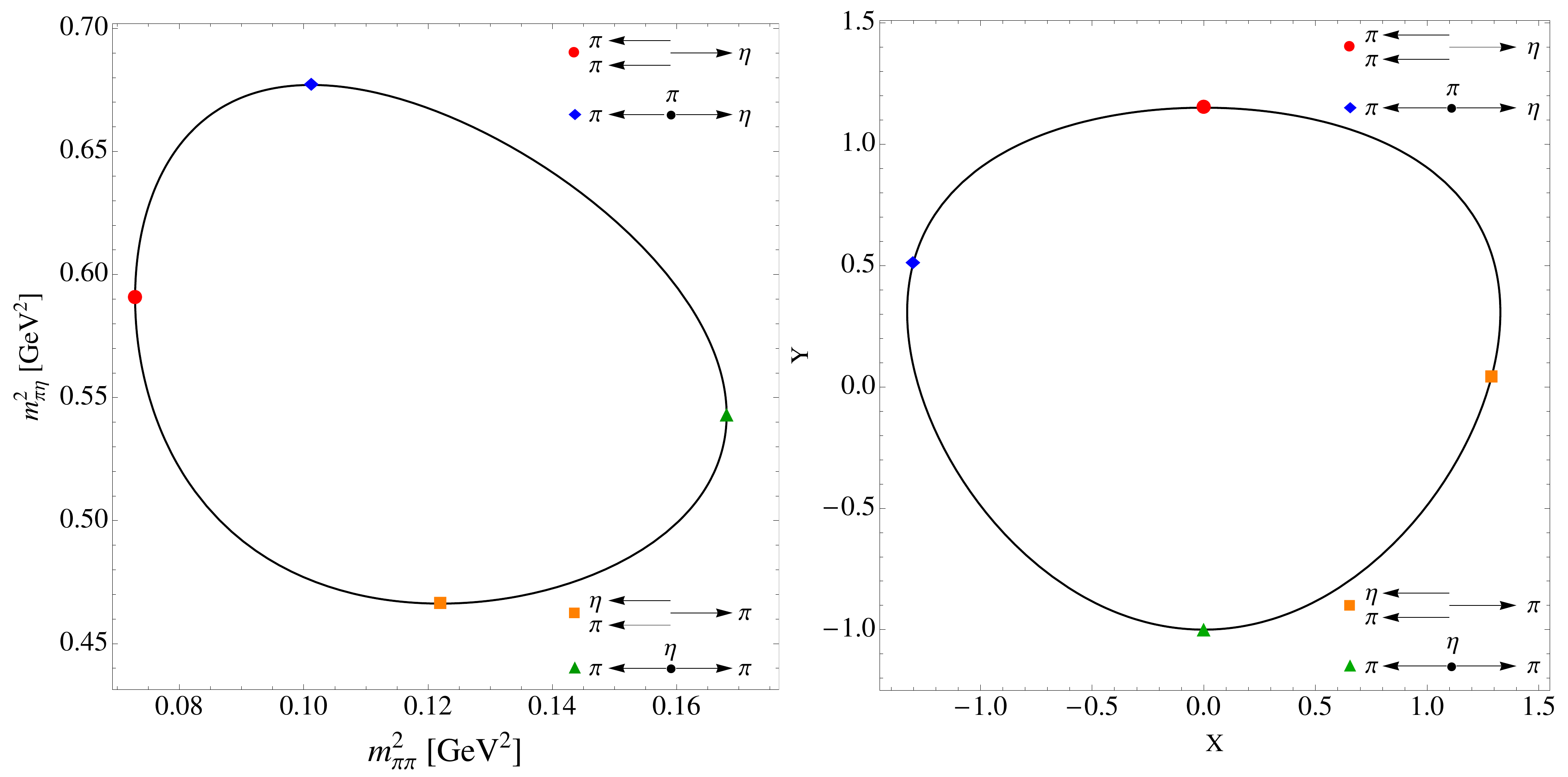}
\caption{\label{DalitzBoundaries}Boundary of the Dalitz plot for $\eta^{\prime}\to\eta\pi\pi$ in terms of the invariant masses $m^{2}_{\pi\pi}$ and $m^{2}_{\pi\eta}$ (left) and in terms of the Dalitz variables $X$ and $Y$ (right).}
\end{center}
\end{figure}

Let us consider the amplitude for the $\eta^{\prime}(p_{\eta^{\prime}})\to\eta(p_{\eta})\pi(p_{1})\pi(p_{2})$ decay, $\mathcal{M}(s,t,u)$. 
It is given in terms of the Mandelstam variables
\begin{eqnarray}
s&=&\left(p_{\eta^{\prime}}-p_{\eta}\right)^{2}=\left(p_{1}+p_{2}\right)^{2}\,,\nonumber\\
t&=&\left(p_{\eta^{\prime}}-p_{1}\right)^{2}=\left(p_{\eta}+p_{2}\right)^{2}\,,\nonumber\\
u&=&\left(p_{\eta^{\prime}}-p_{2}\right)^{2}=\left(p_{\eta}+p_{1}\right)^{2}\,,
\end{eqnarray}
which fulfill the relation
\begin{eqnarray}
s+t+u=m_{\eta^{\prime}}^{2}+m_{\eta}^{2}+2m_{\pi}^{2}\,.
\end{eqnarray}
The partial decay rate reads \cite{Olive:2016xmw}
\begin{eqnarray}
\Gamma\left(\eta^{\prime}\to\eta\pi\pi\right)=\frac{1}{256\pi^{3}m_{\eta^{\prime}}^{3}\mathcal{N}}\int ds\,dt\,|\mathcal{M}(s,t,u)|^{2}\,,
\label{width}
\end{eqnarray}
where $\mathcal{N}$ accounts for the number of identical particles in the final state; $\mathcal{N}=1$ for the charged and $\mathcal{N}=2$ for the neutral decay modes, respectively.
The boundaries of the physical decay region in $t$ lie within $[t_{\rm{min}}(s),t_{\rm{max}}(s)]$ with
\begin{eqnarray}
t_{\rm{max/min}}(s)=\frac{1}{2}\Bigg[m_{\eta^{\prime}}^{2}+m_{\eta}^{2}+2m_{\pi}^{2}-s\pm\frac{\lambda^{1/2}(s,m_{\eta^{\prime}}^{2},m_{\eta}^{2})\lambda^{1/2}(s,m_{\pi}^{2},m_{\pi}^{2})}{s}\Bigg]\,,
\label{tmaxmin}
\end{eqnarray}
where $\lambda(x,y,z)=x^{2}+y^{2}+z^{2}-2xy-2xz-2yz$. The boundaries in $s$ are given by
\begin{eqnarray}
s_{\rm{min}}=4m_{\pi}^{2}\,,\quad s_{\rm{max}}=(m_{\eta^{\prime}}-m_{\eta})^{2}\,.
\end{eqnarray}
However, the experimental measurements are often given as a power expansion in terms of the so-called Dalitz variables $X$ and $Y$.
These two variables are defined by
\begin{eqnarray}
X=\frac{\sqrt{3}}{Q}\left(T_{\pi_{1}}-T_{\pi_{2}}\right)\,,\quad Y=\frac{m_{\eta}+2m_{\pi}}{m_{\pi}}\frac{T_{\eta}}{Q}-1\,,
\end{eqnarray}
where $T_{\pi_{1,2}}$ and $T_{\eta}$ are the kinetic energies of the mesons in the $\eta^{\prime}$ rest frame: 
\begin{eqnarray}
T_{\eta}=\frac{\left(m_{\eta^{\prime}}-m_{\eta}\right)^{2}-s}{2m_{\eta^{\prime}}}\,,\quad T_{\pi_{1}}=\frac{\left(m_{\eta^{\prime}}-m_{\pi}\right)^{2}-t}{2m_{\eta^{\prime}}}\,,\quad T_{\pi_{2}}=\frac{\left(m_{\eta^{\prime}}-m_{\pi}\right)^{2}-u}{2m_{\eta^{\prime}}}\,,
\end{eqnarray}
and $Q=T_{\eta}+T_{\pi_{1}}+T_{\pi_{2}}=m_{\eta^{\prime}}-m_{\eta}-2m_{\pi}$.
\\
\\
The decay width Eq.\,(\ref{width}) can therefore also be written as: 
\begin{eqnarray}
\Gamma\left(\eta^{\prime}\to\eta\pi\pi\right)=\frac{m_{\pi}Q^{2}}{128\sqrt{3}\pi^{3}m_{\eta^{\prime}}(2m_{\pi}+m_{\eta})\mathcal{N}}\int dX\,dY\,|\mathcal{M}(X,Y)|^{2}\,,
\label{width2}
\end{eqnarray}
where now the integration boundaries are given by
\begin{eqnarray}
Y_{\rm{min}}=-1\,,\quad Y_{\rm{max}}=\frac{1}{2m_{\eta^{\prime}}m_{\pi}}\left(m_{\eta}m_{\eta^{\prime}}-m_{\eta}^{2}+4m_{\pi}^{2}\right)\,,
\label{Yvariable}
\end{eqnarray}
and
\begin{eqnarray}
X_{\rm{min}}(s)=-\frac{\sqrt{3}}{2m_{\eta^{\prime}}Q}h\left((m_{\eta^{\prime}}-m_{\eta})^{2}-\frac{2m_{\eta^{\prime}}m_{\pi}Q}{m_{\eta}+2m_{\pi}}(Y+1)\right)\,,\,X_{\rm{max}}(s)=-X_{\rm{min}}(s)\,.
\label{Xvariable}
\end{eqnarray}
The function $h(s)$ is defined as
\begin{eqnarray}
h(s)=\frac{\lambda^{1/2}(s,m_{\eta^{\prime}}^{2},m_{\eta}^{2})\lambda^{1/2}(s,m_{\pi}^{2},m_{\pi}^{2})}{s}\,.
\end{eqnarray}
Fig.\,\ref{DalitzBoundaries} 
shows the boundaries of the Dalitz plot in $m^{2}_{\pi\pi} \equiv s$ and $m_{\pi\eta}^{2} \equiv t,u$, the invariant masses (left panel) and in the Dalitz variables $X$ and $Y$ (right panel).
At the $\pi\pi$ threshold, $m^{2}_{\pi\eta}\sim0.59$ GeV$^{2}$, the two pions move together in the same direction with equal velocities and the $\eta$ moves in opposite direction (red point).
At $m^{2}_{\pi\pi}\sim0.11$ GeV$^{2}$, the range of $m^{2}_{\pi\eta}$ increases going from $\sim0.47$ GeV$^{2}$ to $\sim0.67$ GeV$^{2}$. 
In this last point, one pion is at rest and the other one moves in opposite direction of the $\eta$ (blue diamond).
At $m_{\pi\eta}^{2}\sim0.54$ GeV$^{2}$, the allowed range of $m_{\pi\pi}^{2}$ values is large and 
reaches energies close to the region of influence of the $\sigma$ meson. 
When $m^{2}_{\pi\pi} \sim 0.16$ GeV$^{2}$, the $\eta$ is at rest and the two pions move back-to-back (green triangle). 
We can anticipate that the region around this point will contain the largest number of events of the Dalitz plot decay distribution. 
Finally, at the $\pi\eta$ threshold, $m^{2}_{\pi\pi}\sim0.12$ GeV$^{2}$, the $\eta$ and one pion move in one direction with equal velocities and the other pion in the opposite direction (orange square).

The Dalitz plot parametrisation for $\eta^{\prime}\to\eta\pi\pi$ decays is obtained by expanding the squared of the decay amplitude in powers of $X$ and $Y$ around the center of the Dalitz plot
\begin{eqnarray}
\Gamma(X,Y)=|\mathcal{M}(X,Y)|^{2}=|N|^{2}\Big[1+aY+bY^{2}+cX+dX^{2}+\cdots\Big]\,,
\label{expansionDalitz}
\end{eqnarray}
where $a,b,c$ and $d$ are the real-valued Dalitz parameters and $N$ is an overall normalization\footnote{An alternative parameterization would be the so-called linear expansion
\begin{equation}
|\mathcal{M}(X,Y)|^{2}=|N|^{2}\left(|1+\alpha Y|^{2}+cX+dX^{2}+\cdots\right)\,,
\end{equation}
where $\alpha$ is complex.
A comparison with the parameterization of Eq.\,(\ref{expansionDalitz}) gives $a=2\rm{Re}(\alpha)$ and $b=\rm{Re}(\alpha)^{2}+\rm{Im}(\alpha)^{2}$. 
The two parameterizations are equivalent if $b>a^{2}/4$.}.

Table \ref{DalitzParamStatus} contains the current state-of-the-art Dalitz-plot parameters extracted from measurements together 
with their theoretical estimates. We can see large discrepancies between the results. 
VES result on the $a$ parameter is $2.6\sigma$ from the result of GAMS-4$\pi$ and $3.8\sigma$ away from the result of BESIII 
(2011 analysis). 
This disagreement persists between the 2017 updated BESIII values and the new A2 measurement. 
The theoretical value for $a$ as obtained in Ref.\,\cite{Borasoy:2005du} is in agreement with VES result.
This is not surprising since this data set was fitted in their analysis. 
In Ref.\,\cite{Escribano:2010wt} the parameter $a$ is not predicted but rather fixed from an average of the VES and GAMS-4$\pi$ values.
From fits to the 2011 BESIII and VES data, the recent dispersive analysis of Ref.\,\cite{Isken:2017dkw} obtains values ranging from $-0.041$ to $-0.148$, which agree with the corresponding measured values reported by these two collaborations. 
The parameter $b$ determined by VES is $\sim1\sigma$ away from the measurements of GAMS-4$\pi$, BESIII and A2.
A deviation of $\sim(1$-$2)\sigma$ is also seen with respect to the analyses of Refs.\,\cite{Escribano:2010wt,Borasoy:2005du}.
However, this discrepancy disappears in the analysis of Ref.\,\cite{Isken:2017dkw}.
Both recent experimental measurements and theoretical predictions seem to indicate an unambiguously negative value for $b$  
in clear disagreement with a vanishing $b$ obtained in Refs.\,\cite{Fariborz:1999gr,Fariborz:2014gsa}.
Regarding the parameter $c$, the symmetry of the wave function forbids such a term in the neutral channel. In the charged channel the odd terms in $X$ are forbidden since $C$-parity is conserved by strong interactions. So its value is predicted theoretically to be zero in 
the Standard Model. 
And the measured values of this parameter are consistent with zero.
Finally, for the value of the parameter $d$, experimental results seem to favour the predictions of Refs.\,\cite{Escribano:2010wt,Isken:2017dkw} with respect to those of Refs.\,\cite{Borasoy:2005du,Fariborz:2014gsa}. 
In conclusion, from Table~\ref{DalitzParamStatus}, we observe an inconsistent picture so far. 
The theoretical determinations of the Dalitz parameters are 
intricately linked to the data used to constrain the corresponding theories or models. 
 
\begin{table}
\footnotesize
\centering
\begin{tabular}{llllll}
\hline
$\eta^{\prime}\to\eta\pi^{0}\pi^{0}$&$a[Y]$&$b[Y^{2}]$&$c[X]$&$d[X^{2}]$&$\#$ events\\
\hline
GAMS-4$\pi$ \cite{Blik:2009zz}& $-0.067(16)(4)$& $-0.064(29)(5)$& $=0$& $-0.067(20)(3)$&$15000$\\ 
GAMS-4$\pi$ \cite{Blik:2009zz}& $-0.066(16)(4)$& $-0.063(28)(4)$& $-0.107(96)(3)$& $0.018(78)(6)$&$15000$\\ 
A2 \cite{Adlarson:2017wlz} & $-0.074(8)(6)$&$-0.063(14)(5)$&---&$-0.050(9)(5)$&$\sim1.23\cdot10^{5}$\\
BESIII \cite{Ablikim:2017irx}& $-0.087(9)(6)$& $-0.073(14)(5)$& $0$& $-0.074(9)(4)$&$56249$\\
\hline                           
Borasoy {\rm{\it{et.al.}}}\,\cite{Borasoy:2005du}&$-0.127(9)$&$-0.049(36)$&---&$0.011(21)$\\
\hline             
\hline
$\eta^{\prime}\to\eta\pi^{+}\pi^{-}$&$a[Y]$&$b[Y^{2}]$&$c[X]$&$d[X^{2}]$\\
\hline
VES \cite{Dorofeev:2006fb}& $-0.127(16)(8)$& $-0.106(28)(14)$& $0.015(11)(14)$& $-0.082(17)(8)$\\
BESIII \cite{Ablikim:2010kp}& $-0.047(11)(3)$& $-0.069(19)(9)$& $0.019(11)(3)$& $-0.073(12)(3)$&$43826$\\
BESIII \cite{Ablikim:2017irx}& $-0.056(4)(3)$& $-0.049(6)(6)$& $2.7(2.4)(1.8)\cdot10^{-3}$& $-0.063(4)(4)$&$351016$\\
\hline                           
Borasoy {\rm{\it{et.al.}}}\,\cite{Borasoy:2005du}&$-0.116(11)$&$-0.042(34)$&$0$&$0.010(19)$\\
Escribano {\rm{\it{et.al.}}}\,\cite{Escribano:2010wt}&$-0.098(48)$&$-0.050(1)$&$0$&$-0.092(8)$\\
Escribano {\rm{\it{et.al.}}}\,\cite{Escribano:2010wt}&$-0.098(48)$&$-0.033(1)$&$0$&$-0.072(1)$\\
Kubis {\rm{\it{et.al.}}}\,\cite{Isken:2017dkw}&$-0.041(9)$&$-0.088(7)$&$0$&$-0.068(11)$\\
Kubis {\rm{\it{et.al.}}}\,\cite{Isken:2017dkw}&$-0.148(18)$&$-0.082(14)$&$0$&$-0.086(22)$\\
\hline     
\end{tabular}
\caption{Experimental and theoretical determinations of the Dalitz slope parameters associated to $\eta^{\prime}\to\eta\pi^{0}\pi^{0}$ (up panel) and $\eta^{\prime}\to\eta\pi^{+}\pi^{-}$(down panel) decays. For the parameter values of Ref.\,\cite{Isken:2017dkw}, we only show the results obtained with one of the solutions presented in the paper; the other solution leads to similar values. Moreover, while systematic uncertainties are ascribed to the parameter values of this reference, here only the statistical uncertainties are shown.}
\label{DalitzParamStatus}
\end{table}
 
\section{Formalism}\label{section3}

The Large-$N_{C}$ ChPT is the effective field theory of QCD in the chiral and large-$N_{C}$ (number of colour) limits.
Within this framework the singlet field $\eta_{1}$, absent in $SU(3)$ ChPT, becomes a new degree of freedom of the effective theory i.e.\,the ninth Goldstone boson associated with the spontaneous breaking of $U(3)_{L}\times U(3)_{R}\to U(3)_{V}$.
The theory is usually called $U(3)$ ChPT and the same counting is assigned to the squared momenta $p^{2}$, to the light quark masses $m_{q}$ and to the inverse of the number of colors $1/N_{C}$, giving rise to a combined triple expansion in $\delta\sim p^{2}\sim m_{q}\sim1/N_{C}$.
With the use of the single power-counting parameter $\delta$ the expansion of the effective Lagrangian is given by \cite{Kaiser:2000gs} 
\begin{equation}
\mathcal{L}_{\rm{eff}}=\sum_{i=0}^{\infty}\mathcal{L}^{\delta^{i}}\,.
\end{equation}
In this notation, the contributions $\mathcal{L}^{\delta^{i}}$ are of order $\mathcal{O}(\delta^{i})$.
For example, the leading and next-to-leading order Lagrangians, $\mathcal{L}^{\delta^{0}}$ and $\mathcal{L}^{\delta^{1}}$, are of order $\mathcal{O}(\delta^{0})=\mathcal{O}(1)$ and $\mathcal{O}(\delta)$, respectively. 
One interesting feature of the combined power counting is that meson loop diagrams with vertices from $\mathcal{L}^{\delta^{0}}$ count as $\mathcal{O}(\delta^{2})$. They are $1/N_{C}$ suppressed with respect to the NLO diagrams from $\mathcal{L}^{\delta^{1}}$.
Therefore at $\mathcal{O}(\delta)$ only tree level diagrams from $\mathcal{L}^{\delta^{0}}$ and $\mathcal{L}^{\delta^{1}}$ need to be taken into account. The loop diagrams are higher order in the counting. 
In our analysis, we include the one-loop corrections for the first time for 
describing $\eta' \to \eta \pi \pi$ and work at $\mathcal{O}(\delta^2)$ in order to match the high level of precision of the experimental measurements. We will study the impact of such inclusion. 

Contrary to $SU(3)$ or $U(3)$ ChPT, the inclusion of resonances into the description of the effective field theory spoils the power counting.
This is why, while RChT includes ChPT at $\mathcal{O}(p^{2})=\mathcal{O}(\delta^{0})$, it does not include the next order either in the chiral or in the combined triple expansion scheme. It rather substitutes it by a Lagrangian accounting for the interactions between pseudoscalar mesons and resonances.
In fact, resonance exchanges saturate 
the higher order local contributions of the effective expansion. 
As a consequence, 
the systematic effective field theory with a rigorous power counting scheme is lost. 
It is replaced by a model based on the large-$N_{C}$ limit as a guiding principle.
This model allows in general for reasonable descriptions of QCD processes at low-energies.

The relevant Lagrangian for our work including scalar resonances is written as \cite{Ecker:1988te,Guo:2011pa}
\begin{equation}
\mathcal{L}_{\rm{R\chi T}}=\mathcal{L}_{\chi}^{(2)}+\mathcal{L}_{S}+\mathcal{L}_{\rm{kin}}^{S}+\cdots+\mathcal{L}_{\Lambda}\,,
\label{RChTLag}
\end{equation}
where the dots denote operators with three or more resonance fields which we neglect in this analysis. 
The first term on the right-hand side of Eq.~(\ref{RChTLag}) is the chiral Lagrangian at leading order in $U(3)$ ChPT, it is $\mathcal{O}(\delta^{0})$ and reads
\begin{eqnarray}
\mathcal{L}_{\chi}^{(2)}=\frac{F^{2}}{4}\langle u_{\mu}u^{\mu}\rangle+\frac{F^{2}}{4}\langle\chi_{+}\rangle+\frac{F^{2}}{3}m_{0}^{2}\ln^{2}\det u\,,
\label{leadingorder}
\end{eqnarray}
where the last term accounts for the $U(1)_{A}$ anomaly contribution to the pseudoscalar singlet $\eta_{1}$ of mass $m_{0}^{2}$. {\color{blue} 
}
The chiral building blocks are defined by
\begin{eqnarray}
&&\nonumber U=u^{2}=e^{i\frac{\sqrt{2}\Phi}{F}}\,,\quad\chi=2B(s+ip)\,,\quad\chi_{\pm}=u^{\dagger}\chi u^{\dagger}\pm u\chi^{\dagger}u\,,\\[1ex]
&&u_{\mu}=iu^{\dagger}D_{\mu}Uu^{\dagger}\,,\quad D_{\mu}U=\partial_{\mu}U-i(v_{\mu}+a_{\mu})U+iU(v_{\mu}-a_{\mu})\,.
\end{eqnarray} 
$F$ is the axial 
decay constant of the pseudo-Goldstone bosons in the chiral and Large-$N_{C}$ limits. $s,p,v_{\mu},a_{\mu}$ stand for external fields and the parameter $B$ is related to the quark condensate $\langle0|\bar{q}^{i}q^{j}|0\rangle=-F^{2}B\delta^{ij}$.
In the absence of external fields, i.e. $v^{\mu}=a^{\mu}=p=0$, we have $s={\rm{diag}}(m_{u},m_{d},m_{s})$, with $m_{q}$ the light quark masses, encoding the explicit chiral symmetry breaking.
The pseudo-Goldstone bosons are collected in the matrix 

\begin{eqnarray}
\Phi=\begin{pmatrix}\frac{1}{\sqrt{2}}\pi_{3}+\frac{1}{\sqrt{6}}\eta_{8}+\frac{1}{\sqrt{3}}\eta_{1}&\pi^{+}&K^{+}\cr
\pi^{-}&-\frac{1}{\sqrt{2}}\pi_{3}+\frac{1}{\sqrt{6}}\eta_{8}+\frac{1}{\sqrt{3}}\eta_{1}&K^{0}\cr
K^{-}&\bar{K}^{0}&-\frac{2}{\sqrt{6}}\eta_{8}+\frac{1}{\sqrt{3}}\eta_{1}
\end{pmatrix}\,.
\label{matrix}
\end{eqnarray}
Expanding $u(\Phi)$ in Eq.\,(\ref{leadingorder}) in terms of $\Phi$, one obtains the kinetic term plus a tower of derivative interactions increasing in an even number of pseudoscalar fields. This together with the quadratic mass term plus additional interactions proportional to the quark masses gives
\begin{eqnarray}\nonumber
&&\mathcal{L}_{\chi}^{(2)}=\frac{1}{2}\langle\partial_{\mu}\Phi\partial^{\mu}\Phi\rangle+\frac{1}{12F^{2}}\langle\left(\Phi(\partial_{\mu}\Phi)-(\partial_{\mu}\Phi)\Phi\right)\left(\Phi(\partial^{\mu}\Phi)-(\partial^{\mu}\Phi)\Phi\right)\rangle\\
&&+B_{0}\left\{-\langle\mathcal{M}\Phi^{2}\rangle+ \left(\frac{1}{6F^{2}}\right)\langle\mathcal{M}\Phi^{4}\rangle\right\}+\mathcal{O}\left(\frac{\Phi^{6}}{F^{4}}\right)\,.
\label{leadingorderexpansion}
\end{eqnarray}
The mass term $\mathcal{M}$ in the previous expression induces a $\eta_{8}$-$\eta_{1}$ mixing.
At leading order, the physical $\eta$ and $\eta^{\prime}$ mass eigenstates are then obtained after diagonalising the mass matrix with the following orthogonal transformation:
\begin{eqnarray}
\begin{pmatrix}
\eta\cr\eta^{\prime}
\end{pmatrix}
=
\begin{pmatrix}
\cos\theta&-\sin\theta\cr
\sin\theta&\cos\theta
\end{pmatrix}
\begin{pmatrix}
\eta_{8}\cr\eta_{1}
\end{pmatrix}\,,
\end{eqnarray}
where the mixing angle $\theta$ stands for the degree of admixture. 

When higher order terms are to be considered, the transformations accounting for the mixing are more involved 
since not only the mass terms do mix but also the kinetic ones.
In this work, we treat the $\eta$-$\eta^{\prime}$ mixing as described 
in section III of Ref.\,\cite{Guo:2011pa} 
and refer the reader to this reference for more details and the expression of the higher order terms.
In particular, we use 
the next-to-leading order 
expression given in Eq.\,(15) of Ref.\,\cite{Guo:2011pa} based on the one-angle scheme approximation\footnote{The so-called two-step mixing procedure makes the single mixing angle at the lowest-order to be split in two mixing angles at next-to-leading order. 
One can thus express their associated parameters either in the form of two mixing angles and two decay constants or one mixing angle, the one entering at the lowest-order $\theta$, and three wave-function renormalization corrections appearing only at NLO. 
We follow the second option in this work.
See also Refs.\,\cite{Guo:2015xva,Bickert:2016fgy} for recent studies of the mixing phenomenon at higher orders.}.

The second term in Eq.~(\ref{RChTLag}) corresponds to the interaction terms of two pseudo-Goldstone-bosons with one resonance and is given by
\begin{eqnarray}
\mathcal{L}_{S}=c_{d}\langle S_{8}u_{\mu}u^{\mu}\rangle+c_{m}\langle S_{8}\chi_{+}\rangle+\tilde{c_{d}}S_{1}\langle u_{\mu}u^{\mu}\rangle+\tilde{c_{d}}S_{1}\langle \chi_{+}\rangle\,.
\label{resonanceLag}
\end{eqnarray}
The resonance state building blocks are 
\begin{eqnarray}
&&S_{8}=\begin{pmatrix}\frac{1}{\sqrt{2}}a_{0}^{0}+\frac{1}{\sqrt{6}}\sigma_{8}&a_{0}^{+}&\kappa^{+}\cr
a_{0}^{-}&-\frac{1}{\sqrt{2}}a_{0}^{0}+\frac{1}{\sqrt{6}}\sigma_{8}&\kappa^{0}\cr
\kappa^{-}&\bar{\kappa}^{0}&-\frac{2}{\sqrt{6}}\sigma_{8}\end{pmatrix}\,,\\
&&S_{1}=\sigma_{1}\,.
\label{scalarres}
\end{eqnarray}
Similarly to Eq.\,(\ref{leadingorderexpansion}), expanding the Lagrangian in Eq.~(\ref{resonanceLag}) in terms of $\Phi$ we get
\begin{eqnarray}
\nonumber\mathcal{L}_{S}&=&\frac{2c_{d}}{F^{2}}\langle S_{8}(\partial_{\mu}\Phi)(\partial^{\mu}\Phi)\rangle+4B_{0}c_{m}[\langle S_{8}\mathcal{M}\rangle-\frac{1}{4F^{2}}\langle S_{8}(\Phi^{2}\mathcal{M}+\mathcal{M}\Phi^{2}+2\Phi \mathcal{M}\Phi)\rangle]\\
\nonumber&&+\frac{2\tilde{c}_{d}}{F^{2}}S_{1}\langle(\partial_{\mu}\Phi)(\partial^{\mu}\Phi)\rangle+4B_{0}\tilde{c}_{m}S_{1}[\langle \mathcal{M}\rangle-\frac{1}{4F^{2}}\langle(\Phi^{2}\mathcal{M}+\mathcal{M}\Phi^{2}+2\Phi \mathcal{M}\Phi)\rangle]\,,\\
\end{eqnarray}
where we have used $\chi=2B_{0}\mathcal{M}$ and neglected other external fields $(v^{\mu}=a^{\mu}=p=0)$.
Note that the interaction terms proportional to $c_{d}$ and $\tilde{c}_{d}$ enter only with derivatives while $c_{m}$ and $\tilde{c}_{m}$ are proportional to the quark masses.

The third term in Eq.~(\ref{RChTLag}) contains the kinetic term,
\begin{eqnarray}
\mathcal{L}_{\rm{kin}}^{S}=\frac{1}{2}\langle\bigtriangledown^{\mu}S_{8}\bigtriangledown_{\mu}S_{8}-M_{S_{8}}^{2}S_{8}^{2}\rangle+\frac{1}{2}\langle\partial^{\mu}S_{1}\partial_{\mu}S_{1}-M_{S_{1}}^{2}S_{1}^{2}\rangle\,,
\end{eqnarray}
where
\begin{equation}
\bigtriangledown_{\mu}S=\partial_{\mu}S+[\Gamma_{\mu},S]\,,\quad \Gamma_{\mu}=\frac{1}{2}\{u^{+}[\partial_{\mu}-ir_{\mu}]u+u[\partial_{\mu}-i\ell_{\mu}]u^{\dagger}\}\,.
\end{equation}

Finally, the last term in Eq.~(\ref{RChTLag}) is a local operator of $\mathcal{O}(\delta)$, influencing only the singlet sector. 
It cannot be generated from the exchange of the scalar resonance discussed above. It is obtained by integrating out pseudoscalar resonances instead.
It only involves pseudo-Goldstone bosons and reads 
\begin{eqnarray}
\mathcal{L}_{\Lambda}=\Lambda_{1}\frac{F^{2}}{12}D_{\mu}\psi D^{\mu}\psi-i\Lambda_{2}\frac{F^{2}}{12}\langle U^{\dagger}\chi-\chi^{\dagger}U\rangle\,,
\label{LagLambda}
\end{eqnarray} 
where
\begin{eqnarray}
\psi=-i\ln\det U\,,\quad D_{\mu}\psi=\partial_{\mu}\psi-2\langle a_{\mu}\rangle\,,
\end{eqnarray}
with $a_{\mu}=(r_{\mu}-l_{\mu})/2$.

\section{Structure of the decay amplitude}\label{section4}

\begin{figure}[h!]
\begin{center}
\includegraphics[scale=0.375]{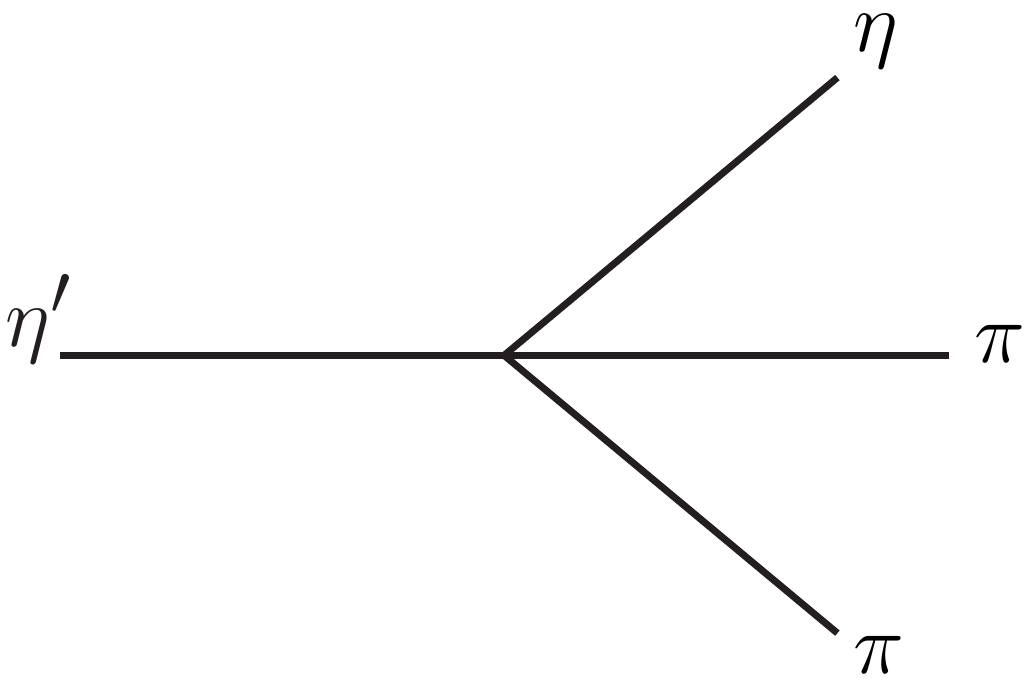}\qquad\includegraphics[scale=0.325]{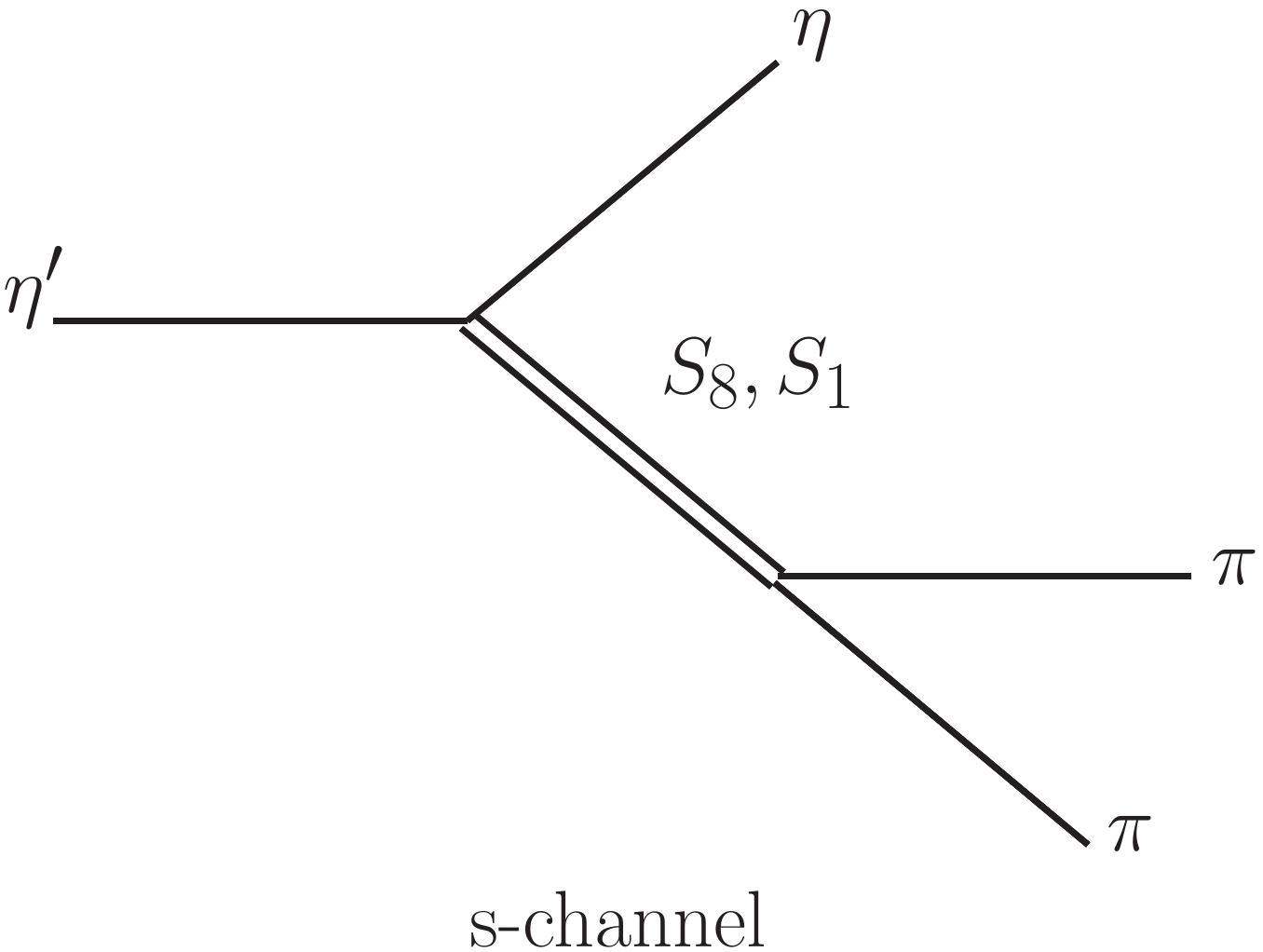}\qquad\includegraphics[scale=0.325]{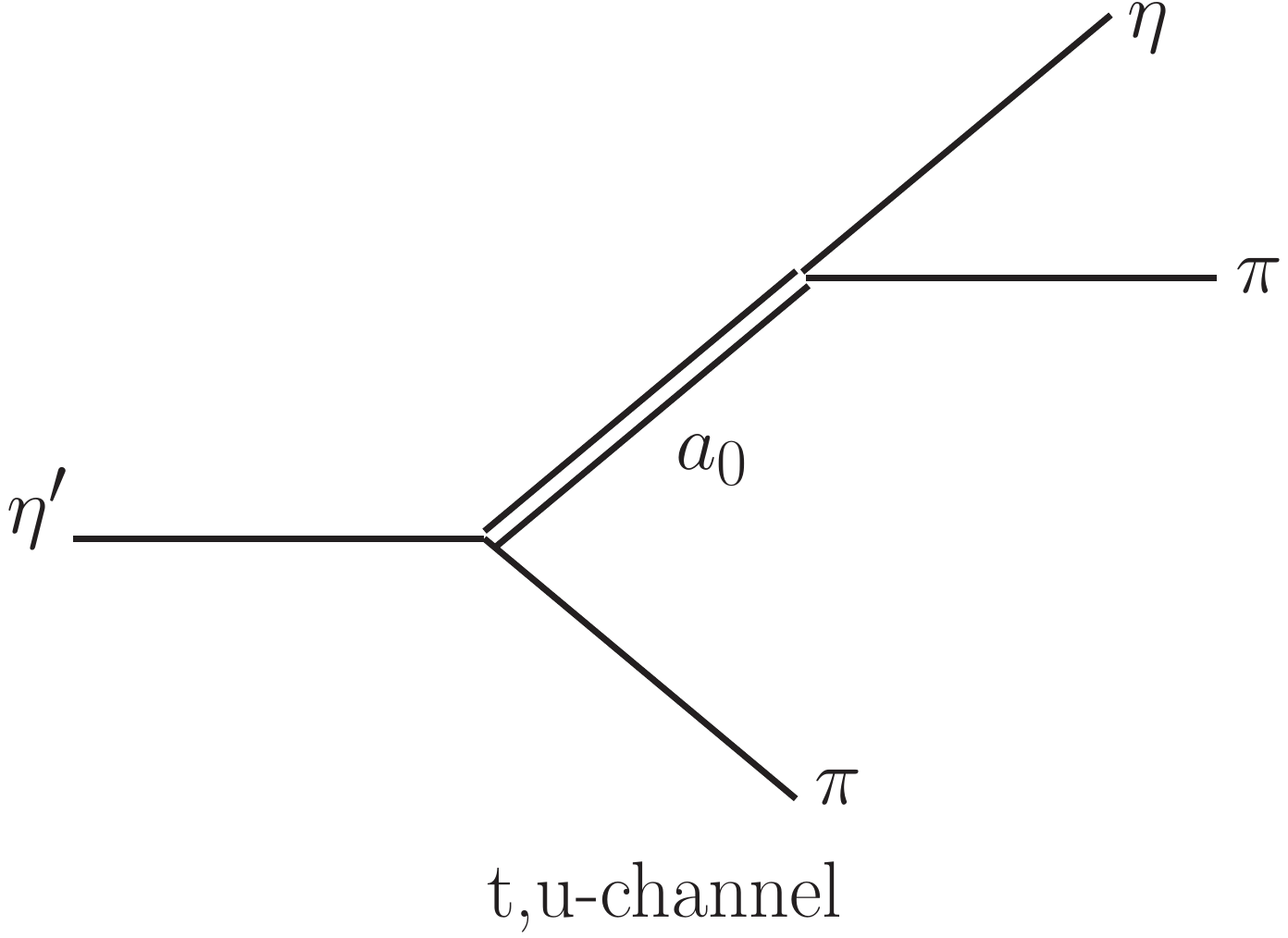}\\[3ex]
\includegraphics[scale=0.4]{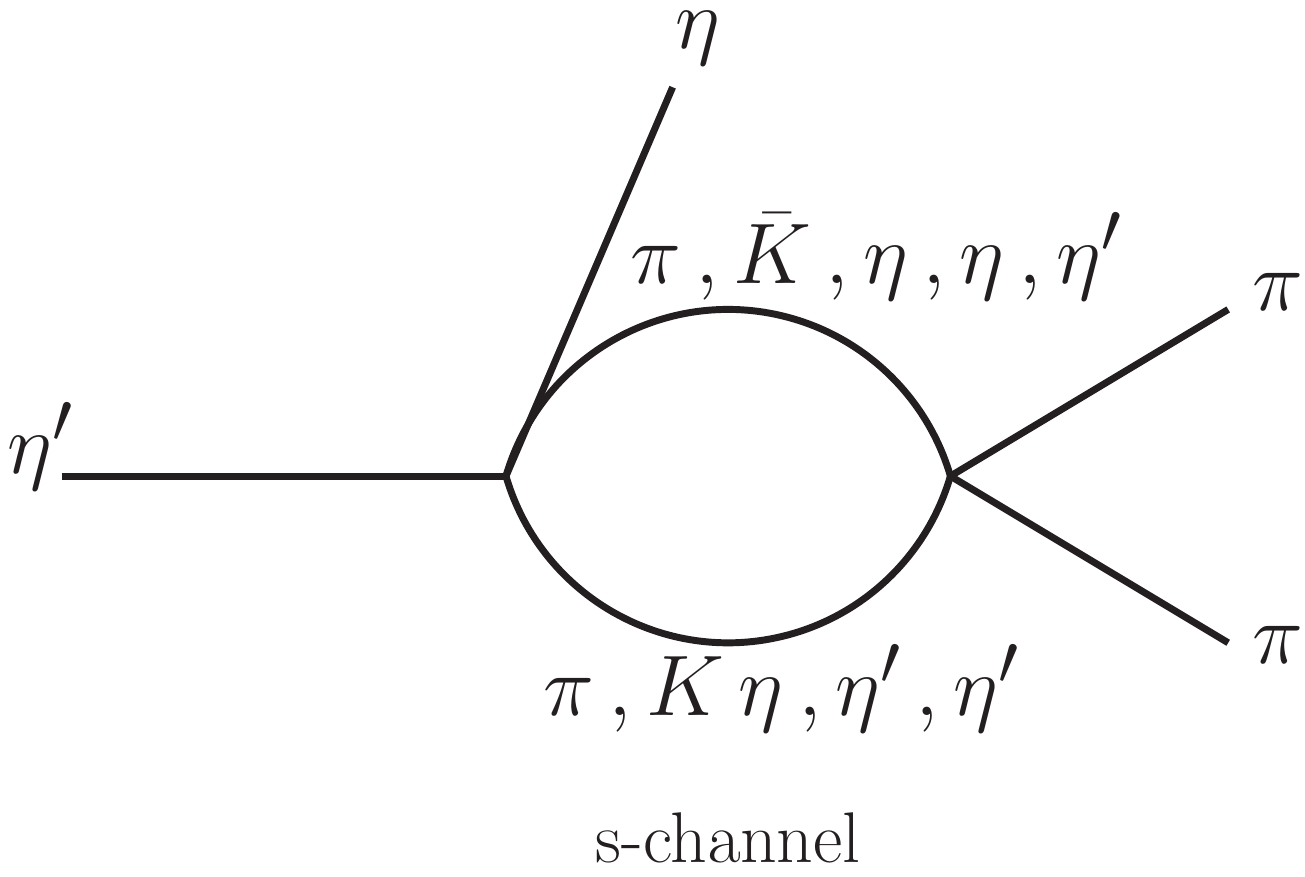}\quad\includegraphics[scale=0.4]{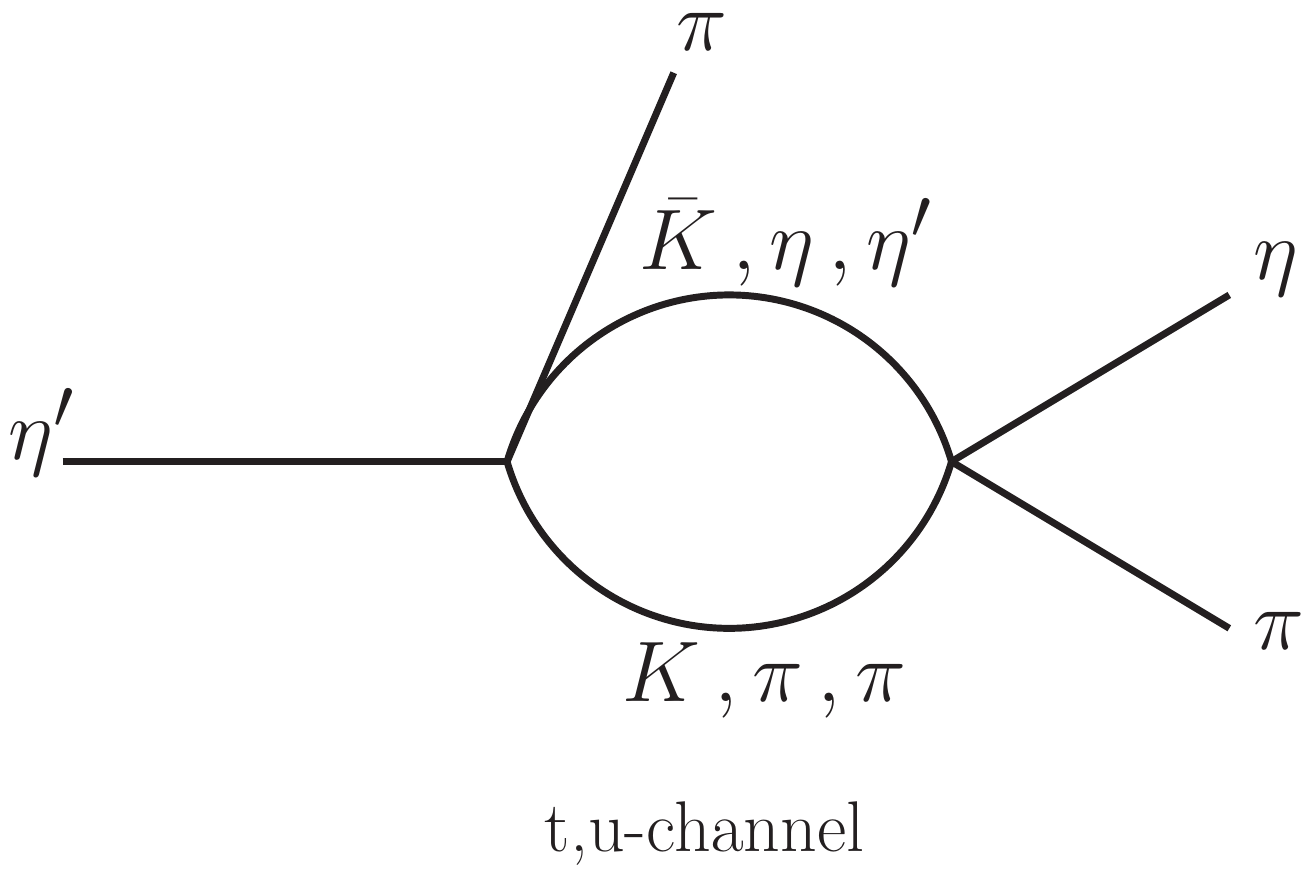}\quad\includegraphics[scale=0.4]{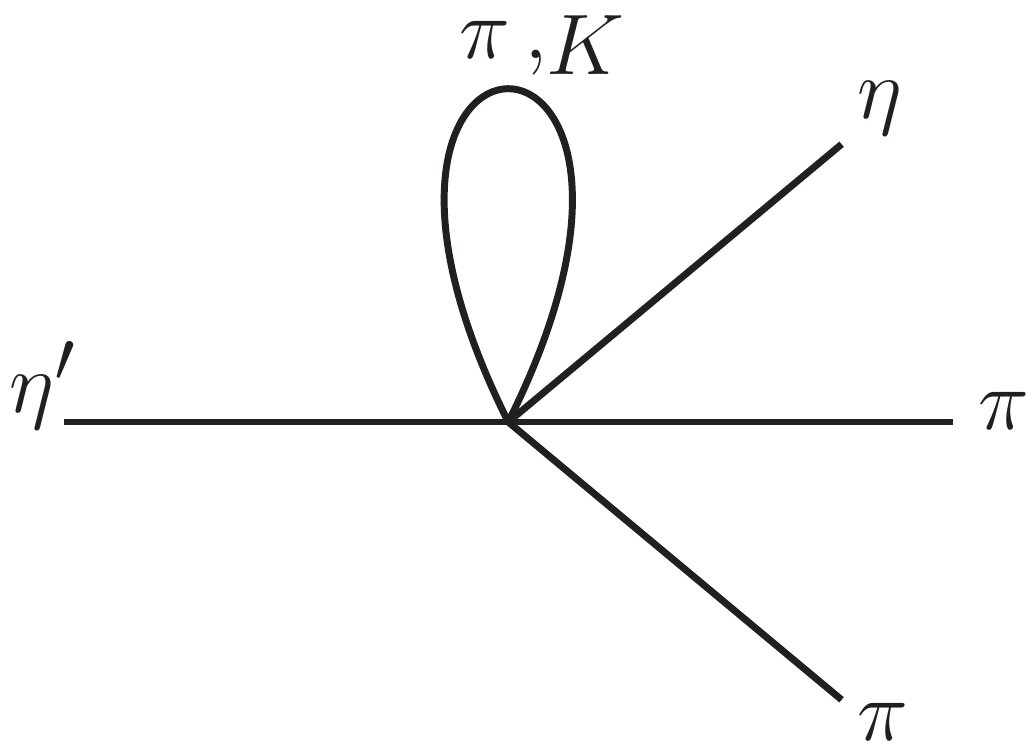}
\caption{\label{Diagrams}Diagrams contributing to the $\eta^{\prime}\to\eta\pi\pi$ decay amplitude within $U(3)$ Large-N$_{C}$ ChPT at one-loop including resonance states. From left to right and top to bottom we have: $i)$ lowest-order; $ii)$ exchange of scalar resonances in the $s,t$ and $u$ channels; $iii)$ one-loop corrections in the $s,t$ and $u$-channels; $iv)$ tadpole contributions.}
\end{center}
\end{figure}

The calculation of the $\eta^{\prime}\to\eta\pi\pi$ decay amplitude includes the diagrams depicted in Fig.\,\ref{Diagrams} and can be gathered as
\begin{eqnarray}
\mathcal{M}(s,t,u)=\mathcal{M}^{(2)}+\mathcal{M}^{\rm{Res}}+\mathcal{M}^{\rm{Loop}}+\mathcal{M}^{\Lambda}\,.
\label{structureamplitude}
\end{eqnarray}
$\mathcal{M}^{(2)}$ is the amplitude at lowest order in ChPT. $\mathcal{M}^{\rm{Res}}$ represents the amplitude involving the exchange of resonances, $\mathcal{M}^{\rm{Loop}}$ the loop contributions,  and $\mathcal{M}^{\Lambda}$ the $\Lambda$ term\footnote{The graphs corresponding to the mass and wave-function renormalizations are not shown but have also been included in the calculation.
In addition, the axial decay constant $F$ is also modified to one loop giving rise to higher order contributions.
We provide our expressions in terms of a single decay constant, that we have chosen to be the pion decay constant $F_{\pi}$.
The relation between $F$ and $F_{\pi}$ can be found in Eq.\,(C2) of Ref.\,\cite{Guo:2011pa}.}. 
The lowest order ChPT contribution is constant  
\begin{eqnarray}
\mathcal{M}^{(2)}=\left(2\sqrt{2}\cos(2\theta)-\sin(2\theta)\right)\frac{m_{\pi}^{2}}{6F_{\pi}^{2}}\,,
\label{LO}
\end{eqnarray}
where isospin symmetry $\left( m_u = m_d = \hat{m} \equiv \frac{m_u + m_d}{2} \right)$ has been assumed. 
The origin of this term stems entirely from the interactions proportional to quark masses appearing in the last term of Eq.\,(\ref{leadingorderexpansion}) i.e.\,no derivative interactions contribute at this order.
It is thus chirally suppressed explaining the smallness of this contribution.

The (zero-width) resonance exchange contribution $\mathcal{M}^{\rm{Res}}$ reads
\begin{eqnarray}
&&\mathcal{M}^{\rm{Res}}(s,t,u)=\frac{2}{9F_{\pi}^{4}}\left(\sqrt{2}\cos^{2}\theta-\cos\theta\sin\theta-\sqrt{2}\sin^{2}\theta\right)\times\Bigg\lbrace\frac{12c_{d}c_{m}m_{\pi}^{2}}{M_{S_{8}}^{2}}\left(m_{\pi}^{2}-m_{K}^{2}\right)\nonumber\\[1ex]
&&-\frac{\left(2c_{m}\left(m_{\pi}^{2}-4m_{K}^{2}\right)+3c_{d}\left(m_{\eta}^{2}+m_{\eta^{\prime}}^{2}-s\right)\right)\left(2c_{m}m_{\pi}^{2}+c_{d}\left(s-2m_{\pi}^{2}\right)\right)}{M_{S_{8}}^{2}-s}\nonumber\\[1ex]
&&-\frac{24\tilde{c}_{m}\left(m_{K}^{2}-m_{\pi}^{2}\right)\left(2\tilde{c}_{m}m_{\pi}^{2}+\tilde{c}_{d}\left(s-2m_{\pi}^{2}\right)\right)}{M_{S_{1}}^{2}-s}\nonumber\\[1ex]
&&+\frac{3\left(4c_{m}^{2}m_{\pi}^{4}-2c_{d}c_{m}m_{\pi}^{2}\left(m_{\eta}^{2}+m_{\eta^{\prime}}^{2}+2m_{\pi}^{2}-2t\right)+c_{d}^{2}\left(m_{\eta}^{2}+m_{\pi}^{2}-t\right)\left(m_{\eta^{\prime}}^{2}+m_{\pi}^{2}-t\right)\right)}{M_{a_{0}}^{2}-t}\nonumber\\[1ex]
&&+(t\leftrightarrow u)\Bigg\rbrace\,.
\end{eqnarray}

Unitarity loop corrections to the decay amplitude can occur either in the $s$ or in the $t$ and $u$ channels.
In the $s$-channel, the meson pairs $\pi\pi,K\bar{K},\eta\eta,\eta\eta^{\prime}$ and $\eta^{\prime}\eta^{\prime}$ enter within the loop while $K\bar{K},\pi\eta$ and $\pi\eta^{\prime}$ pairs contribute in the $t$ and $u$-channels.
Tadpole contributions with $\pi$ and $K$ also appear.
The complete expression for the unitarity loop corrections $\mathcal{M}^{\rm{Loop}}$, including the tadpoles
can be found in appendix \ref{loopcorrections}.

Finally, the term from $\mathcal{L}_{\Lambda}$ is constant 
\begin{eqnarray}
\mathcal{M}^{\Lambda}=\frac{m_{\pi}^{{2}}\Lambda_{2}}{3F_{\pi}^{2}}\left(\sqrt{2}\cos^{2}\theta-4\cos\theta\sin\theta-\sqrt{2}\sin^{2}\theta\right)\,.
\label{Lambdaterm}
\end{eqnarray} 

Moreover, there is a contribution from the mixing~\cite{Guo:2011pa}
\begin{eqnarray}
\mathcal{M}^{\rm{mixing}}=-\frac{m_{\pi}^{2}}{6F_{\pi}^{2}}\Big[8\sqrt{2}\cos\theta\sin\theta\sin\theta_{\delta}+3\delta_{K}+2\sin\theta_{\delta}\left(\cos^{2}\theta-\sin^{2}\theta\right)\Big]\,,
\label{mixingterm}
\end{eqnarray}
with $\sin\theta_{\delta}\simeq0$ \cite{Guo:2011pa} and 
\begin{eqnarray}
\delta_{K}=\frac{\sin\theta\cos\theta}{F_{\pi}^{2}}\mu_{K}+\frac{1}{3F_{\pi}^{2}M_{S_{8}}^{2}}\Big[16c_{d}c_{m}(m_{K}^{2}-m_{\pi}^{2})(\sqrt{2}\sin^{2}\theta+\cos\theta\sin\theta-\sqrt{2}\cos^{2}\theta)\Big]\,.\nonumber\\
\end{eqnarray}
$\mu_{K}$ is defined in Eq.\,(\ref{mutadpole}).
Although the mixing contribution is omitted all along the formulae presented in the next sections, we have included it in our analysis.
Note that both the $\Lambda$ and mixing terms, Eqs.\,(\ref{Lambdaterm}) and (\ref{mixingterm}), are constant and their contributions are found to be small. Indeed they are both proportional to $m_{\pi}^{2}$ and hence chirally suppressed. 

\section{Partial waves and unitarisation of the amplitude}\label{section5}

Within the ChPT framework described in sections \ref{section3} and \ref{section4}, the loop contributions 
are unitary order-by-order in the perturbative expansion.
Nearby the resonance region, however, the perturbative chiral amplitudes violate unitarity.
This inherent limitation of the theory is addressed by using a unitarization procedure.
For this, we rely on the $N/D$ method.
A detailed account of this method can be found in Refs.\,\cite{Guo:2011pa,Oller:2000ma,Oller:1998zr,Oller:2000fj}. 
In the following, we recall the main features of this approach that are relevant for our analysis. 
Another method relying on the Khuri-Treiman framework has been followed in Ref.~\cite{Isken:2017dkw}. 

We start by writing the most general unitarity relation for $\eta^{\prime}(p_{\eta^{\prime}})\to\eta(p_{\eta})\pi(p_{1})\pi(p_{2})$ decay
\begin{eqnarray}
{\rm{Im}}\mathcal{M}_{\eta^{\prime}\to\eta\pi\pi}=\frac{1}{2}\sum_{n}\left(2\pi\right)^{4}\delta^{4}\left(p_{\eta}+p_{1}+p_{2}-p_{n}\right)\mathcal{T}_{n\to\eta\pi\pi}^{*}\mathcal{M}_{\eta^{\prime}\to n}\,. 
\label{unitarityrelation0}
\end{eqnarray}
$\mathcal{M}_{\eta^{\prime}\to n}$ denotes the $\eta^{\prime}\to n$ decay amplitude and $\mathcal{T}_{n\to\eta\pi\pi}$ the transition $n\to\eta\pi\pi$.  
If for simplicity we restrict $n \leq 3$ 
then the scattering matrix element $\mathcal{T}_{n\to\eta\pi\pi}$ contains pure 3-body $\to\eta\pi\pi$ contributions as well as two-body final-state interactions.
The three-body final-state interactions 
are suppressed by power counting and 
phase-space compared to the two-body final-state interactions. 
In our analysis, we include only the dominant two-body final-state interactions. 
In a three-body decay, two-body final-state interactions can occur either by means of a rescattering where two out of the three final state particles rescatter an arbitrary number of times in each of the two-particle channels considering the third particle as a spectator or by interactions among one of the two rescattering particles together with the third spectating-particle.
While the former will be fully accounted for in our study only portions of the later will be incorporated. 

In the following, we limit the sum over $n$ in Eq.\,(\ref{unitarityrelation0}) to $\pi\pi$ and $\pi\eta$ intermediate states.
The unitarity condition for the $\eta^{\prime}\to\eta\pi\pi$ decay in terms of the $\pi\pi$ and $\pi\eta$ scattering amplitudes can be written as
\begin{eqnarray}
\nonumber&&{\rm{Im}}\mathcal{M}^{I}_{\eta^{\prime}\to\eta\pi\pi}(s,t,u)=\\[1ex]
\nonumber&& \frac{1}{2(2\pi)^{2}\mathcal{N}}\int\frac{dq_{b}^{3}}{2q_{b}^{0}}\frac{dq_{c}^{3}}{2q_{c}^{0}}\delta^{4}(q_{b}+q_{c}-p_{1}-p_{2})\mathcal{T}_{\pi\pi\to\pi\pi}^{I}(s,\theta_{s}^{\prime\prime})^{*}\mathcal{M}^{I}_{\eta^{\prime}\to\eta\pi\pi}(s,\theta_{s}^{\prime},\phi_{s}^{\prime})\\[1ex]
\nonumber&&+\frac{1}{2(2\pi)^{2}\mathcal{N}}\int\frac{dq_{a}^{3}}{2q_{a}^{0}}\frac{dq_{b}^{3}}{2q_{b}^{0}}\delta^{4}(q_{a}+q_{b}-p_{1}-p_{\eta})\mathcal{T}_{\pi\eta\to\pi\eta}^{I}(t,\theta_{t}^{\prime\prime})^{*}\mathcal{M}^{I}_{\eta^{\prime}\to\eta\pi\pi}(s,\theta_{t}^{\prime},\phi_{t}^{\prime})\\[1ex]
&&+\frac{1}{2(2\pi)^{2}\mathcal{N}}\int\frac{dq_{a}^{3}}{2q_{a}^{0}}\frac{dq_{c}^{3}}{2q_{c}^{0}}\delta^{4}(q_{a}+q_{c}-p_{2}-p_{\eta})\mathcal{T}_{\pi\eta\to\pi\eta}^{I}(u,\theta_{u}^{\prime\prime})^{*}\mathcal{M}^{I}_{\eta^{\prime}\to\eta\pi\pi}(u,\theta_{u}^{\prime},\phi_{u}^{\prime})\,.
\label{unitaritydecay}
\end{eqnarray}
The symmetry factor is $\mathcal{N}=2$ in case of identical $(\pi\pi)$ and $\mathcal{N}=1$ for distinguishable $(\pi\eta)$ particles. 
$\theta_{s,t,u}^{\prime}$ stands for the center-of-mass scattering angle between the initial and intermediate state and $\theta_{s,t,u}^{\prime\prime}$ denotes the center-of-mass scattering angle between the intermediate and final state.  

The decay and scattering amplitudes, $\mathcal{M}^{I}$ and $\mathcal{T}^{I}$, can be decomposed in partial waves with definite isospin $I$ and angular momentum $J$ through
\begin{eqnarray}
\mathcal{M}^{I}(s,\cos\theta^{\prime},\phi^{\prime})&=&\sum_{J}32\pi(2J+1)P_{J}(\cos\theta^{\prime})m^{IJ}(s)\,,\label{partialwaveM}\\[1ex]
\mathcal{T}^{I}(s,\cos\theta^{\prime\prime})&=&\sum_{J}16\pi\mathcal{N}(2J+1)P_{J}(\cos\theta^{\prime\prime})t^{IJ}(s)\,,
\label{partialwaveT}
\end{eqnarray}
where $P_{J}$ is the Legendre polynomial of the $J^{\rm{th}}$ degree.

Isospin conservation constrains the total isospin of the final state $\pi\pi$ and $\pi\eta$ pairs to be $I=0$ and $I=1$, respectively. We limit our study to the $S$ $(J=0)$ and $D$ $(J=2)$ waves for the $\pi\pi$ scattering and to the $S$-wave for the $\pi\eta$ scattering\footnote{The $P$-wave $(J=1)$ of the $\pi\eta$ scattering is found to be strongly suppressed \cite{Borasoy:2005du,Kubis:2009sb} and will not be considered in this work.}. 
Inserting Eqs.\,(\ref{partialwaveM}) and (\ref{partialwaveT}) in Eq.\,(\ref{unitaritydecay}) and integrating over the momentum using the relation
\begin{equation}
\int d\Omega^{\prime}P_{J}(\cos\theta^{\prime\prime})P_{J^{\prime}}(\cos\theta^{\prime})=\frac{4\pi}{2J+1}\delta_{JJ^{\prime}}P_{J}(\cos\theta)\,,
\end{equation}
the following unitarity relations  for each partial-wave of the decay amplitude of definite isospin can be derived:
\begin{eqnarray}
\nonumber{\rm{Im}}\left(m^{00}_{\eta^{\prime}\to\eta\pi\pi}(s)\right)&=&\sigma_{\pi}(s) \left(t_{\pi\pi\to\pi\pi}^{00}(s)\right)^{*}m_{\eta^{\prime}\to\eta\pi\pi}^{00}(s)\theta(s-4m_{\pi}^{2})\,,\\[1ex]
\nonumber{\rm{Im}}\left(m^{02}_{\eta^{\prime}\to\eta\pi\pi}(s)\right)&=&\sigma_{\pi}(s) \left(t_{\pi\pi\to\pi\pi}^{02}(s)\right)^{*}m_{\eta^{\prime}\to\eta\pi\pi}^{02}(s)\theta(s-4m_{\pi}^{2})\,,\\[1ex]
\nonumber{\rm{Im}}\left(m^{10}_{\eta^{\prime}\to\eta\pi\pi}(t)\right)&=&\nonumber\frac{\lambda^{1/2}(t,m_{\pi}^{2},m_{\eta}^{2})}{t}\left( t_{\pi\eta\to\pi\eta}^{10}(t)\right)^{*}m_{\eta^{\prime}\to\eta\pi\pi}^{10}(t)\theta(t-(m_{\pi}+m_{\eta})^{2})\,,\\[1ex]
{\rm{Im}}\left(m^{10}_{\eta^{\prime}\to\eta\pi\pi}(u)\right)&=&\frac{\lambda^{1/2}(u,m_{\pi}^{2},m_{\eta}^{2})}{u}
\left(t_{\pi\eta\to\pi\eta}^{10}(u)\right)^{*}m_{\eta^{\prime}\to\eta\pi\pi}^{10}(u)\theta(u-(m_{\pi}+m_{\eta})^{2})\,,\nonumber\\
\label{unitarityrelation}
\end{eqnarray}
where
\begin{equation}
\sigma_{\pi}(s)=\sqrt{1-\frac{4m_{\pi}^{2}}{s}}\,.
\end{equation}

\subsection{$\pi\pi$ final-state interactions}\label{section51}

Let us consider first the unitarity relation in the $s$-channel for illustrating the $N/D$ method applied to the $\eta^{\prime}\to\eta\pi\pi$ decay.
For a well-defined isospin $I$ and angular momentum $J$ from Eq.\,(\ref{unitarityrelation}) we have
\begin{eqnarray}
{\rm{Im}}\left(m_{\eta^{\prime}\to\eta\pi\pi}^{IJ}(s)\right)=\sigma_{\pi}(s) \left(t_{\pi\pi}^{IJ}(s)\right)^{*}m_{\eta^{\prime}\to\eta\pi\pi}^{IJ}(s)\theta(s-4m_{\pi}^{2})\,.
\end{eqnarray}

A possible way to fulfill the previous equation is as follows.
We 
write the partial wave associated to the perturbative decay amplitude, Eq.\,(\ref{structureamplitude}), as
{\small
\begin{eqnarray}
m^{IJ,\,{\rm{pert}}}_{\eta^{\prime}\to\eta\pi\pi}(s)=\left( m^{IJ}_{\eta^{\prime}\to\eta\pi\pi}(s)\right)^{(2)}+{\rm{Res+Loop+\Lambda}}-16\pi \left(m^{IJ}_{\eta^{\prime}\to\eta\pi\pi}(s)\right)^{(2)}g_{\pi\pi}(s) \left(t^{IJ}_{\pi\pi}(s)\right)^{(2)}\,, \nonumber\\
\label{Amplitude}
\end{eqnarray}}
where \textquotedblleft${\rm{pert}}$\textquotedblright\,stands for perturbative. The amplitudes with superscript $(2)$ correspond to 
the tree-level amplitudes, Res to the resonance exchanges in the $s,t$ and $u$ channels, Loop to the loop contributions in the $t$ and $u$ channels and finally $\Lambda$ represents the contribution from the $\Lambda$ term. 
The function $g_{\pi\pi}(s)$ entering the second term of Eq.\,(\ref{Amplitude}) accounts for the discontinuity along the right-hand cut 
due to the two-pion intermediate states 
\begin{eqnarray}
g_{\pi\pi}(s)=\frac{1}{16\pi^{2}}\left(a_{\pi\pi}(\mu)+\log\frac{m^{2}_{\pi}}{\mu^{2}}-\sigma_{\pi}(s)\log\frac{\sigma_{\pi}(s)-1}{\sigma_{\pi}(s)+1}\right)\,.
\label{gpipi}
\end{eqnarray}
$a_{\pi\pi}(\mu)\equiv a_{\pi\pi}$ is a subtraction constant that is not directly determined by the unitarization procedure but should be fixed from elsewhere. 
$g_{\pi\pi}(s)$ satisfies ${\rm{Im}}g_{\pi\pi}(s)=-\sigma_{\pi}(s)/16\pi$ \cite{Guo:2011pa}. It is related to the standard one-loop function given in Eq.\,(\ref{loopfunctionidenticalmasses}) through
\begin{eqnarray}
g_{\pi\pi}(s)=-B_{0}^{\rm{eq}}(s,m_{\pi})+a_{\pi\pi}^{\prime}(\mu)\,,
\label{eq:gpipi}
\end{eqnarray}
where $a_{\pi\pi}^{\prime}(\mu)$ is an arbitrary subtraction constant. 

The basic idea of the $N/D$ unitarisation method consists in collecting the left-and right-hand cuts in two different functions. 
A $N/D$ representation of  the decay amplitude, Eq.~(\ref{Amplitude}), can be obtained rewriting it as~\cite{Guo:2012yt} 
\begin{eqnarray}
m^{IJ}_{\eta^{\prime}\to\eta\pi\pi}(s)=[1+16\pi N^{IJ}_{\pi\pi}(s)g_{\pi\pi}(s)]^{-1}R^{IJ}_{\eta^{\prime}\to\eta\pi\pi}(s)\,,
\label{NDrepresentation}
\end{eqnarray}
where
\begin{equation}
\begin{array}{rcl}
N^{IJ}_{\pi\pi}(s)&=&\left(t^{IJ}_{\pi\pi}(s)\right)^{(2)+\rm{Res+Loop}}\,,\\[1ex]
R^{IJ}_{\eta^{\prime}\to\eta\pi\pi}(s)&=& \left(m^{IJ}_{\eta^{\prime}\to\eta\pi\pi}(s)\right)^{(2)+\rm{Res+Loop+\Lambda}}\,.\\[0.5ex]
\label{NDexpressions}
\end{array}
\end{equation}
$\left(t^{IJ}_{\pi\pi}(s)\right)^{(2)+\rm{Res+Loop}}$ and $\left(m^{IJ}_{\eta^{\prime}\to\eta\pi\pi}(s)\right)^{(2)+\rm{Res+Loop+\Lambda}}$ contain, respectively, the corresponding perturbative calculation of the partial wave $\pi\pi$ scattering and the ${\eta^{\prime}\to\eta\pi\pi}$ decay amplitudes. 
\begin{figure}[h!]
\begin{center}
\includegraphics[scale=0.375]{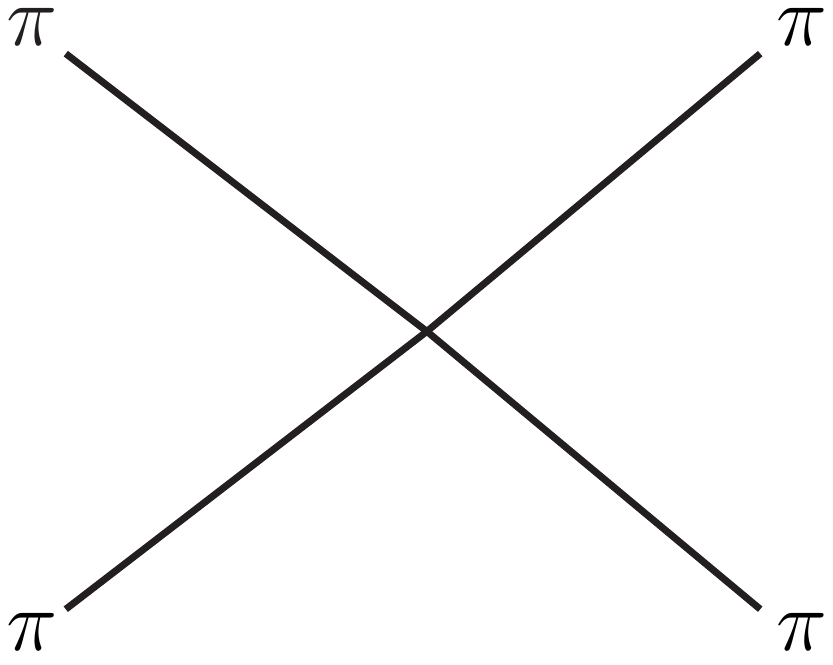}\qquad\includegraphics[scale=0.375]{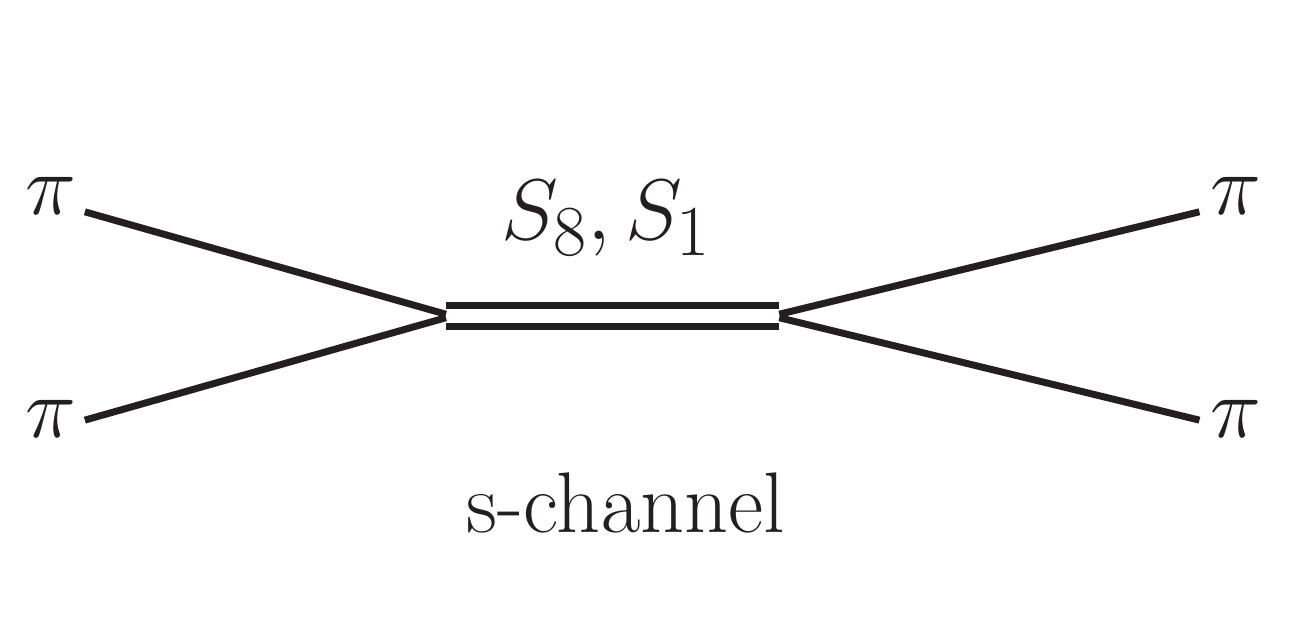}\\[3ex]
\includegraphics[scale=0.4]{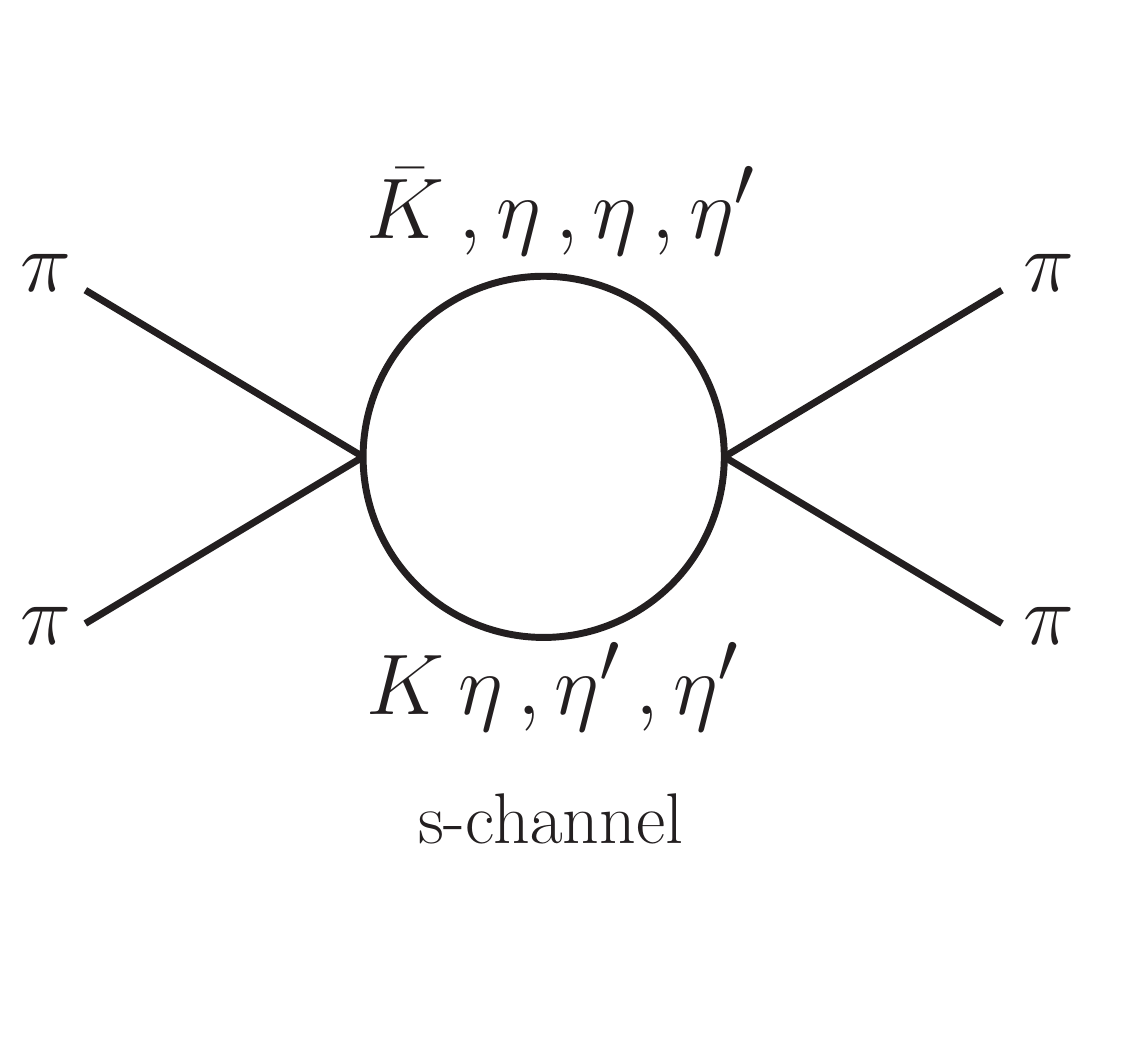}\qquad\includegraphics[scale=0.4]{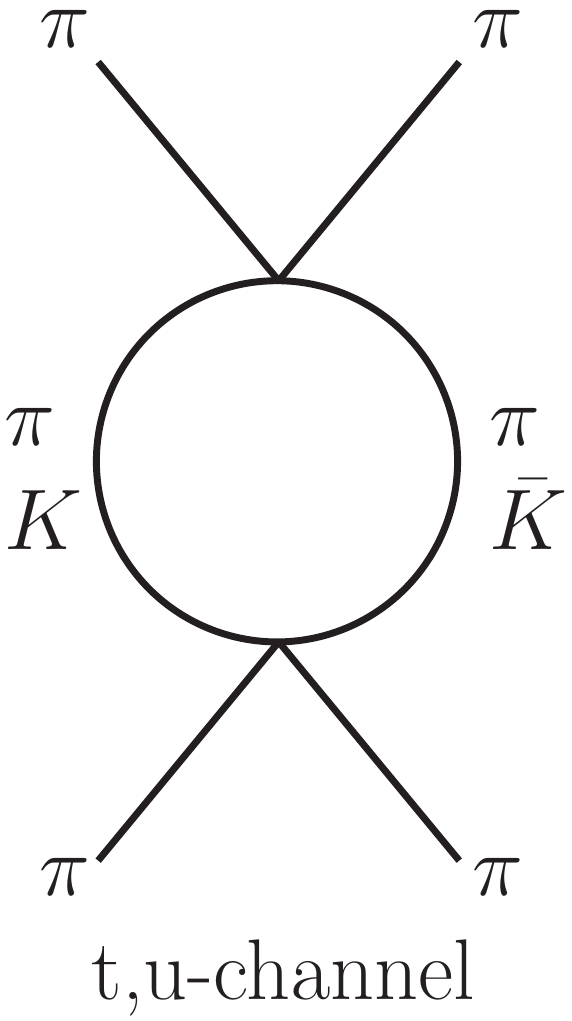}\qquad\includegraphics[scale=0.4]{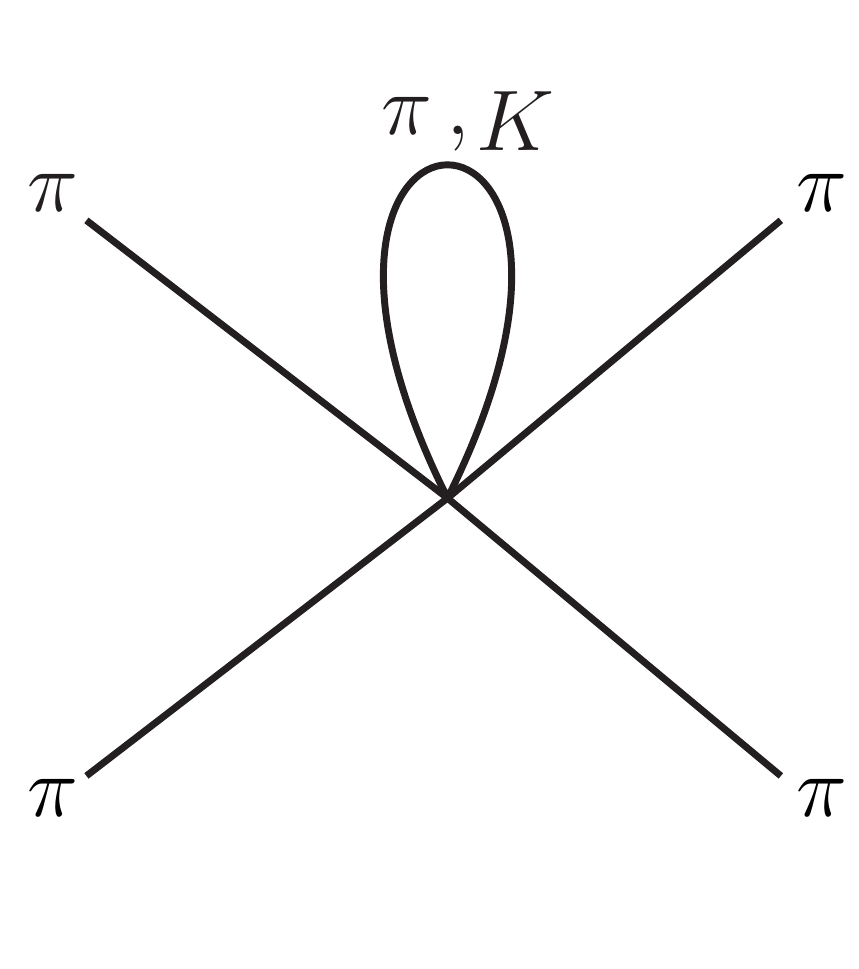}
\caption{\label{DiagramsScattering}Diagrams contributing to the $\pi\pi\to\pi\pi$ scattering amplitude within $U(3)$ Large-N$_{C}$ ChPT at one-loop including resonance states. From left to right and top to bottom we have: $i)$ lowest-order; $ii)$ exchange of scalar resonances in the $s,t$ and $u$ channels; $iii)$ one-loop corrections in the $s,t$ and $u$-channels; $iv)$ tadpole contributions.}
\end{center}
\end{figure}
As explained previously, in $\left(t^{IJ}_{\pi\pi}(s)\right)^{(2)+\rm{Res+Loop}}$, Eq.\,(\ref{NDexpressions}), the part with superscript $(2)$ corresponds to the tree level amplitude, the one with superscript Res to the exchange of resonances in the $s,t$ and $u$ channels and the Loop one denotes the loop contributions in the $t$ and $u$ channels as well as the inelastic loop contributions in the $s$-channel. 
The corresponding diagrams are depicted in Fig.\,\ref{DiagramsScattering} recovering Eq.\,(\ref{Amplitude}) up to higher orders. 
Note that a chiral expansion of Eq.\,(\ref{NDrepresentation}) leads to {\small
\begin{equation}
\begin{array}{rcl}
m^{IJ}_{\eta^{\prime}\to\eta\pi\pi}(s)&=&R^{IJ}_{\eta^{\prime}\to\eta\pi\pi}(s)-16\pi N^{IJ}_{\pi\pi}(s)g_{\pi\pi}(s)R^{IJ}_{\eta^{\prime}\to\eta\pi\pi}(s)+\cdots\\[2ex]
&=&\left(m^{IJ}_{\eta^{\prime}\to\eta\pi\pi}(s)\right)^{(2)+\rm{Res+Loop}+\Lambda}-16\pi \left(t^{IJ}_{\pi\pi}(s)\right)^{(2)} g_{\pi\pi}(s)
\left(m^{IJ}_{\eta^{\prime}\to\eta\pi\pi}(s)\right)^{(2)}+\cdots\,.
\end{array}
\end{equation}}
We would like to emphasize that the function $R(s)$ entering Eq.\,(\ref{NDexpressions}) does contain the left-hand cut (LHC) of the decay amplitude, perturbatively treated, but does not contain the $\pi\pi$ right-hand cut which is treated non-perturbatively using the function $g_{\pi\pi}(s)$. 

Thus, the absorptive part of the unitarised partial wave decay amplitude $m^{IJ}(s)$ as given in Eq.\,(\ref{NDrepresentation}) satisfies $s$-channel unitarity\footnote{In principle, the unitarity relation given in Eq.\,(\ref{unitarityfulfillment}) is valid up to the first inelastic threshold i.e. the $K\bar{K}$ threshold. 
However, there is a spurious contribution coming from the imaginary part of the $t$-and $u$-channel $\pi\eta$ left-hand cut loops. 
They sit on the elastic region and induces a violation of unitarity.
We have checked that this unitarity violation is numerically tiny and therefore acceptable for our purposes.
There is, moreover, another source of unitarity violation coming from the unitarization method itself.
Spurious singularities, as for example the ones given by the diagram below,
\begin{center}
\includegraphics[scale=0.35]{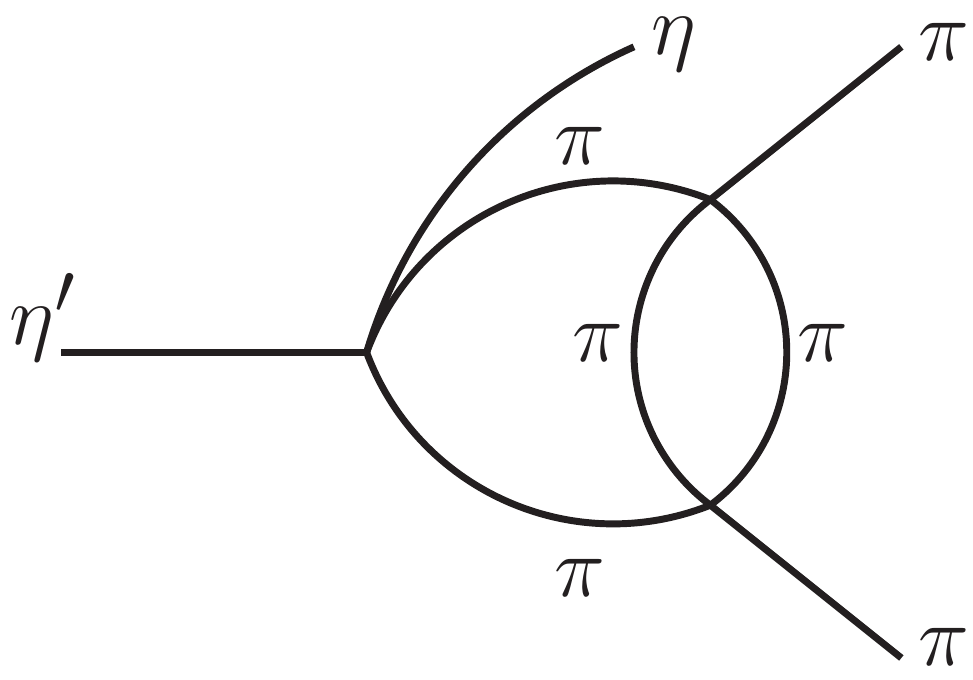}
\end{center}
\vspace{-0.1cm}
are generated from the on-shell approximation within the $N/D$ method by the $t$-and $u$-channel $\pi\pi$ loops entering the $\pi\pi$ scattering amplitude.
In fact, this a drawback of the unitarization method (see Refs.\,\cite{GomezNicola:2001as,Ledwig:2014cla} for more details about this pathology). This violation of unitarity turns out to be acceptable since these effects are usually found to be numerically small. 
See section \ref{section62} for a discussion of the size of these constributions in the present case.}:
\vspace{-0.5cm}
\begin{eqnarray}
{\rm{Im}}\left(m^{IJ}_{\eta^{\prime}\to\eta\pi\pi}(s)\right)&=&-16\pi \left(t^{IJ}_{\pi\pi}(s)\right)^{(2)}{\rm{Im}}g_{\pi\pi}(s) \left(m^{IJ}_{\eta^{\prime}\to\eta\pi\pi}(s)\right)^{(2)}\nonumber\\
&=& \left(t^{IJ}_{\pi\pi}(s)\right)^{(2)}\sigma_{\pi}(s) \left(m^{IJ}_{\eta^{\prime}\to\eta\pi\pi}(s)\right)^{(2)}\,.
\label{unitarityfulfillment}
\end{eqnarray}
The unitarized $\eta^{\prime}\to\eta\pi\pi$ $I=0$ decay amplitude written in terms of the $S$-and $D$-waves is
\begin{eqnarray}
\nonumber\mathcal{M}^{I=0}_{\eta^{\prime}\to\eta\pi\pi}(s,\cos\theta_{s})&=&\sum_{J}32\pi(2J+1)P_{J}(\cos\theta_{s})m^{IJ}_{\eta^{\prime}\to\eta\pi\pi}(s)\\
&=&32\pi P_{0}(\cos\theta_{s})m^{00}_{\eta^{\prime}\to\eta\pi\pi}(s)+160\pi P_{2}(\cos\theta_{s})m^{02}_{\eta^{\prime}\to\eta\pi\pi}(s)\,.
\end{eqnarray}
Using Eq.\,(\ref{NDrepresentation}) it becomes 
\begin{eqnarray}
\nonumber\mathcal{M}^{I=0}_{\eta^{\prime}\to\eta\pi\pi}(s,\cos\theta_{s})&=&32\pi P_{0}(\cos\theta_{s})
\frac{\left(m^{00}_{\eta^{\prime}\to\eta\pi\pi}(s)\right)^{(2)+\rm{Res+Loop}+\Lambda}}{1+16\pi g_{\pi\pi}(s)\left(t^{00}_{\pi\pi}(s)\right)^{(2)+\rm{Res+Loop}}}\\[1ex]
&&+160\pi P_{2}(\cos\theta_{s})\frac{\left(m^{02}_{\eta^{\prime}\to\eta\pi\pi}(s)\right)^{(2)+\rm{Res+Loop}+\Lambda}}{1+16\pi g_{\pi\pi}(s)\left(t^{02}_{\pi\pi}(s)\right)^{(2)+\rm{Res+Loop}}}\,.
\label{schannelunitarityamplitude}
\end{eqnarray}

The corresponding partial waves $m^{IJ}(s)$ are obtained through
\begin{eqnarray}
\hspace{-0.45cm}m^{IJ}_{\eta^{\prime}\to\eta\pi\pi}(s)=\frac{1}{32\pi}\frac{s}{\lambda^{1/2}(s,m_{\eta^{\prime}}^{2},m_{\eta}^{2})\lambda^{1/2}(s,m_{\pi}^{2},m_{\pi}^{2})}\int^{t_{\rm{max}}}_{t_{\rm{min}}}dtP_{J}(\cos\theta_{s})\mathcal{M}^{I}_{\eta^{\prime}\to\eta\pi\pi}(s,t,u)\,,
\end{eqnarray}
with $t_{\rm{max/min}}$ defined in Eq.\,(\ref{tmaxmin}) and where $\theta_{s}$ is the angle of $\bf{p}_{\pi}$ with respect to $\bf{p}_{\eta}$ in the rest frame of the pion pair. It is given by
\begin{eqnarray}
\cos\theta_{s}=-\frac{s\left(m_{\eta^{\prime}}^{2}+m_{\eta}^{2}+2m_{\pi}^{2}-s-2t\right)}{\lambda^{1/2}(s,m_{\eta^{\prime}}^{2},m_{\eta}^{2})\lambda^{1/2}(s,m_{\pi}^{2},m_{\pi}^{2})}\,.
\end{eqnarray}

The $\pi\pi$-scattering amplitude and the calculation of the $S$-and $D$-waves, $\left(t^{00}_{\pi\pi}(s)\right)^{(2)+\rm{Res+Loop}}$ and $\left(t^{02}_{\pi\pi}(s)\right)^{(2)+\rm{Res+Loop}}$, are detailed in appendix \ref{pipiscattering}. 

\subsection{$\pi\eta$ final-state interactions}\label{section52}

By analogy with Eq.\,(\ref{NDrepresentation}), the $N/D$ representation of the decay amplitude can be written as
\begin{eqnarray}
&&\hspace{-0.5cm}m^{IJ}_{\eta^{\prime}\to\eta\pi\pi}(t,u)=\nonumber\\[0.25cm]
&&\hspace{-0.5cm}=[1+16\pi N^{IJ}_{\pi\eta}(t)g_{\pi\eta}(t)]^{-1}R^{IJ}_{\eta^{\prime}\to\eta\pi\pi}(t)+[1+16\pi N^{IJ}_{\pi\eta}(u)g_{\pi\eta}(u)]^{-1}R^{IJ}_{\eta^{\prime}\to\eta\pi\pi}(u)\,,
\label{NDpieta}
\end{eqnarray}
where 
the $\pi\eta$ final-state interactions in the $t$-and $u$-channels have been resummed. We have 
\begin{equation}
\begin{array}{rcl}
N^{IJ}_{\pi\eta}(t)&=& \left(t^{IJ}_{\pi\eta}(t)\right)^{(2)+\rm{Res+Loop}+\Lambda}\,,\\[1ex]
R^{IJ}_{\eta^{\prime}\to\eta\pi\pi}(t)&=& \left(m^{IJ}_{\eta^{\prime}\to\eta\pi\pi}(t)\right)^{(2)+\rm{Res+Loop+\Lambda}}\,.\\[0.25cm]
\label{NDexpressions2}
\end{array}
\end{equation}
$\left(t^{IJ}_{\pi\eta}(s)\right)^{(2)+\rm{Res+Loop}+\Lambda}$ contains the perturbative calculation of the partial wave $\pi\eta$ scattering and $\left(m^{IJ}_{\eta^{\prime}\to\eta\pi\pi}(s)\right)^{(2)+\rm{Res+Loop+\Lambda}}$ the ${\eta^{\prime}\to\eta\pi\pi}$ decay amplitude one. 
The diagrammatic structure of the $\pi\eta\to\pi\eta$ scattering amplitude resembles the ones shown in Fig.\,\ref{DiagramsScattering} for $\pi\pi$. The notation here follows the convention adopted for $\pi \pi$ see Eq.~(\ref{NDexpressions}) with $\pi \pi$ replaced by the $\pi \eta$-system.  

The two-particle discontinuity along the right-hand cut due to the $\pi\eta$ intermediate states reads

\begin{equation}
g_{\pi\eta}(t)=\frac{1}{16\pi^{2}}\left(a_{\pi\eta}(\mu)+\log\frac{m^{2}_{\pi}}{\mu^{2}}-x_{+}\log\frac{x_{+}-1}{x_{+}}-x_{-}\log\frac{x_{-}-1}{x_{-}}\right)\,,
\label{gpieta}
\end{equation}
with
\begin{eqnarray}
x_{\pm}=\frac{t+m_{\eta}^{2}-m_{\pi}^{2}}{2t}\pm\frac{1}{-2t}\sqrt{-4t(m_{\eta}^{2}-i0^{+})+(t+m_{\eta}^{2}-m_{\pi}^{2})^{2}}\,,
\label{xplusminus}
\end{eqnarray}
and analogously for $g_{\pi\eta}(u)$ with $t\leftrightarrow u$.
However, we slightly modify the expression of Eq.\,(\ref{NDpieta}) and construct the amplitude such that the perturbative terms of the decay amplitude are kept. They are supplemented by the inclusion of the $S$-wave $\pi\eta$ final-state interactions.
In this way, we are able to quantify the importance of the $\pi\eta$ rescattering contribution with respect to the perturbative calculation. 
Since Eq.\,(\ref{NDpieta}) generates the $S$-wave projection of the lowest-order, 
the resonance exchanges, the loop contributions and the $\Lambda$ term~\footnote{twice (one for each channel).}, we need 
to remove these contributions and add by hand the term $\mathcal{M}(s,t,u)^{(2)+\rm{Res+Loop+\Lambda}}$, see section \ref{section4}.
Thus, the unitarized decay amplitude for $\eta^{\prime}\to\eta\pi\pi$ taking into account the $I=1$ $S$-wave $\pi\eta$ final-state interactions can be written as {\small
\begin{eqnarray}\label{tchannelunitarityamplitude}
&&\hspace{-0.75cm}\mathcal{M}_{\eta^{\prime}\to\eta\pi\pi}(s,t,u,\cos\theta_{t},\cos\theta_{u})= \left(\mathcal{M}(s,t,u)\right)^{(2)+\rm{Res+Loop+\Lambda}}\\[2ex]
&&\hspace{-0.75cm}\nonumber+32\pi P_{0}(\cos\theta_{t})\frac{\left(m_{\eta^{\prime}\to\eta\pi\pi}^{10}(t)\right)^{(2)+\rm{Res+Loop}+\Lambda}}{1+16\pi g_{\pi\eta}(t) \left(t^{10}_{\pi\eta}(t)\right)^{(2)+\rm{Res+Loop}+\Lambda}} \\[2ex]
&&\hspace{-0.75cm}\nonumber +32\pi P_{0}(\cos\theta_{u})\frac{\left(m_{\eta^{\prime}\to\eta\pi\pi}^{10}(u)\right)^{(2)+\rm{Res+Loop}+\Lambda}}{1+16\pi g_{\pi\eta}(u)\left(t^{10}_{\pi\eta}(u)\right)^{(2)+\rm{Res+Loop}+\Lambda}}\nonumber\\[2ex]
&&\hspace{-0.75cm}-32\pi P_{0}(\cos\theta_{t}) \left(m_{\eta^{\prime}\to\eta\pi\pi}^{10}(t)\right)^{(2)+\rm{Res+Loop+\Lambda}}-32\pi P_{0}(\cos\theta_{u}) \left(m_{\eta^{\prime}\to\eta\pi\pi}^{10}(u)\right)^{(2)+\rm{Res+Loop+\Lambda}}\,.\nonumber 
\end{eqnarray}}
Note that $\mathcal{M}(s,t,u)^{(2)+\rm{Res+Loop+\Lambda}}$ is not projected into partial waves and the last two terms are introduced to avoid double counting.
The $m_{\eta^{\prime}\to\eta\pi\pi}^{10}$ partial waves are derived through
\begin{eqnarray}
m_{\eta^{\prime}\to\eta\pi\pi}^{10}(t)=-\frac{1}{64\pi}\frac{t}{\lambda^{1/2}(s,m_{\eta^{\prime}}^{2},m_{\pi}^{2})\lambda^{1/2}(s,m_{\eta}^{2},m_{\pi}^{2})}\int^{s_{\rm{max}}}_{s_{\rm{min}}}dsP_{0}(\cos\theta_{t})\mathcal{M}^{1}(s,t,u)\,,
\label{partialwavecrossedchannels}
\end{eqnarray}
where 
{\small
\begin{eqnarray}
s_{\rm{max/min}}(s)=\frac{1}{2}\Bigg[m_{\eta^{\prime}}^{2}+m_{\eta}^{2}+2m_{\pi}^{2}-t-\frac{\Delta_{\eta^{\prime}\pi}\Delta_{\eta\pi}}{t}\pm\frac{\lambda^{1/2}(t,m_{\eta^{\prime}}^{2},m_{\pi}^{2})\lambda^{1/2}(t,m_{\eta}^{2},m_{\pi}^{2})}{t}\Bigg]\,,
\label{smaxmin}
\end{eqnarray}}
with $\Delta_{PQ}=m_{P}^{2}-m_{Q}^{2}$.
In Eq.\,(\ref{partialwavecrossedchannels}), $\theta_{t}$ is the angle of $\bf{p_{\eta}}$ with respect to $\bf{p_{\pi}}$ in the $\pi\eta$ center-of-mass frame given by
\begin{eqnarray}
\cos\theta_{t}=\frac{t\left(u-s\right)-\Delta_{\eta^{\prime}\pi}\Delta_{\eta\pi}}{\lambda^{1/2}(t,m_{\eta^{\prime}}^{2},
m_{\pi}^{2})\lambda^{1/2}(t,m_{\eta}^{2},m_{\pi}^{2})}\,.
\end{eqnarray}
Analogous expressions are valid in the $u$-channel with the replacements $t\leftrightarrow u$ and $\cos\theta_{t}\leftrightarrow-\cos\theta_{u}$.

The required $\pi\eta$-scattering amplitude and the calculation of the corresponding $S$-wave projection $\left(t^{10}_{\pi\eta}(s)\right)^{(2)+\rm{Res+Loop}+\Lambda}$ are given in appendix \ref{pietascattering}. 

\subsection{$\pi\pi$ and $\pi\eta$ final-state interactions}\label{pipipietafsi}

While in sections \ref{section51} and \ref{section52} the individual effects of the $I=0$ $S$-and $D$-wave $\pi\pi$ and $I=1$ $\pi\eta$ $S$-wave final-state interactions have been considered, we now account for both effects simultaneously. 
The unitarized decay amplitude is written as
\begin{eqnarray}
m_{\eta^{\prime}\to\eta\pi\pi}(s,t,u)&=&
[1+N_{\pi\pi}^{00}(s)g_{\pi\pi}(s)]^{-1}R_{\eta^{\prime}\to\eta\pi\pi}^{00}(s)+[1+N_{\pi\pi}^{02}(s)g_{\pi\pi}(s)]^{-1}R_{\eta^{\prime}\to\eta\pi\pi}^{02}(s)\nonumber\\[2ex]
&+&[1+N_{\pi\eta}^{10}(t)g_{\pi\eta}(t)]^{-1}R_{\eta^{\prime}\to\eta\pi\pi}^{10}(t)+[1+N_{\pi\eta}^{10}(t)g_{\pi\eta}(u)]^{-1}R_{\eta^{\prime}\to\eta\pi\pi}^{10}(u)\,,\nonumber\\[1ex]
\label{NDfsifull}
\end{eqnarray}
where the functions $g_{\pi\pi}(s)$ and $g_{\pi\eta}(t)$ are defined in Eqs.\,(\ref{gpipi}) and $(\ref{gpieta})$, respectively. 
The $\pi\pi$ and $\pi\eta$ scattering amplitudes as well as the $\eta^{\prime}\to\eta\pi\pi$ decay are defined 
along the lines of the previous sections. 
In an analogous fashion to section $\ref{section52}$, we build the full decay amplitude keeping its perturbative 
part that we supplement by the inclusion the $S$-and $D$-wave $\pi\pi$ and the $S$-wave $\pi\eta$ final-state interactions. 
This allows us to have a direct access to the rescattering effects that can be compared to the contribution coming from the 
perturbative calculation. 
Similarly to what is described in section $\ref{section52}$, the unitarization procedure, Eq.\,(\ref{NDfsifull}), generates 
redundant terms that need to be removed. 
Moreover, $\left( \mathcal{M}(s,t,u)\right)^{(2)+\rm{Res+Loop+\Lambda}}$ perturbatively calculated in section \ref{section4} needs to be added. 
This leads to {\small
\begin{eqnarray}
&&\hspace{-0.8cm}\mathcal{M}(s,t,u,\cos\theta_{s,t,u})=
\left(\mathcal{M}(s,t,u)\right)^{(2)+\rm{Res+Loop+\Lambda}}\nonumber\\[1ex]
&&\hspace{-0.8cm}+32\pi P_{0}(\cos\theta_{s})\frac{\left(m^{00}_{\eta^{\prime}\to\eta\pi\pi}(s)\right)^{(2)+\rm{Res+Loop}+\Lambda}}{1+16\pi g_{\pi\pi}(s) \left(t^{00}_{\pi\pi}(s)\right)^{(2)+\rm{Res+Loop}}}\nonumber+160\pi P_{2}(\cos\theta_{s})\frac{\left(m^{02}_{\eta^{\prime}\to\eta\pi\pi}(s)\right)^{(2)+\rm{Res+Loop}+\Lambda}}{1+16\pi g_{\pi\pi}(s) \left(t^{02}_{\pi\pi}(s)\right)^{(2)+\rm{Res+Loop}}}\nonumber
\end{eqnarray}
\begin{eqnarray}
&&\hspace{-0.8cm}-32\pi P_{0}(\cos\theta_{s}) \left(m_{\eta^{\prime}\to\eta\pi\pi}^{00}(t)\right)^{(2)+\rm{Res+Loop+\Lambda}}-160\pi P_{2}(\cos\theta_{s})\left(m_{\eta^{\prime}\to\eta\pi\pi}^{02}(u)\right)^{(2)+\rm{Res+Loop+\Lambda}}\nonumber\\[1ex]
&&\hspace{-0.8cm}+32\pi P_{0}(\cos\theta_{t})\frac{\left(m_{\eta^{\prime}\to\eta\pi\pi}^{10}(t)\right)^{(2)+\rm{Res+Loop}+\Lambda}}{1+16\pi g_{\pi\eta}(t) \left(t^{10}_{\pi\eta}(t)\right)^{(2)+\rm{Res+Loop}+\Lambda}}\nonumber+32\pi P_{0}(\cos\theta_{u})\frac{\left(m_{\eta^{\prime}\to\eta\pi\pi}^{10}(u)\right)^{(2)+\rm{Res+Loop}+\Lambda}}{1+16\pi g_{\pi\eta}(u)\left(t^{10}_{\pi\eta}(u)\right)^{(2)+\rm{Res+Loop}+\Lambda}}\nonumber\\[2ex]
&&\hspace{-0.8cm}-32\pi P_{0}(\cos\theta_{t}) \left(m_{\eta^{\prime}\to\eta\pi\pi}^{10}(t)\right)^{(2)+\rm{Res+Loop+\Lambda}}-32\pi P_{0}(\cos\theta_{u}) \left(m_{\eta^{\prime}\to\eta\pi\pi}^{10}(u)\right)^{(2)+\rm{Res+Loop+\Lambda}}\,.\nonumber \\
\end{eqnarray}}

\section{Fits to experimental data}\label{section6}

We relate the theoretical expression for the differential decay rate of $\eta^{\prime}\to\eta\pi^{0}\pi^{0}$, Eq.\,(\ref{width2}), to 
the Dalitz distribution of the measured number of events through 

\begin{equation}
\label{theory_to_experiment}
\frac{{\rm d}^{2}N_{\rm events}}{{\rm d}X{\rm d}Y} \,=2\,
\frac{N_{\rm events}}{\Gamma_{\eta^{\prime}}\bar{B}(\eta^{\prime}\to\eta\pi^{0}\pi^{0})}\frac{{\rm d}\Gamma(\eta^{\prime}\to\eta\pi^{0}\pi^{0})}{{\rm d}X{\rm d}Y}
\,\Delta X\Delta Y\,.
\end{equation}
For our study, we analyze the acceptance corrected $\eta^{\prime}\to\eta\pi^{0}\pi^{0}$ Dalitz distribution recently released by the A2 collaboration \cite{Adlarson:2017wlz}.
The factor of $2$ accounts for the fact that the data is given for half of the (symmetric) Dalitz distribution.
$N_{\rm events}$ is the total number of events for the considered process. $\Gamma_{\eta^{\prime}}$ is the total decay width of the $\eta^{\prime}$ meson. $\Delta X$ and $\Delta Y$ are the bin width of the $X$ and $Y$ variables. 
$\bar{B}(\eta^{\prime}\to\eta\pi^{0}\pi^{0})\equiv \bar{B}$ 
is a normalisation constant that, for a perfect description of the spectrum, would be equal to the corresponding branching fraction.
For our analysis, we fix this normalisation to the PDG reported value $\bar{B}=22.8(8)\%$ \cite{Olive:2016xmw}\footnote{The A2 collaboration does not provide a measurement for the branching ratio. 
Otherwise, we would fix this constant to the A2 measured value for consistency. 
Another possibility would be to let this constant float and infer its value from fits to the data. 
However, in order to reduce the number of free parameters to fit, we prefer to fix this constant to the PDG average.}. 
Two analyses of the same data set called analysis I and analysis II have been performed  in Ref.\,\cite{Adlarson:2017wlz}. They correspond to different analysis frameworks and to different selections of data samples. 
The corresponding efficiency corrected numbers of events ($N_{\rm events}$) are $463066$ for analysis I and $473044$ for analysis II.  The width of the bins is $\Delta X=\Delta Y=0.10$~MeV.
The central value of the results of our analysis are obtained by considering the data set of analysis I.  
The results obtained with the data set of analysis II is used to assess the systematic uncertainties of our fit results presented in section \ref{section64}. 

The $\chi^{2}$ function we minimize is 
\begin{equation}
\chi^{2}=\sum_{X,Y}\left(\frac{\mathcal{N}_{XY}^{\rm{th}}-\mathcal{N}_{XY}^{\rm{exp}}}{\sigma_{XY}^{\rm{exp}}}\right)^{2}\,,
\end{equation} 
where $\mathcal{N}_{XY}^{\rm{exp}}$ is the experimental number of events and $\sigma_{XY}^{\rm{exp}}$ the corresponding uncertainties in the $XY$-th bin. 
The number of data points to be fitted is 200.

Our fitting strategy is organized in a bottom-up approach guided by step-by-step implementation of two-body unitarity. 
The free parameters to fit are $M_{S_{8}},M_{S_{1}},M_{a_{0}},c_{m},\tilde{c}_{m},c_{d}$ and $\tilde{c}_{d}$.
However, 
in order to reduce the number of free parameters we 
invoke the large-$N_{C}$ relations for the couplings and masses of the octet and singlet. 
We set $\tilde{c}_{d}=c_{d}/\sqrt{3}$, $\tilde{c}_{m}=c_{m}/\sqrt{3}$ and $M_{S}=M_{S_{8}}=M_{S_{1}}=M_{a_{0}}$ \cite{Ecker:1988te}. 
For the $\eta$-$\eta^{\prime}$ mixing angle we take $\theta=-13.3(5)^{\circ}$ \cite{Ambrosino:2009sc}\footnote{In Ref.~\cite{Ambrosino:2009sc}, the value $\phi_{\eta\eta^\prime}=(41.4\pm 0.5)^\circ$ is obtained in the quark-flavor basis.
However, at the lowest order, this value is equivalent in the octet-singlet basis to
$\theta_{\eta\eta^\prime}=\phi_{\eta\eta^\prime}-\arctan\sqrt{2}=(-13.3\pm 0.5)^\circ$.} while we use $\Lambda_{2}=-0.37$ \cite{Guo:2012yt} and set the regularization scale to $\mu=700$ MeV. 
Note that by imposing the large-$N_{C}$ constraints we are introducing large correlations between the fit parameters. 
This implies that different values for the couplings and masses can lead to a similar fit quality underestimating the true statistical errors. 
Therefore the fit results presented below should be taken with a word of caution. 
 
\subsection{ChPT fits including resonances and one-loop corrections}\label{section61}

We start by fitting Eq.\,(\ref{theory_to_experiment}) to A2 data \cite{Adlarson:2017wlz} with the amplitude described in section \ref{section4}.
For our first fit we impose the restriction $c_{d}=c_{m}$ for the couplings 
which comes from the requirement of the $K\pi$ scalar form factor to vanish at high energies \cite{Jamin:2001zq}. 
The resulting fit parameters take the values
\begin{equation}
M_{S}=973(5)\,{\rm{MeV}}\,,\quad c_{d}=c_{m}=30.1(4)\,{\rm{MeV}}\,,
\label{fit1}
\end{equation}
with a $\chi^{2}/{\rm{dof}}=242.2/198=1.22$, which leads to $\tilde{c}_{d,m}=c_{d,m}/\sqrt{3}=17.4(2)$ MeV for the singlet couplings while the associated Dalitz-plot parameters are found to be
\begin{equation}
a=-0.095(6)\,,\quad b=0.005(1)\,,\quad d=-0.037(5)\,.
\label{Dalitzparam1}
\end{equation}

For our second fit we allow the couplings $c_{d}$ and $c_{m}$ to float. We obtain
\begin{equation}
M_{S}=954(47)\,{\rm{MeV}}\,,\quad c_{d}=28.0(4.6)\,{\rm{MeV}}\,,\quad c_{m}=53.4(52.0)\,{\rm{MeV}}\,,
\label{fit2}
\end{equation}
with a $\chi^{2}/{\rm{dof}}=242.0/197=1.23$. The corresponding Dalitz parameters in this case are
\begin{equation}
a=-0.093(45)\,,\quad b=0.004(3)\,,\quad d=-0.039(18)\,.
\end{equation}

The fitted parameters are particularly strongly correlated in this case leading to a large error for the coupling $c_{m}$.
This is not a surprise since $c_{m}$ always enters with $m_{\pi}^{2}$ in the amplitude, and hence is chirally suppressed, indicating that its influence is small.
This is in agreement with previous estimates of this coupling suffering from a large uncertainty. 
For example, $c_{m}=31.5^{+19.5}_{-22.5}$ MeV in \cite{Guo:2011pa}, $c_{m}=15(30)$ MeV in \cite{Oller:1998zr} and $c_{m}=80(21)$ MeV in \cite{Guo:2009hi}.
In order to alleviate this correlation, we also consider fits where we fix the couplings $c_{m}$ and $\tilde{c}_{m}$. 
For instance, we first fix them to $c_{m}=41.1(1)$ MeV and $\tilde{c}_{m}=18.9(9)$ MeV \cite{Ledwig:2014cla} using results obtained from meson-meson scattering (see Refs.\,\cite{Guo:2011pa,Guo:2012yt} for other possible values for these couplings).
Imposing in addition the constraint $\tilde{c}_{d}=c_{d}/\sqrt{3}$ we obtain
\begin{equation}
M_{S}=992(7)\,{\rm{MeV}}\,,\quad c_{d}=31.9(5)\,{\rm{MeV}}\,,
\label{fit3}
\end{equation}
with a $\chi^{2}/{\rm{dof}}=246.4/198=1.24$. This leads to $\tilde{c}_{d}=18.4(3)$ MeV for the singlet coupling and
\begin{equation}
a=-0.083(6)\,,\quad b=-0.0002(1)\,,\quad d=-0.057(5)\,,
\label{DalitzParamfit3}
\end{equation}
for the associated Dalitz parameters. If now we let the coupling $\tilde{c}_{d}$ float, we get
\begin{equation}
M_{S}=968(11)\,{\rm{MeV}}\,,\quad c_{d}=29.8(9)\,{\rm{MeV}}\,\quad\tilde{c}_{d}=21.2(1.2)\,{\rm{MeV}}\,,
\label{fit4}
\end{equation}
with a $\chi^{2}/{\rm{dof}}=241.9/197=1.23$. The corresponding Dalitz-plot parameters are found to be
\begin{equation}
a=-0.092(5)\,,\quad b=0.004(2)\,,\quad d=-0.041(11)\,.
\end{equation}   

On the contrary, if we take $c_{m}=80(21)$ MeV from Ref.\,\cite{Guo:2009hi} and impose the constraint $\tilde{c}_{d,m}=c_{d,m}/\sqrt{3}$, we get
\begin{equation}
M_{S}=926(5)(25)\,{\rm{MeV}}\,,\quad c_{d}=25.7(4)(1.9)\,{\rm{MeV}}\,,
\label{fit5}
\end{equation}
with a $\chi^{2}/{\rm{dof}}=242.3/198=1.22$. The first error is the fit uncertainty and the second is the systematic uncertainty due to $c_{m}$.
The corresponding singlet couplings are found to be $\tilde{c}_{d}=14.8(2)(1.1)$ MeV and $\tilde{c}_{m}=46.2(12.1)$ MeV. The Dalitz parameters read
\begin{equation}
a=-0.090(7)(3)\,,\quad b=0.004(0)(0)\,,\quad d=-0.041(7)(1)\,.
\end{equation}

We have also tried fits having all couplings i.e.\,$c_{m},\tilde{c}_{m},c_{d}$ and $\tilde{c}_{d}$ as free parameters but these fits are unstable because there are too many free parameters to fit. 

\begin{figure}[h!]
\begin{center}
\includegraphics[scale=0.35]{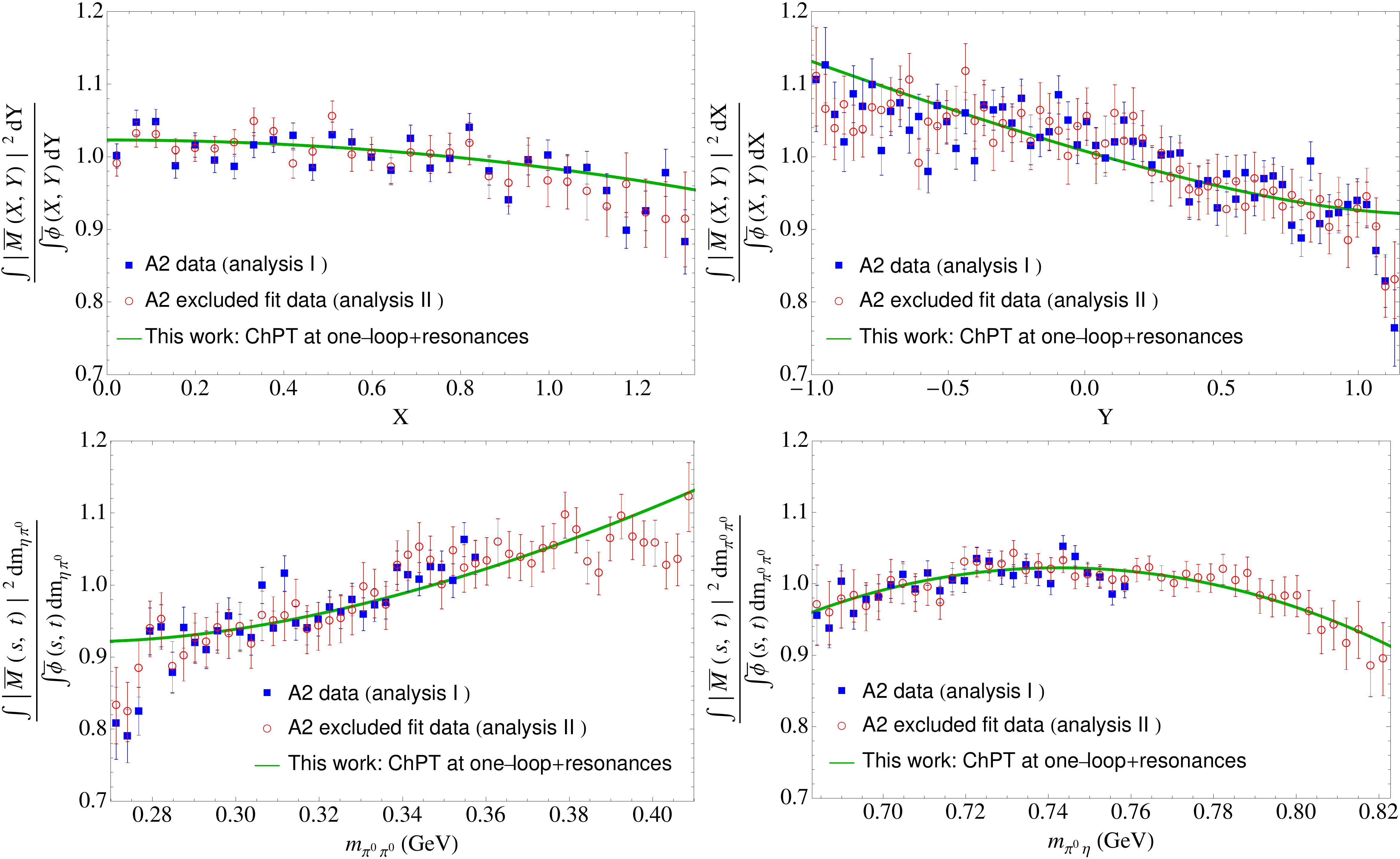}
\caption{\label{distribution1}Differential decay rate distribution for $\eta^{\prime}\to\eta\pi^{0}\pi^{0}$ divided by the phase-space, both individually normalized, for $X$ (left-top panel), $Y$ (right-top panel), $m_{\pi^{0}\pi^{0}}$, (left-down panel) and $m_{\pi^{0}\eta}$ (right-down panel) using the fit results of Eq.\,(\ref{fit1}). 
The data are taken from Ref.\,\cite{Adlarson:2017wlz}.}
\end{center}
\end{figure}

In Fig.\,\ref{distribution1}, we provide a graphical account of the ratio of the differential decay width distributions of $\eta^{\prime}\to\eta\pi^{0}\pi^{0}$ as a function of $X,Y,m_{\pi^{0}\pi^{0}}$ and $m_{\pi^{0}\eta}$ over the phase-space obtained from the fit results of Eq.\,(\ref{fit1}).
In order to compare with the experimental data, both expressions are normalized such that the individual integrated branching ratio is $1$. 
The corresponding normalized amplitudes are denoted as $\bar{M}$ and $\bar{\phi}$ in the figure. 
A cusp effect at the $\pi^{+}\pi^{-}$ mass threshold is neatly visible in the data (see top-right and bottom-left panels of the figure).
This is the first cusp structure observed in $\eta^{\prime}\to\eta\pi^{0}\pi^{0}$ and is not accommodated by the 
theoretical description given here. In the next section \ref{section62} we will improve the theoretical description in order to 
obtain a better agreement with the experimental data and to try to describe the cusp effect. 

\begin{figure}[h!]
\begin{center}
\includegraphics[scale=0.35]{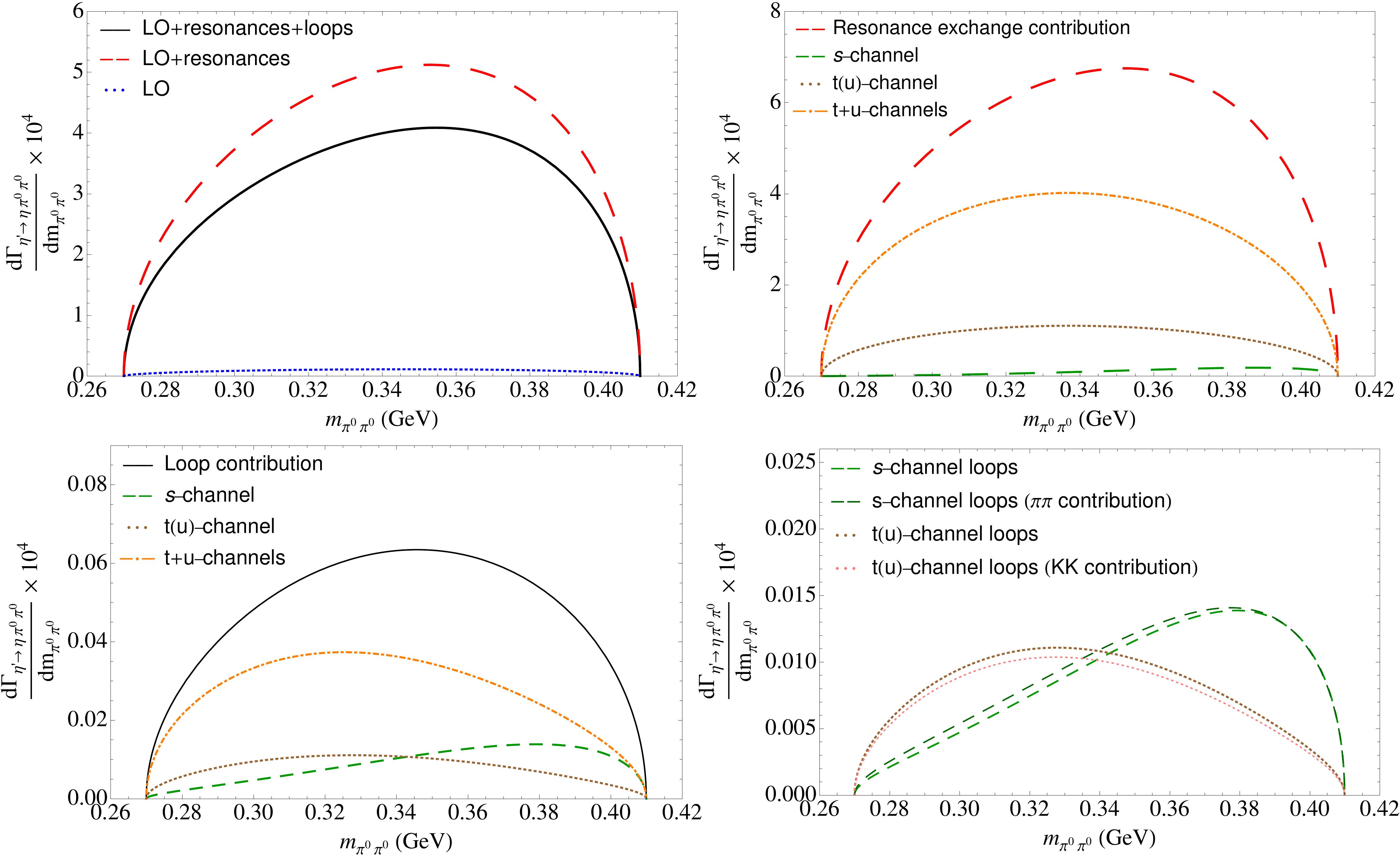}
\caption{\label{distributionhierarchy}Individual contributions to the differential decay rate distribution for $\eta^{\prime}\to\eta\pi^{0}\pi^{0}$. See main text for details.}
\end{center}
\end{figure}

In Fig.\,\ref{distributionhierarchy} we display graphically the different contributions entering the decay amplitude as a function of the $m_{\pi^{0}\pi^{0}}$ invariant mass distribution. 
The hierarchy between the resonance exchange and loop contributions is shown.  
From the top-left panel, we observe that the lowest-order contribution is tiny while the decay is largely dominated by the resonance exchanges with the loop contributions interfering destructively. 
The integrated branching ratio associated to these curves is of $0.6\%$ for  the lowest-order (blue dotted curve), 
$28.2\%$ for the lowest-order plus resonance exchanges (red dashed curve) and of $22.5\%$ for the lowest-order including resonance exchanges and loop contributions (black solid curve). 
The individual resonance exchange contributions are shown in the top-right panel.
The $t(u)$ channel (brown dotted curve) dominates over the $s$-channel contribution (green dashed curve) 
in the entire allowed phase space.
Further, in the bottom-left panel we classify the loop contributions which are indeed subleading.
In this case, the $t(u)$ channel loops dominate the first half of the spectrum while the loop contributions in the $s$-channel dominate the second half.
A decomposition into the individual loop contributions is provided on the bottom-right panel. 
This reveals that the $s$-channel loops (green dashed curve) are largely dominated by the $\pi\pi$ contribution (light green dashed curve) while the $KK$ loops (light pink dotted curve) dominate the crossed channels ones (brown dotted curve).  

Finally, we would also like to provide an estimate of the chiral couplings $L_{5}$ and $L_{8}$ as well as the sum $3L_{2}+L_{3}$ as 
output of our fits. 
Assuming resonance saturation for the order $p^{4}$ ChPT couplings constants \cite{Ecker:1988te}, the following relations can be derived
\begin{equation}
3L_{2}+L_{3}=\frac{c_{d}^{2}}{2M_{S}^{2}}\,,\quad L_{5}=\frac{c_{d}c_{m}}{M_{S}^{2}}\,,\quad L_{8}=\frac{c_{m}^{2}}{2M_{S}^{2}}\,.
\label{resonancesaturation}
\end{equation}
Using the results of the fit, Eq.\,(\ref{fit1}), we obtain
\begin{equation}
3L_{2}+L_{3}=0.47\cdot10^{-3}\,,\quad L_{5}=0.95\cdot10^{-3}\,,\quad L_{8}=0.47\cdot10^{-3}\,.
\label{LECS1}
\end{equation}
With the results of the fit Eq.\,(\ref{fit3}), we get:
\begin{equation}
3L_{2}+L_{3}=0.51\cdot10^{-3}\,,\quad L_{5}=1.03\cdot10^{-3}\,,\quad L_{8}=0.51\cdot10^{-3}\,,
\label{LECS2}
\end{equation}
Considering the results of the fit Eq.\,(\ref{fit5}), we obtain
\begin{equation}
3L_{2}+L_{3}=0.38\cdot10^{-3}\,,\quad L_{5}=2.39\cdot10^{-3}\,,\quad L_{8}=3.73\cdot10^{-3}\,.
\label{LECS3}
\end{equation}

The values of Eqs.\,(\ref{LECS1}) and (\ref{LECS2}) are in reasonable agreement with the chiral couplings determinations at $\mathcal{O}(p^{4})$ (see Ref.\,\cite{Bijnens:2014lea} for a recent review) while the values for $L_{5,8}$ in Eq.\,(\ref{LECS3}) 
disagree.
However, the first relation in Eq.\,(\ref{resonancesaturation}) is not well fulfilled for most of the values of $L_{2}$ and $L_{3}$ obtained at $\mathcal{O}(p^{6})$.
For example, the most recent analysis of $K_{\ell4}$ decays gives $L_{2}^{r}=0.63(13)\cdot10^{-3}$ and $L_{3}^{r}=-2.63(46)\cdot10^{-3}$~\cite{Colangelo:2015kha} leading to a negative value for the sum $3L_{2}+L_{3}$. 
So the left-hand side of the first equality of Eq.~(\ref{LECS3}) is negative while the right-hand side of the equation is definite positive. 
This inconsistency shows 
that resonance saturation of low-energy constants by scalar resonances 
should be taken with a word of caution.

\subsection{Fits including individual $\pi\pi$ and $\pi\eta$ final-state interactions}\label{section62}

In order to improve our fits to the data, we unitarize the parameterization of the decay amplitude. As derived in Eq.\,(\ref{schannelunitarityamplitude}) we first include the $\pi\pi$ final-state interactions only.
By analogy with section \ref{section61}, we first perform a fit imposing the relations $\tilde{c}_{d,m}=c_{d,m}/\sqrt{3}$ and 
$c_{d}=c_{m}$. In this case we obtain
\begin{equation}
M_{S}=1001(24)\,{\rm{MeV}}\,,\quad c_{d}=c_{m}=29.5(1.8)\,{\rm{MeV}}\,,\quad a_{\pi\pi}=0.73(25)\,,
\label{fit6}
\end{equation}
with a $\chi^{2}/{\rm{dof}}=220.4/197=1.12$. $\tilde{c}_{d,m}=17.0(1.0)$ MeV for the singlet couplings and \begin{equation}
a=-0.075(9)\,,\quad b=-0.051(1)\,,\quad d=-0.049(14)\,,
\label{fit6Dalitzaparam}
\end{equation}
for the associated Dalitz-plot parameters. 
Note that, with respect to the previous section, there is one more free parameter to fit since we have the subtraction constant, 
Eq.~(\ref{eq:gpipi}) to determine. 
Due to the important correlations between the fit parameters, the statistical uncertainties have increased while the total $\chi^{2}/{\rm{dof}}$ is slightly improved. 
Comparing these results to the ones of Eq.\,(\ref{Dalitzparam1}) obtained within ChPT including resonance exchanges and one-loop corrections without considering any final-state interactions, shows that the inclusion of the $\pi\pi$ final-state interactions has a large effects on the Dalitz plot parameters. 
In particular, the $a$ and $b$ parameters associated to powers of $Y$, have been substantially shifted downwards while the change on the  $d$ parameter related to $X^{2}$ is slightly less severe. 
We would like to note that while the $S$-wave only affects the parameters $a$ and $b$ associated to powers of $Y$, the $D$-wave 
affects the variable $X^{2}$ but also the determination of $a$ and $b$. 
For example, if we consider only the $S$-wave for the $\pi\pi$ final-state interaction, we obtain 
\begin{equation}
a=-0.094(7)\,,\quad b=-0.034(2)\,,
\label{fit6Dalitzaparambis}
\end{equation}
instead of the results of Eq.~(\ref{fit6Dalitzaparam}). 
Comparing to Eq.\,(\ref{Dalitzparam1}) $a$ seems unaffected by the inclusion of the $S$-wave $\pi\pi$ final-state interaction 
while $b$ is clearly moved down.
We therefore conclude that while a precise determination of the $b$ parameter requires to take into account the 
the $S$-wave $\pi\pi$ final-state interaction the determination of the $a$ and $d$ parameters are dominated by the $D$-wave. 

\begin{figure}[h!]
\begin{center}
\includegraphics[scale=0.55]{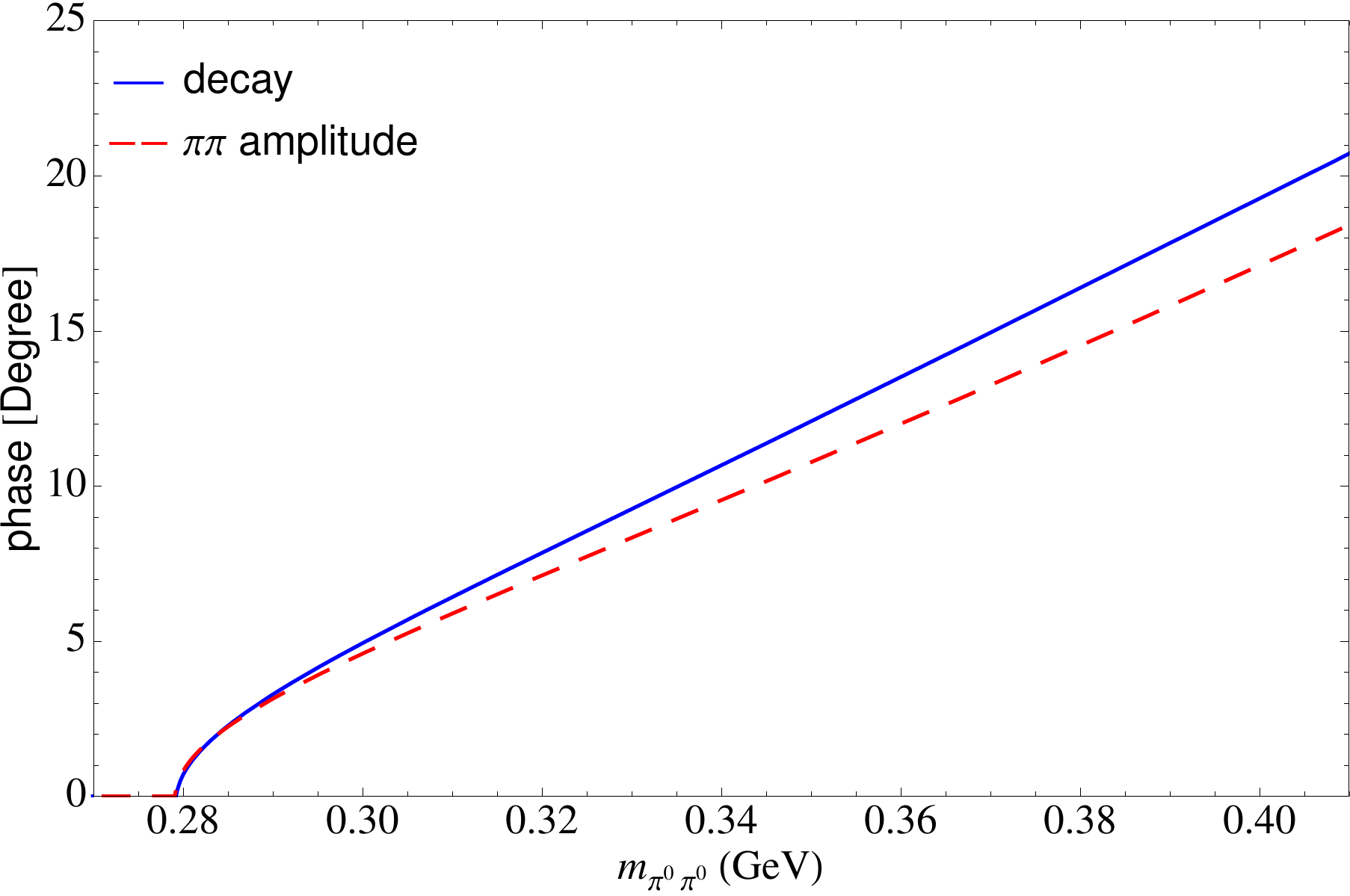}
\caption{\label{plotunitatiryviolation} $S$-wave phases of the $\eta^{\prime}\to\eta\pi^{0}\pi^{0}$ decay amplitude (blue line) and of the $\pi\pi$ scattering (red dashed line). The phases differ from one another due to the appearance of spurious contributions that induce (small) unitarity violations. See main text for details.}
\end{center}
\end{figure}

We shall now return to the discussion on the spurious singularities introducing unitarity violations, see section \ref{section51}.
According to the Watson's theorem \cite{Watson:1954uc}, the phase of the $\eta^{\prime}\to\eta\pi^{0}\pi^{0}$ decay amplitude equals the $\pi\pi$ scattering phase-shift in the elastic region. 
On figure \ref{plotunitatiryviolation} we compare the $S$-wave phase of the decay amplitude (blue line) to the $\pi\pi$ scattering phase-shift (red dashed line) using the results of the fits Eq.\,(\ref{fit6}). 
The difference between the two phases in the elastic region is 
due to such spurious contributions. This implies that unitarity is no longer fulfilled exactly.
However, this effect is numerically quite small (for the $D$-wave this effect is completely negligible) and acceptable for our analysis.
We have checked that by removing these spurious contributions (i.e. removing the $t$-and $u$-channel from the $\pi\pi$ loop contributions) entering $\pi\pi$ scattering, 
unitarity is restored. The changes in the fits are negligible. 

We also perform the two other fits corresponding to Eqs.~(\ref{fit2}), (\ref{fit3}) and (\ref{fit4}) 
with the inclusion of the $\pi \pi$ final-state interactions. 
If we let the couplings $c_{d}$ and $c_{m}$ as free parameters we do not manage to get acceptable fits. 
Fixing $c_{m}=41.1(1)$ MeV and $\tilde{c}_{m}=18.9(9)$ MeV \cite{Ledwig:2014cla} together with the relation $\tilde{c}_{d}=c_{d}/\sqrt{3}$, we obtain
\begin{equation}
M_{S}=988(10)\,{\rm{MeV}}\,,\quad c_{d}=28.6(7)\,{\rm{MeV}}\,,\quad a_{\pi\pi}=0.24(12)\,,
\label{fit7}
\end{equation}
with a $\chi^{2}/{\rm{dof}}=220.1/197=1.12$. $\tilde{c}_{d}=16.5(4)$ MeV for the singlet coupling and 
the associated Dalitz-plot parameters are found to be
\begin{equation}
a=-0.075(7)\,,\quad b=-0.056(1)\,,\quad d=-0.050(4)\,.
\label{fit7Dalitzaparam}
\end{equation}
On the contrary, if we take $c_{m}=80(21)$ MeV \cite{Guo:2009hi} the fit leads to
\begin{equation}
M_{S}=930(7)(39)\,{\rm{MeV}}\,,\quad c_{d}=23.5(6)(2.4)\,{\rm{MeV}}\,,\quad a_{\pi\pi}=0.41(11)(19)\,,
\label{fit8}
\end{equation}
with a $\chi^{2}/{\rm{dof}}=220.1/197=1.12$. The associated Dalitz-plot parameters are found to be
\begin{equation}
a=-0.074(7)(1)\,,\quad b=-0.053(1)(1)\,,\quad d=-0.049(4)(1)\,,
\label{fit8Dalitzaparam}
\end{equation}
where the first error is the fit uncertainty while the second is due to $c_{m}$.
In this case the singlet couplings read $\tilde{c}_{d}=13.6(4)(1.4)$ MeV and $\tilde{c}_{m}=46.2(12.1)$ MeV.

Note that the Dalitz parameters as obtained in Eqs.\,(\ref{fit6Dalitzaparam}), (\ref{fit7Dalitzaparam}) and (\ref{fit8Dalitzaparam}) 
do not change between the different fits. 
If we allow more parameters to float in the fits such as i.e. $c_{m},\tilde{c}_{m},c_{d}$ and $\tilde{c}_{d}$ the fits become 
unstable due to the large number of free parameters. 

In this case we find for the chiral couplings: 
\begin{equation}
3L_{2}+L_{3}=0.43\cdot10^{-3}\,,\quad L_{5}=0.87\cdot10^{-3}\,,\quad L_{8}=0.43\cdot10^{-3}\,,
\label{LECS4}
\end{equation}
for the fit Eq.\,(\ref{fit5}).
\begin{equation}
3L_{2}+L_{3}=0.42\cdot10^{-3}\,,\quad L_{5}=1.20\cdot10^{-3}\,,\quad L_{8}=0.87\cdot10^{-3}\,,
\label{LECS5}
\end{equation}
for the the fit Eq.\,(\ref{fit6}). 
\begin{equation}
3L_{2}+L_{3}=0.31\cdot10^{-3}\,,\quad L_{5}=2.17\cdot10^{-3}\,,\quad L_{8}=3.70\cdot10^{-3}\,,
\label{LECS6}
\end{equation}
for the the fit of Eq.\,(\ref{fit7}).

Similarly to what have been done for $\pi \pi$, we can take into account 
only the individual $\pi\eta$ final-state interaction effects using the representation 
Eq.\,(\ref{tchannelunitarityamplitude}).
The corresponding results can be found in table\,\ref{DalitzParampieta} 
where different fit settings have been considered. Note that the $\pi\eta$ subtraction constant is fixed to $a_{\pi\eta}=2.0^{+3.1}_{-3.4}$ \cite{Guo:2011pa} for these fits
\footnote{The subtraction constant $a_{\pi\eta}$ is not well determined by the fits. We therefore prefer to fix its value.}.
$i)$ Fit A corresponds to imposing the restriction $c_{d}=c_{m}$ and using the relation $\tilde{c}_{d,m}=c_{d,m}/\sqrt{3}$; $ii)$ Fit B let the couplings $c_{d}$ and $c_{m}$ to float and uses the relation $\tilde{c}_{d,m}=c_{d,m}/\sqrt{3}$; $iii)$ Fit C fixes $c_{m}=41.1(1)$ MeV and $\tilde{c}_{m}=18.9(9)$ MeV \cite{Ledwig:2014cla} using the relation $\tilde{c}_{d}=c_{d}/\sqrt{3}$ where $c_{d}$ is a free parameter of the fit; $iv)$ Fit D takes $c_{m}=80(21)$ MeV \cite{Guo:2009hi} and uses the relation $\tilde{c}_{d,m}=c_{d,m}/\sqrt{3}$ with $c_{d}$ a free parameter of the fit. 
\begin{table}
\centering
\begin{tabular}{llllll}
\hline
Parameter&Fit A&Fit B&Fit C&Fit D\\
\hline
$M_{S}$& $985(7)(20)$& $913(12)(32)$& $999(9)(14)$&$940(9)(47)$\\ 
$c_{d}$& $30.6(5)(7)$& $24.4(1.0)(4.6)$& $32.3(6)(9)$&$26.3(5)(2.5)$\\ 
$c_{m}$& $=c_{d}$& $100.5(3.0)(50.0)$& $=41.1(1)$&$=80(21)$\\ 
$\tilde{c}_{d}$& $17.7(2)(2)$& $14.1(6)(2.7)$& $18.7(4)(5)$&$15.2(3)(1.4)$\\ 
$\tilde{c}_{m}$& $=\tilde{c}_{d}$& $58.0(1.7)(28.9)$& $=18.9(9)$&$46.2(12.1)$\\ 
$\chi^{2}_{\rm{dof}}$&$243.2/198\sim1.23$&$242.9/197\sim1.23$&$244.5/198\sim1.24$&$242.7/197\sim1.23$\\ 
\hline     
$a[Y]$&$-0.094(6)(9)$&$-0.086(9)(14)$&$-0.083(7)(7)$&$-0.089(8)(15)$\\
$b[Y^{2}]$&$0.005(1)(1)$&$0.004(1)(1)$&$0.001(1)(1)$&$0.004(1)(1)$\\
$d[X^{2}]$&$-0.031(5)(4)$&$-0.035(8)(9)$&$-0.049(5)(13)$&$-0.035(7)(9)$\\
\hline 
\end{tabular}
\caption{Results for the parameters of different fits and their associated Dalitz parameters after resumming the $\pi\eta$ final-state interactions.
The first error corresponds to the statistical fit uncertainty, the second error is due to the systematic uncertainty 
coming from the subtraction constant $a_{\pi\eta}$ and the corresponding couplings that are fixed. 
The masses and couplings are given in MeV. 
See the main text for details.}
\label{DalitzParampieta}
\end{table}
Comparing the results of table~\ref{DalitzParampieta} to the ones given in section \ref{section61} obtained using ChPT including resonances and one-loop corrections without resumming the final-state interaction effects, we observe that the inclusion of the $\pi\eta$ rescattering effects has small effects on the result of the fits. We therefore conclude that $\pi\eta$ rescattering effects are small as 
previously observed, see Ref.\,\cite{Kubis:2009sb}.  
Note that Fit B allowing the coupling $c_{m}$ to float carries a large error bar for the same reasons as discussed in the previous section.

\begin{figure}[h!]
\begin{center}
\includegraphics[scale=0.35]{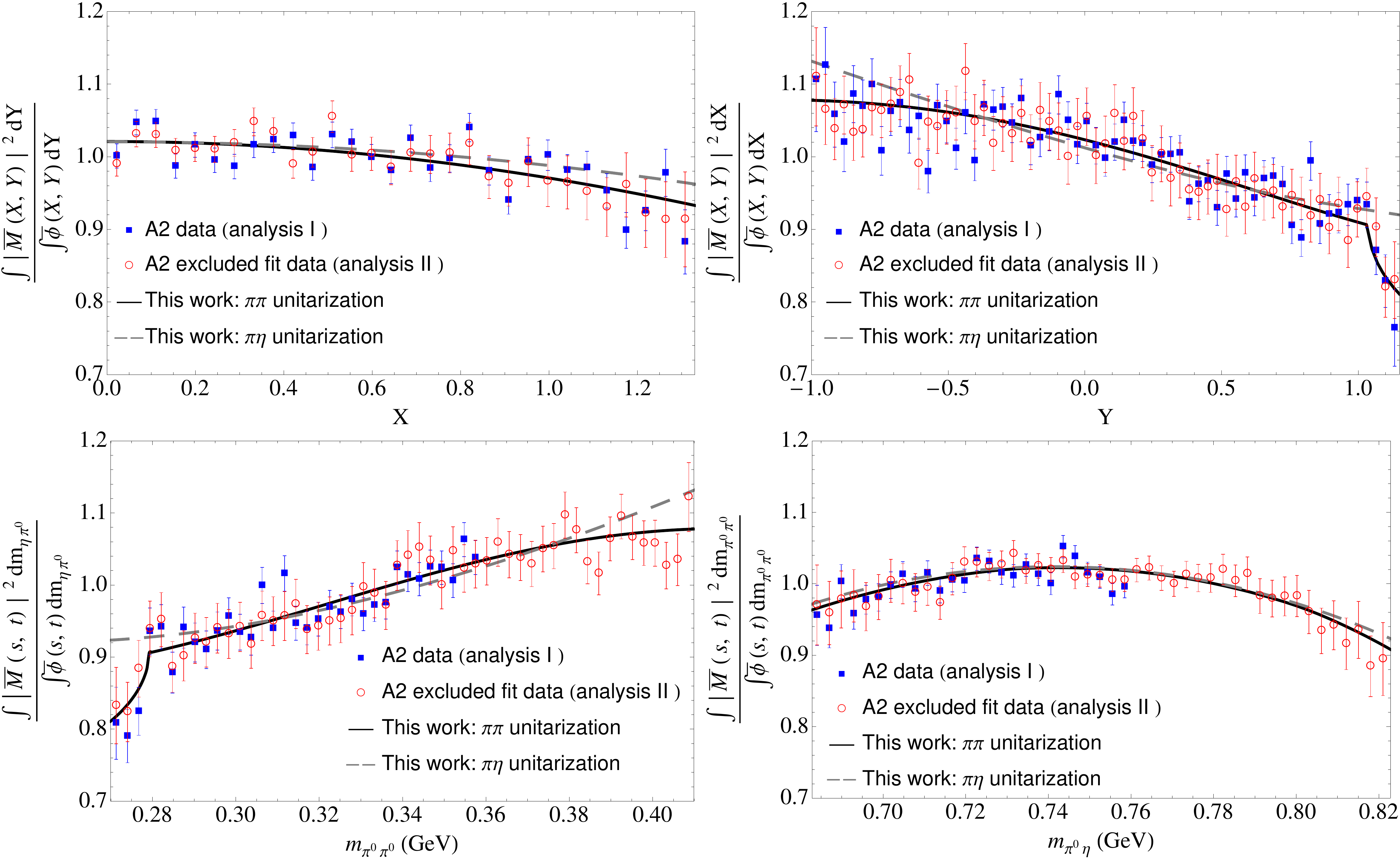}
\caption{\label{distribution3}Differential decay rate distribution for $\eta^{\prime}\to\eta\pi^{0}\pi^{0}$ divided by the phase-space, both individually normalized, for $X$ (left-top panel), $Y$ (right-top panel), $m_{\pi^{0}\pi^{0}}$, (left-down panel) and $m_{\pi^{0}\eta}$ (right-down panel) associated to the fit Eq.\,(\ref{fit5}) and Fit A of table \ref{DalitzParampieta}. These results are obtained after the individual resummation of $\pi\pi$ (black solid curve) and $\pi\eta$ (gray dashed curve) final-state interactions effects. 
Data is taken from Ref.\,\cite{Adlarson:2017wlz}.}
\end{center}
\end{figure}

As in section \ref{section61}, we compare the fit results to the experimental data in Fig.\,\ref{distribution3}. 
The black solid curve corresponds to the fit results Eq.\,(\ref{fit6}) where the $\pi\pi$ final-state interactions have been taken into account. 
The gray dashed curve represents the resulting amplitude obtained from Fit A of table\,\ref{DalitzParampieta} for which 
the $\pi\eta$ final-state interactions have been resummed. 
Contrary to the fit results shown in Fig.\,\ref{distribution1}, the cusp structure at the $\pi^{+}\pi^{-}$ mass threshold (see top-right and bottom-left panels of the figure) is now nicely accounted for within our description. 
This is possible after the inclusion of the $\pi\pi$ unitarization\footnote{We have considered the charged pion mass in the function $g_{\pi\pi}(s)$, Eq.\,(\ref{gpipi}), and in the $t$-and-$u$ channel pion loop contributions entering the $\pi\pi$ scattering (see appendix \ref{pipiscattering}). The other 
pion masses are set to the neutral ones.}.
Note that the theory describes much better the data once $\pi\pi$ rescattering effects have been included compared to $\pi\eta$\footnote{The cusp effect at the $\pi^{+}\pi^{-}$ mass threshold is not included into the description represented by the gray dashed curve. 
While this effect would affect a little the $s$-channel $\pi\pi$ loop of the decay amplitude, the resulting fit parameters w/o including the cusp effect remains unchanged.}. 
This is also in part reflected in the corresponding $\chi^{2}_{\rm{dof}}$.

\subsection{Fits including both $\pi\pi$ and $\pi\eta$ final-state interactions}\label{section64}

This section contains our central results for the corresponding fit parameters as well as for the associated Dalitz-plot slope parameters.
They are obtained by building a representation for the amplitude that takes into account the $\pi\pi$ and $\pi\eta$ final-state interactions simultaneously as described in section \ref{pipipietafsi}. 
In the previous sections, we presented fits to the analysis I of the A2 data set. Here we fit our theoretical representation to analysis I and II of the A2 data set allowing us to include a systematic 
uncertainty on our results. The systematic uncertainty is taken to be the difference between the results of the fits to the two 
different experimental analyses. 
The fit results 
are collected in table \ref{DalitzParam} using two different settings and with fixing the $\pi\eta$ subtraction constant to $a_{\pi\eta}=2.0^{+3.1}_{-3.4}$ \cite{Guo:2011pa}. $i)$ In Fit 1 we impose two conditions $\tilde{c}_{d,m}=c_{d,m}/\sqrt{3}$ and $c_{d}=c_{m}$; $ii)$ In Fit 2 we fix $c_{m}=41.1(1)$ MeV and $\tilde{c}_{m}=18.9(9)$ MeV \cite{Ledwig:2014cla} and impose $\tilde{c}_{d}=c_{d}/\sqrt{3}$.

\begin{table}
\centering
\begin{tabular}{lllll}
\hline\noalign{\smallskip}
\multirow{2}{*}{Parameter} & 
\multicolumn{2}{l}{\qquad Analysis I}& 
\multicolumn{2}{l}{\qquad Analysis II}\\[0.5ex]
&Fit 1&Fit 2& Fit 1&Fit 2\\
\noalign{\smallskip}\hline\noalign{\smallskip}
$M_{S}$& $1017(68)(24)$& $999(33)(23)$& $1040(79)(28)$& $1020(48)(28)$\\ 
$c_{d}$& $30.4(4.8)(9)$& $29.1(2.4)(1.6)$& $32.0(5.3)(9)$& $30.9(3.4)(2.2)$\\ 
$c_{m}$& $=c_{d}$& $=41.1(1)$& $=c_{d}$& $=41.1(1)$\\ 
$\tilde{c}_{d}$& $17.6(2.8)(5)$& $16.8(1.4)(9)$& $18.5(2.8)(5)$&$17.8(2.0)(1.3)$\\ 
$\tilde{c}_{m}$& $=\tilde{c}_{d}$& $=18.9(9)$& $=\tilde{c}_{d}$&$=18.9(9)$\\ 
$a_{\pi\pi}$& $0.76(61)(6)$& $0.34(22)(19)$& $0.98(58)(9)$& $0.57(38)(20)$\\ 
$\chi^{2}_{\rm{dof}}$&$1.12$&$1.12$&$1.23$& $1.23$\\ 
\hline    
$a[Y]$&$-0.074(7)(8)$&$-0.073(6)(9)$&$-0.071(6)(8)$&$-0.070(6)(9)$\\
$b[Y^{2}]$&$-0.049(1)(2)$&$-0.054(1)(2)$&$-0.050(2)(1)$&$-0.054(1)(1)$\\
$d[X^{2}]$&$-0.047(8)(4)$&$-0.047(2)(4)$&$-0.055(6)(4)$&$-0.055(6)(4)$\\
\hline
$\kappa_{03}[Y^{3}]$&$0.001$&$0.003$&$0.001$& $0.002$\\
$\kappa_{21}[YX^{2}]$&$-0.004$&$-0.005$&$-0.005$& $-0.005$\\
$\kappa_{22}[Y^{2}X^{2}]$&$0.001$&$0.002$&$0.002$& $0.002$\\  
\noalign{\smallskip}\hline  
\end{tabular}
\caption{Results for the parameters of the fits together with their associated Dalitz parameters for two different fit scenarios and two different analyses (analysis I and II) of the A2 data set. Masses and coupling are given in MeV. The first error is the statistical uncertainty 
coming from the statistical uncertainties on the data, the second error is the systematic uncertainty coming from the uncertainty on the subtraction constant $a_{\pi\eta}$. See main text for details.}
\label{DalitzParam}
\end{table}

\begin{table}
\centering
\begin{tabular}{lllll}
\hline\noalign{\smallskip}
\multirow{2}{*}{Parameter} & 
\multicolumn{2}{l}{\qquad Analysis I}& 
\multicolumn{2}{l}{\qquad Analysis II}\\[0.5ex]
&Fit 1&Fit 2& Fit 1&Fit 2\\
\noalign{\smallskip}\hline\noalign{\smallskip}
$M_{S}$& $996(66)(25)$& $967(29)(3)$& $1016(63)(31)$& $983(51)(3)$\\ 
$c_{d}$& $23.3(3.5)(1.5)$& $21.5(2.1)(2)$& $24.6(3.8)(1.8)$& $22.5(3.1)(2)$\\ 
$c_{m}$& $=c_{d}$& $=41.1(1)$& $=c_{d}$& $=41.1(1)$\\ 
$\tilde{c}_{d}$& $13.5(2.0)(9)$& $12.4(1.2)(2)$& $14.2(2.2)(1.0)$&$13.0(1.8)(2)$\\ 
$\tilde{c}_{m}$& $=\tilde{c}_{d}$& $=18.9(9)$& $=\tilde{c}_{d}$&$=18.9(9)$\\ 
$a_{\pi\pi}$& $2.01(1.61)(71)$& $0.16(12)(12)$& $2.74(2.18)(90)$& $0.66(1.35)(10)$\\ 
$\chi^{2}_{\rm{dof}}$&$1.24$&$1.16$&$1.39$& $1.29$\\ 
\hline    
$a[Y]$&$-0.091(9)(4)$&$-0.091(8)(2)$&$-0.090(6)(4)$&$-0.089(9)(2)$\\
$b[Y^{2}]$&$-0.013(1)(5)$&$-0.029(1)(1)$&$-0.009(2)(5)$&$-0.024(1)(1)$\\
$d[X^{2}]$&$-0.031(6)(3)$&$-0.030(4)(7)$&$-0.037(6)(3)$&$-0.036(5)(6)$\\
\hline
$\kappa_{03}[Y^{3}]$&$0.001$&$0.003$&$0.001$& $0.002$\\
$\kappa_{21}[YX^{2}]$&$-0.001$&$-0.001$&$-0.001$& $-0.001$\\
$\kappa_{22}[Y^{2}X^{2}]$&$0.0004$&$0.003$&$0.001$& $0.001$\\  
\noalign{\smallskip}\hline  
\end{tabular}
\caption{Same as table \ref{DalitzParam} but without the $D$-wave $\pi\pi$ final-state interactions.}
\label{DalitzParamDwave}
\end{table}

Contrary to what was done in the previous sections, we report here not only the results for the Dalitz plot parameters $a$, $b$ and $d$ but also the results for the higher order ones, $\kappa_{03},~\kappa_{21}$ and $\kappa_{22}$. They are found to be very small 
as reported in previous theoretical analyses~\cite{Escribano:2010wt,Isken:2017dkw}. 

Note that the values found for the Dalitz plot parameters do not change much with the different fit scenarios and 
are very similar to the ones obtained in section \ref{section62} where only the $\pi\pi$ rescattering effects were taken into account. 
This is expected since we saw that the $\pi\pi$ rescattering dominates the final-state interactions. The stability of our fit results 
makes us very confident in the robustness of the results. 

In order to illustrate the overall effects of the $D$-wave $\pi\pi$ final-state interactions, we have also performed fits to the Dalitz plot experimental distribution without the $D$-wave contribution.
The resulting fit results are gathered in table \ref{DalitzParamDwave}. They show a substantial shift of the Dalitz parameters with respect to the ones collected in table \ref{DalitzParam} that include the $D$-wave.
In particular, when the $D$-wave is omitted the value for $a[Y]$ is shifted downwards while the parameters $b[Y^{2}]$ and $d[X^{2}]$ are shifted upwards. This demonstrates the importance of the $D$-wave $\pi\pi$ final-state interactions.

In the following, we study the dependence of the Dalitz parameters with respect to the numerical values of the mass and couplings of the participating scalar multiplets. 
For this exercise, we take different values for the mass and couplings from the literature and make some "crude" predictions.
The resulting estimates are gathered in table \ref{DalitzParamPredictions} where we have used the constraints $\tilde{c}_{d}=c_{d}/\sqrt{3}$ and $\tilde{c}_{m}=c_{m}/\sqrt{3}$, and fixed $a_{\pi\pi}=0.76$ from table \ref{DalitzParam}.
These results show that the Dalitz plot parameters are sensitive mostly to the values of $c_{d}$ and $M_{S}$.
The variation of these parameters has also a sizeable impact on the predicted branching ratio. 
Out of the five predictions shown in this table, the results given in the last column are the most realistic ones.
\begin{table}
\centering
\begin{tabular}{llllllllll}
\hline\noalign{\smallskip}
\multirow{2}{*}{Parameter} & 
\multicolumn{3}{l}{\qquad\qquad\qquad\qquad\quad  Predictions}&\\[0.5ex]
&Constraint&Ref.\,\cite{Ecker:1988te}&Ref.\,\cite{Jamin:2000wn}&Ref.\,\cite{Jamin:2000wn}&Ref.\,\cite{Guo:2009hi}\\
\noalign{\smallskip}\hline\noalign{\smallskip}
$M_{S}$&$1400^{\dagger}$& $983$& $1400^{\dagger}$&$1190$&$980(40)$\\ 
$c_{d}$&$F_{\pi}/2^{\dagger\dagger}$&$32$& $30(10)$&$45.4$& $26(7)$\\ 
$c_{m}$&$F_{\pi}/2$&$43$& $43(14)$&$=c_{d}$&$80(21)$\\ 
\hline    
$a[Y]$&$-0.201$&$-0.045$&$-0.341$&$-0.136$&$-0.083(8)$\\
$b[Y^{2}]$&$-0.055$&$-0.050$&$-0.041$&$-0.056$&$-0.051(1)$\\
$d[X^{2}]$&$-0.088$&$-0.034$&$-0.140$&$-0.071$&$-0.065(4)$\\
\hline
Branching ratio&$11\%$&$47\%$&$2\%$&$46\%$&$22\%$\\
\noalign{\smallskip}\hline  
\end{tabular}
\caption{Predictions for the Dalitz-plot parameters for different values of the mass and couplings (given in MeV). 
In the first column, the estimate$^{\dagger}$ $M_{S}=1400$ GeV assumes that the $a_{0}(980)$ is dynamically generated and the large-$N_{c}$ restriction$^{\dagger\dagger}$ for the couplings $c_{d}=c_{m}=F_{\pi}/2$ \cite{Jamin:2000wn} is taken.}
\label{DalitzParamPredictions}
\end{table}

\begin{figure}[h!]
\begin{center}
\includegraphics[scale=0.35]{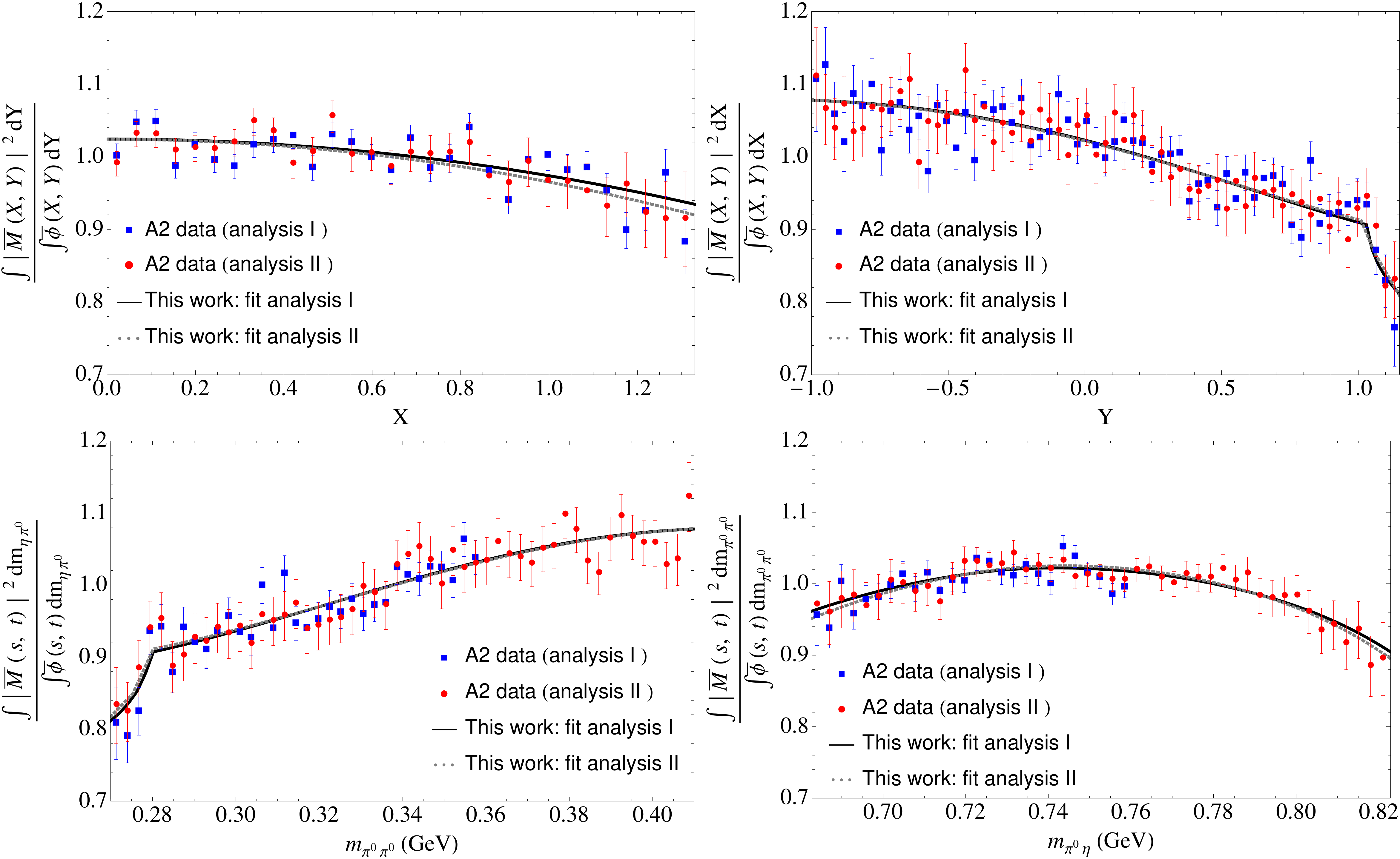}
\caption{\label{distribution4}Differential decay rate distribution for $\eta^{\prime}\to\eta\pi^{0}\pi^{0}$ divided by the phase-space, both individually normalized, for $X$ (left-top panel), $Y$ (right-top panel), $m_{\pi^{0}\pi^{0}}$, (left-down panel) and $m_{\pi^{0}\eta}$ (right-down panel) associated to the results of Fit 1 in table \ref{DalitzParam} (black solid and gray dotted curves for the analyses I and II data sets, respectively). 
They are obtained after resumming both the $\pi\pi$ and the $\pi\eta$ final-state interactions. 
Data is taken from Ref.\,\cite{Adlarson:2017wlz}.}
\end{center}
\end{figure}

\begin{figure}[h!]
\begin{center}
\includegraphics[scale=0.55]{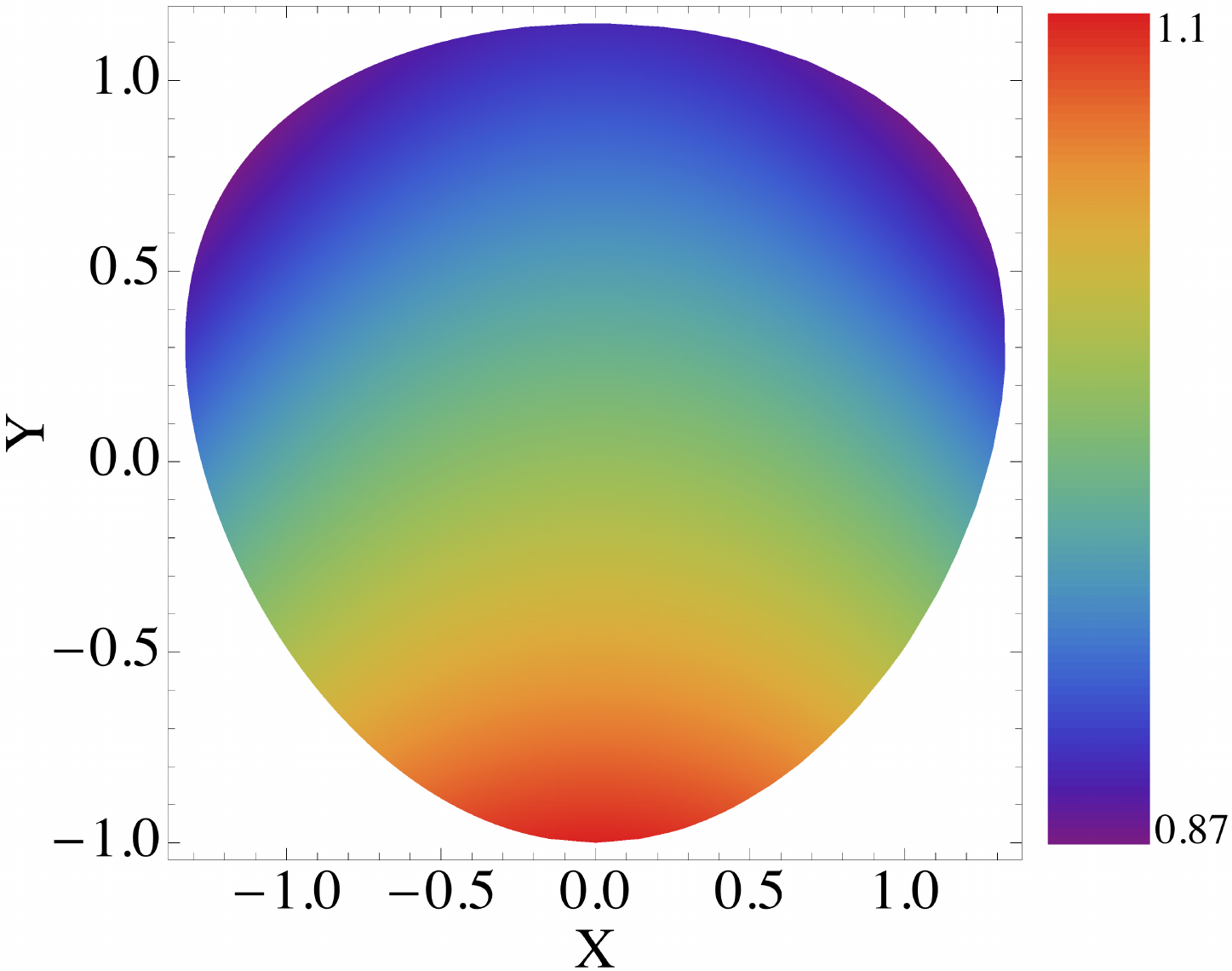}\includegraphics[scale=0.55]{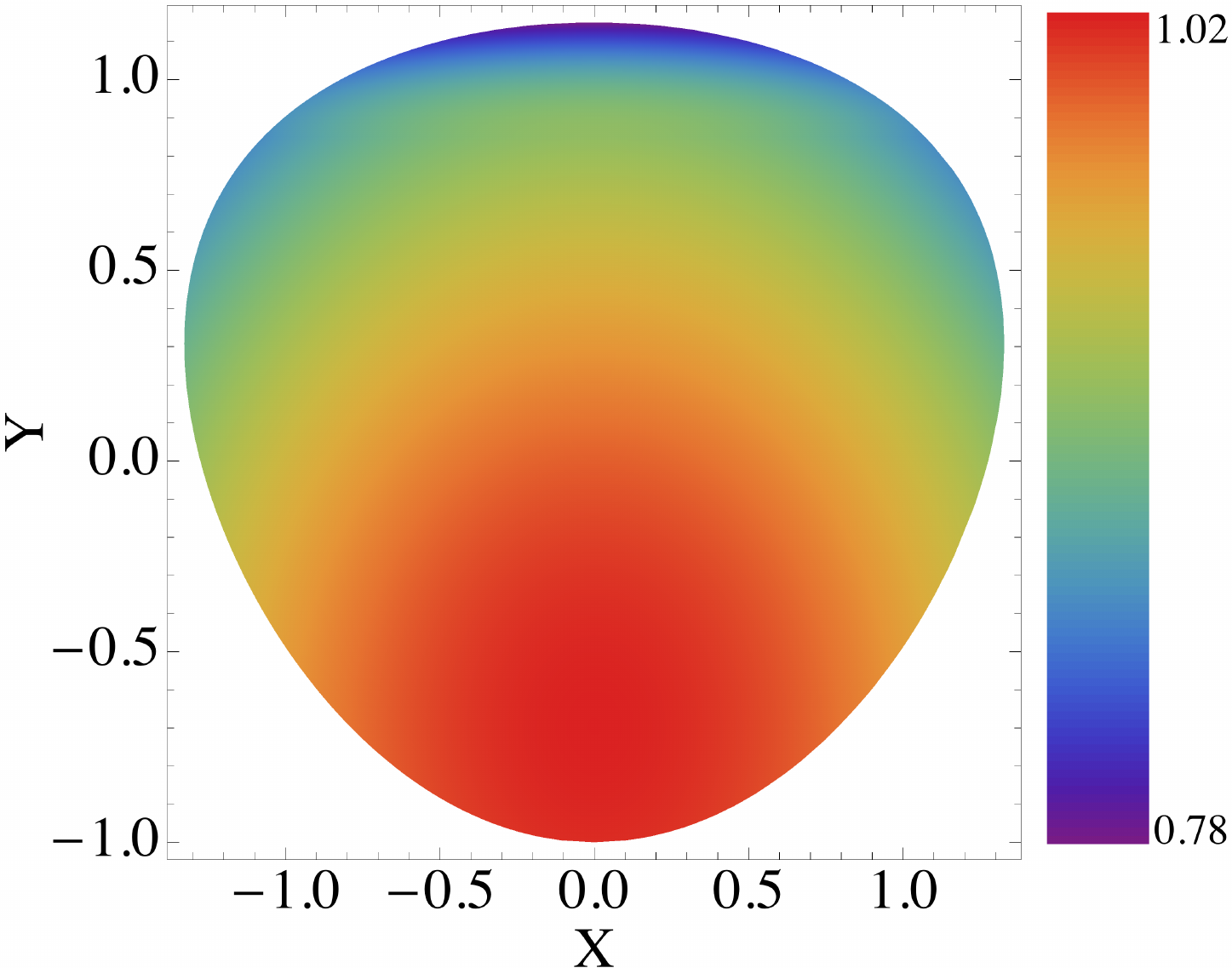}
\includegraphics[scale=0.55]{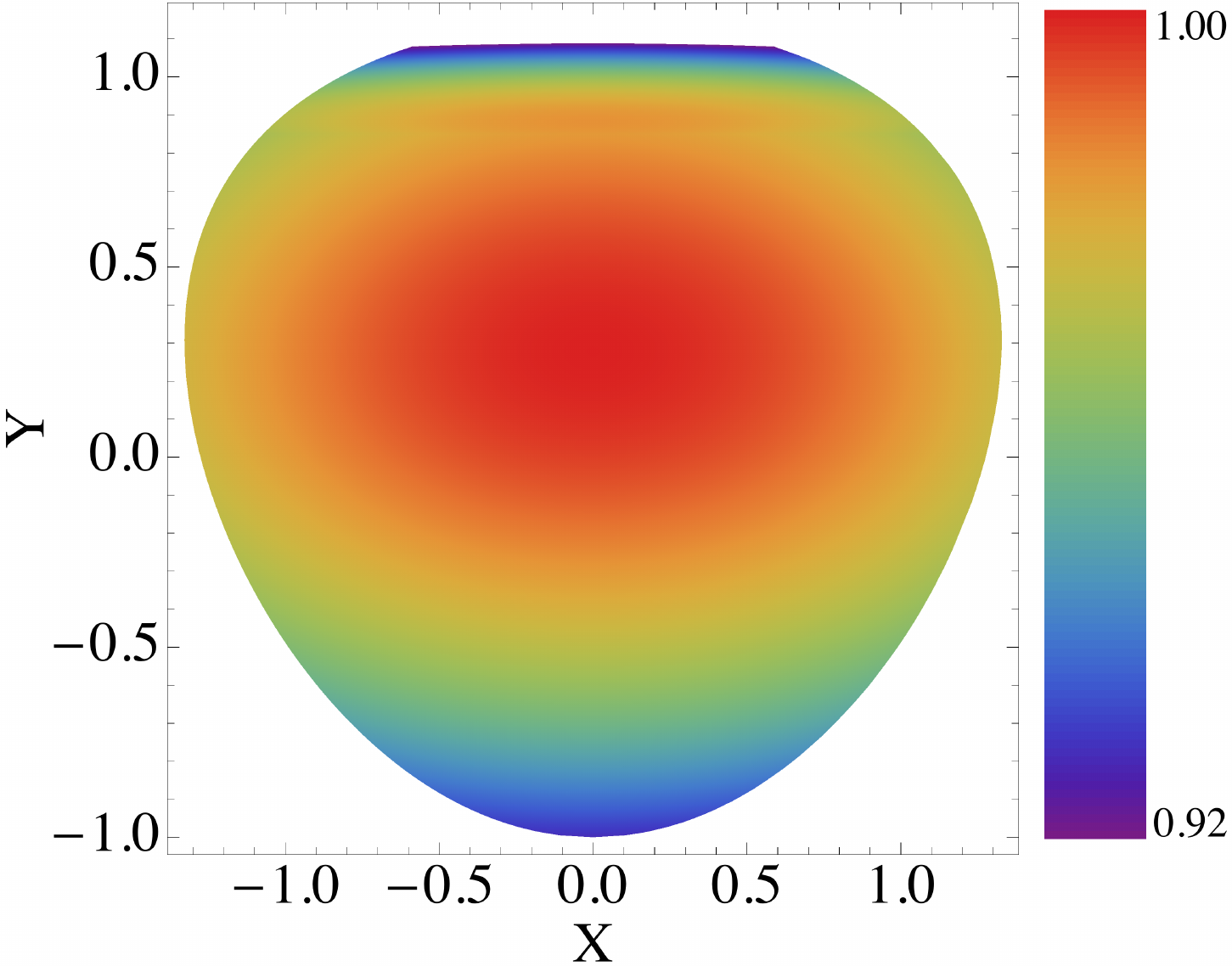}\includegraphics[scale=0.55]{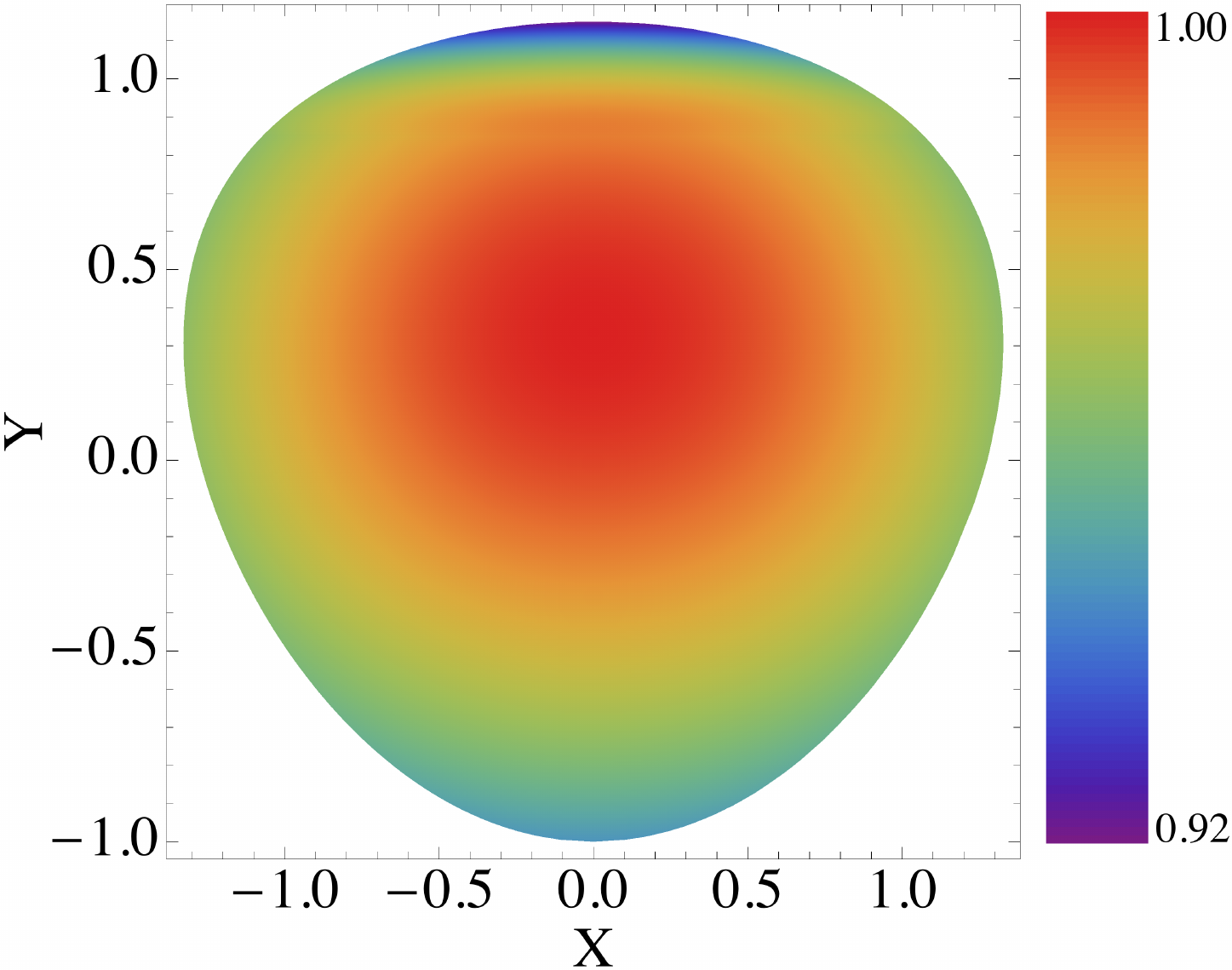}
\caption{\label{Dalitzplot}Dalitz plot distribution for the decay amplitude squared of $\eta^{\prime}\to\eta\pi^{0}\pi^{0}$ normalized to $1$ at $X=Y=0$ as obtained from ChPT including resonances and one-loop corrections $|M(X,Y)_{\rm{ChPT+Res+Loop}}|^{2}$ (top left panel) and after resumming $\pi\pi$ and $\pi\eta$ final-state interactions $|M(X,Y)_{\rm{Full}}|^{2}$ (top right panel). 
The quantities $|M(X,Y)_{\rm{Full}}|^{2}/|M(X,Y)_{\rm{ChPT+Res+Loop}}|^{2}$ and $|M(X,Y)_{\rm{Full}}|^{2}/|M(X,Y)_{\rm{D-wave=0}}|^{2}$ are also show in the bottom left-and right-panel, respectively.}
\end{center}
\end{figure}

In Fig.\,\ref{distribution4} we compare the experimental data to the results corresponding to Fit 1 of table \ref{DalitzParam}.
We observe that the representation of the amplitude obtained from the fit results of analysis I (black solid curve) 
practically overlaps with the one coming from the fit results of analysis II (gray dotted curve). 
The representation of the amplitude built from the results of our fits successfully describes the experimental data including 
the cusp effect at the $\pi^{+}\pi^{-}$ threshold. 
Moreover, in order to compare the decay amplitude with the Dalitz plot experimental measurements, we compute its square in terms of the Dalitz variables $X$ and $Y$ inside the physical decay region.
The shape of the Dalitz distribution, normalized to $1$ in the center of the Dalitz plot, is displayed in Fig.\,\ref{Dalitzplot} for the ChPT results presented in Eq.\,(\ref{fit1}) including resonances and one-loop corrections $|M(X,Y)_{\rm{ChPT+Res+Loop}}|^{2}$ (top left panel) and for the amplitude including $\pi\pi$ and $\pi\eta$ final-state interactions $|M(X,Y)_{\rm{Full}}|^{2}$ as obtained in Fit 1 of the analysis I  of table \ref{DalitzParam} (top right panel). 
The rescattering effects are neatly seen by the enhancement of the distribution in the center of the Dalitz plot, and in the outer up corners to less extent, on the plot of the top right with respect to the plot of the top left.
The top right plot is in good agreement with the experimental results \cite{Adlarson:2017wlz,Ablikim:2017irx}. It shows that the Dalitz distribution is more populated when the pions go back-to-back (cf.\,Fig.\,\ref{DalitzBoundaries}). 
In order to further illustrate the strong effects of all final-state interactions on the Dalitz plot distribution, on the bottom left panel of Fig.\,\ref{Dalitzplot} we plot the quantity $|M(X,Y)_{\rm{Full}}|^{2}$ divided by the same quantity before the unitarization, corresponding to $|M(X,Y)_{\rm{ChPT+Res+Loop}}|^{2}$.
Clearly, the effects of the unitarization of the amplitude, dominated by the $\pi\pi$ rescattering, are very important in the upper central region of the distribution.
Finally, we study the region of the Dalitz plot influenced by the effects of the $D$-wave $\pi\pi$ final-state interactions. 
The answer is given on the bottom right panel of Fig.\,\ref{Dalitzplot} where we show the quantity $|M(X,Y)_{\rm{Full}}|^{2}$ divided by $|M(X,Y)_{\rm{D-wave=0}}|^{2}$. $|M(X,Y)_{\rm{D-wave=0}}|^{2}$ corresponds to $|M(X,Y)_{\rm{Full}}|^{2}$ with the $D$-wave $\pi\pi$ final-state interactions effects set to zero. 
We can see that the $D$-wave effects also appear on the upper central region of the Dalitz plot.

The central values of our final results for the Dalitz-plot parameters associated to the $\eta^{\prime}\to\eta\pi^{0}\pi^{0}$ decay 
correspond to Fit 1 of table \ref{DalitzParam} for the analysis I of A2 data. To assess the "experimental" systematic uncertainty, 
we take the largest variation of the central values with respect to the results considering the analysis II data set of the same table.
We add this uncertainty to the systematic uncertainty coming from the subtraction constant $a_{\pi\eta}$ in quadrature and we obtain   
\begin{equation}
a=-0.072(7)_{\rm{stat}}(8)_{\rm{syst}}\,,\quad b=-0.052(1)_{\rm{stat}}(2)_{\rm{syst}}\,,\quad d=-0.051(8)_{\rm{stat}}(6)_{\rm{syst}}\,.
\label{finalDalitzParam}
\end{equation}

While the values for the Dalitz-plot parameters $a$ and $d$ are in good agreement with the one reported by A2, $a=-0.074(8)(6)$ and $d=-0.050(9)(5)$, the central value of the parameter $b$ is shifted towards a smaller absolute value compared to the A2 one, $b=-0.063(14)(5)$, but in good agreement within errors. 

\begin{figure}[h!]
\begin{center}
\includegraphics[scale=0.8]{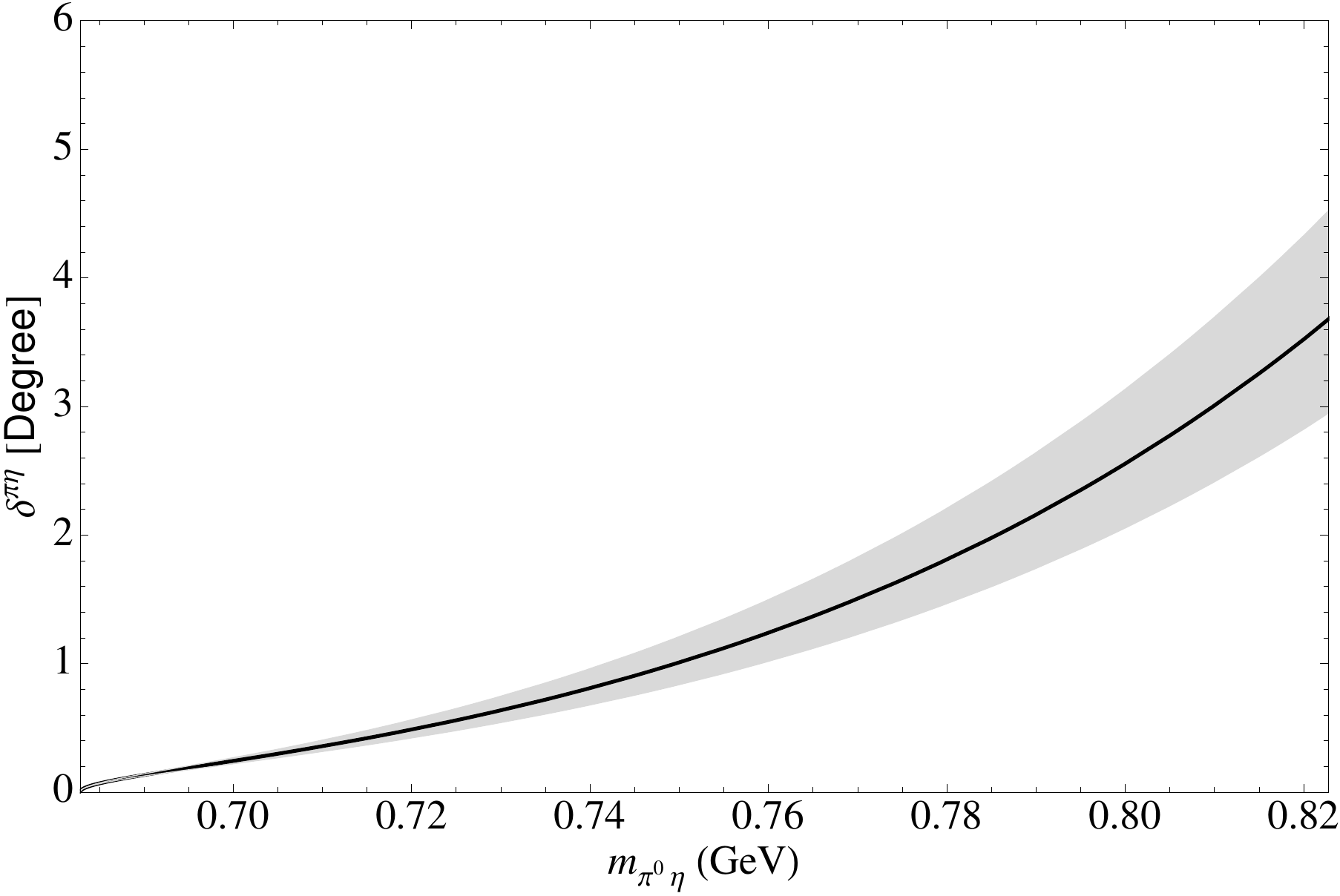}
\caption{\label{pietaphase}Isospin-zero $\pi\eta$ phase shift in the physical region of the decay $\eta^{\prime}\to\eta\pi^{0}\pi^{0}$. The error band is due to the statistical uncertainties associated to the parameters of Fit 1 of table \ref{DalitzParam} and to the subtraction constant $a_{\pi\eta}$.}
\end{center}
\end{figure}

Although our dedicated analysis shows that the $\pi\eta$ rescattering effects are small we can still extract some information about the $I=1$ $\pi\eta$ phase shift as a byproduct of our study.
In Fig.\,\ref{pietaphase}, we display the $\pi\eta$ phase shift in  the physical decay region $t=[(m_{\pi}+m_{\eta})^{2},(m_{\eta^{\prime}}-m_{\pi})^{2}]$\footnote{For a precise extraction of the $\pi\eta$ phase shift at higher energies i.e. reaching the $K\bar{K}$ threshold, a more sophisticated parameterization of the $\pi\eta$ scattering is required. See Ref.\,\cite{Albaladejo:2015aca} for a recent parameterization.} using the results of Fit 1 of table \ref{DalitzParam}. 
The phase shift is calculated as 
\begin{eqnarray}
\tan\delta^{\pi\eta}(s)=\frac{{\rm{Im}}\,T_{\pi\eta}^{10}(s)}{{\rm{Re}}\,T_{\pi\eta}^{10}(s)}\,,
\end{eqnarray}
where
\begin{eqnarray}
T_{\pi\eta}^{10}(s)=\frac{t^{10}_{\pi\eta}(s)}{\left(1+16\pi g_{\pi\eta}(s)t^{10}_{\pi\eta}(s)\right)}\,.
\end{eqnarray}
$g_{\pi\eta}(s)$ is given in Eq.\,(\ref{gpieta}) and $t^{10}_{\pi\eta}(s)$  is defined in appendix \ref{pietascattering}.
We observe that within this energy region the phase shift is small.

\begin{figure}[h!]
\begin{center}
\includegraphics[scale=0.7]{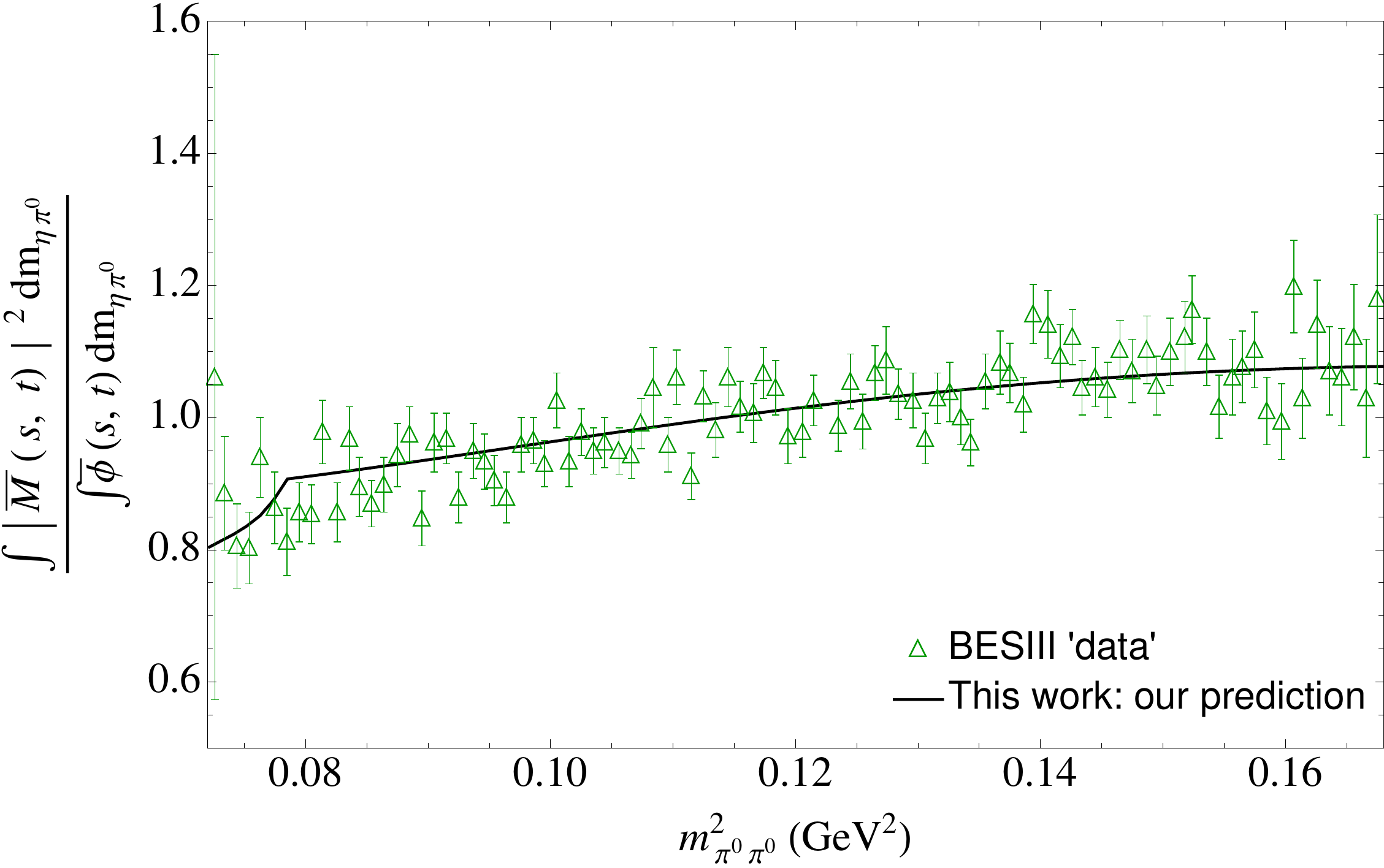}
\caption{\label{predictionBESIII}Differential decay rate distribution for $\eta^{\prime}\to\eta\pi^{0}\pi^{0}$ divided by the phase-space, both individually normalized, associated to the resulting parameters of Fit 1 of table \ref{DalitzParam} as compared with the BESIII experimental data \cite{Ablikim:2017irx}.}
\end{center}
\end{figure}

Before concluding, in Fig.\,\ref{predictionBESIII}, we compare our results on the $\pi^{0}\pi^{0}$ mode to the BESIII measurement \cite{Ablikim:2017irx}. 
Contrary to A2, the BESIII experimental data is not yet publicly available, so we have extracted the data points from figure 7 of 
Ref. \cite{Ablikim:2017irx} for the comparison.
Our prediction is displayed in Fig.\,\ref{predictionBESIII}. It is in very good agreement with the measured data. To show this we 
computed the $\chi^{2}$/dof using the results of Fit 1 of table \ref{DalitzParam} and the BESIII data. We obtain $\chi^{2}$/dof$=101.5/95\sim1.07$. 
Note that contrary to A2 no statistically significant evidence for a cusp at the $\pi^{+}\pi^{-}$ threshold is observed.

Using our representation of the amplitude using the fits to the data on the $\pi^{0}\pi^{0}$ mode from the A2 collaboration, 
we can predict the Decay rate distribution in the charged channel ($\pi^{+}\pi^{-}$). 
To predict the Dalitz-plot parameters of the $\pi^{+}\pi^{-}$ decay mode, one should consider all possible sources of isospin breaking.
In our framework, isospin breaking effects mostly affect the Dalitz variables $X$ and $Y$ if the charged pion mass is used in Eqs.\,(\ref{Yvariable}) and (\ref{Xvariable}).
In Ref.\,\cite{Isken:2017dkw} relations between the Dalitz parameters in the charged and the neutral decay modes have been derived:
\begin{eqnarray}
a^{n}=a^{c}+\varepsilon_{\rm{iso}}(a^{c}+2b^{c})\,,\quad b^{n}=b^{c}(1+2\varepsilon_{\rm{iso}})\,,\quad d^{n}=d^{c}\left(\frac{Q^{n}}{Q^{c}}\right)^{2}\,,
\end{eqnarray}
where the superscripts $c$ and $n$ denote the associated parameters in the charged and neutral systems, respectively, and with $\varepsilon_{\rm{iso}}\sim4.7\%$ \cite{Isken:2017dkw}.
Following this prescription, our estimates for the Dalitz parameters in the charged channel reads 
\begin{eqnarray}
a=-0.065(7)_{\rm{stat}}(8)_{\rm{syst}}\,,\quad b=-0.048(1)_{\rm{stat}}(2)_{\rm{syst}}\,,\quad d=-0.045(7)_{\rm{stat}}(5)_{\rm{syst}}\,.
\label{finalDalitzParamCharged}
\end{eqnarray}

Comparing the above results with the most recent experimental determination of these parameters in the charged system released by BESIII in 2017 \cite{Ablikim:2017irx}, 
$a=-0.056(4)_{\rm{stat}}(3)_{\rm{syst}}\,, b=-0.049(6)_{\rm{stat}}(6)_{\rm{syst}}\,, d=-0.063(4)_{\rm{stat}}(4)_{\rm{syst}}$, 
we observe that our prediction for $b$ is in excellent agreement while $a$ and $d$ are found to be $1\sigma$ and $2\sigma$ away, respectively. 

\begin{figure}[h!]
\begin{center}
\includegraphics[scale=0.525]{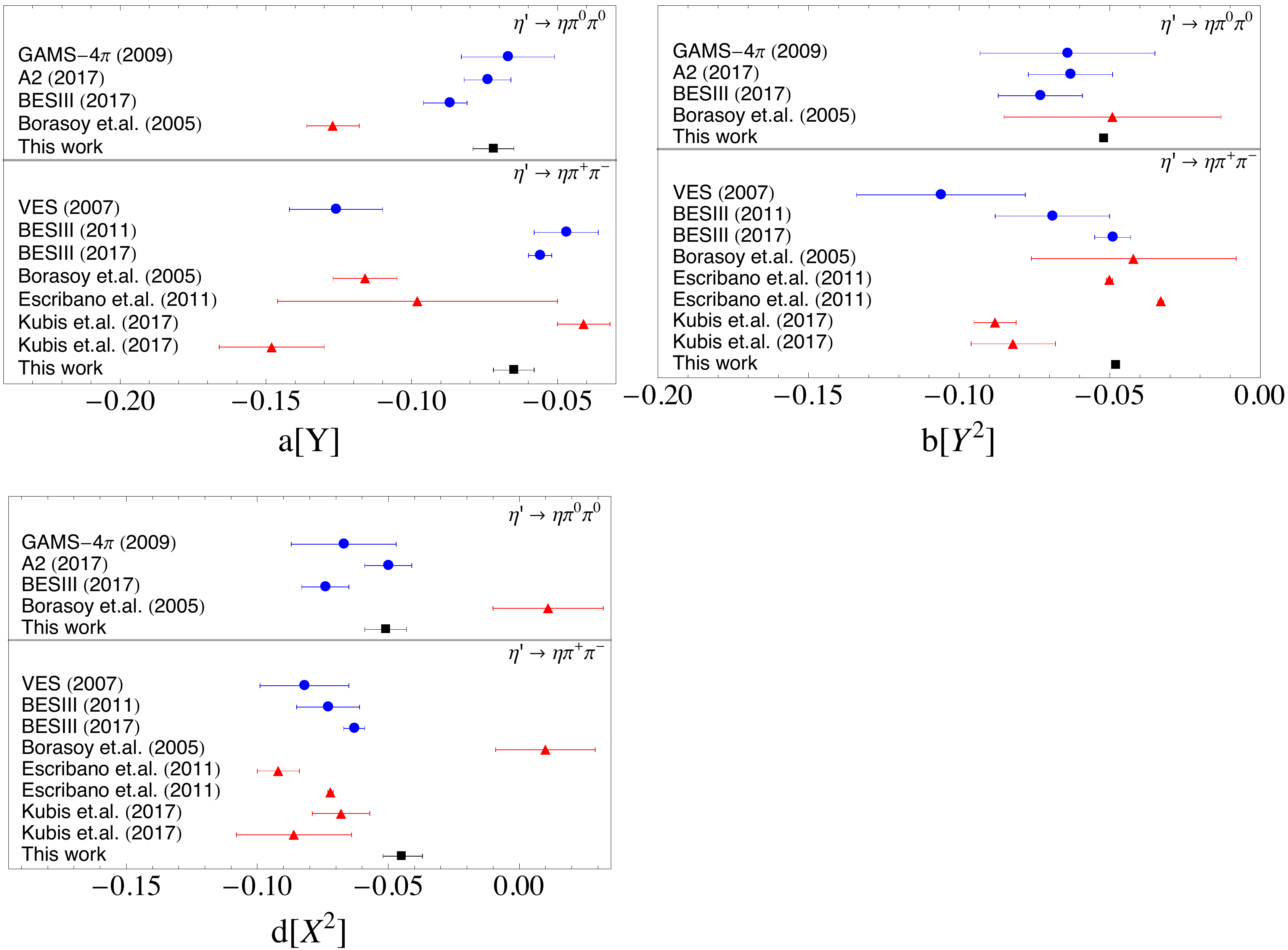}
\caption{\label{plotDalitzParameters}Comparison of experimental $(\textcolor{blue}{\bullet})$ and theoretical $(\textcolor{red}{\blacktriangle})$ determinations of the associated Dalitz-plot slope parameters for $\eta^{\prime}\to\eta\pi\pi$ (cf.\,table \ref{DalitzParamStatus}). 
Our results $(\textcolor{black}{\blacksquare})$ correspond to Eqs.\,(\ref{finalDalitzParam}) and $(\ref{finalDalitzParamCharged})$ for the $\pi^{0}\pi^{0}$ and $\pi^{+}\pi^{-}$ modes, respectively.
Only the statistical uncertainty is shown.}
\end{center}
\end{figure}

Finally, our results given in Eqs.\,(\ref{finalDalitzParam}) and (\ref{finalDalitzParamCharged}) for the neutral and charged decays modes, respectively, are graphically compared to previous experimental and theoretical determinations in Fig.\,\ref{plotDalitzParameters}.

\newpage

\section{Conclusions}\label{conclusions}

Recent measurements of the $\eta$-$\eta^{\prime}$ system
have reached unprecedented precision placing new demands on the 
accuracy of the corresponding theoretical description. 
The $\eta^{\prime}\to\eta\pi\pi$ decays represent a good laboratory to test any extension of $SU(3)$ Chiral Perturbation Theory, the effective field theory of QCD, which has proven to be very successful in describing pion and kaon processes. 
In this work, we have analyzed the $\eta^{\prime}\to\eta\pi\pi$ transition within $U(3)$ ChPT  
at one-loop including scalar resonance states as degrees of freedom. 
The corresponding amplitude has been unitarized 
using the $N/D$ method. Our treatment accounts for simultaneous $\pi\pi$ and $\pi\eta$ final-state interaction effects.

This parametrization has been fitted to the recently released A2 collaboration data on 
the $\eta^{\prime}\to\eta\pi^{0}\pi^{0}$ channel and 
very good agreement has been achieved. The results of the fit show that the Dalitz plot parameter $b$ is shifted downwards compared to $U(3)$ ChPT predictions. 
We demonstrate that this is attributed to the $S$-wave resummation of the $\pi\pi$ final-state interactions. 
Moreover, to match the A2 experimental accuracy requires the inclusion of the $D$-wave contribution. 
On the contrary, the $S$-wave $\pi\eta$ rescattering has shown to be small in agreement with previous studies. 

To further improve the description of the rescattering effects, one can consider a more sophisticated unitarization procedure in coupled channels including inelastic $K\bar{K}$ scattering. 
We postpone it for a future analysis when new measurements, e.g. by GlueX experiment, become available. 

In summary, from our analysis we extract the following Dalitz-plot parameters $a=-0.072(7)_{\rm{stat}}(8)_{\rm{syst}}\,, b=-0.052(1)_{\rm{stat}}(2)_{\rm{syst}}\,, d=-0.051(8)_{\rm{stat}}(6)_{\rm{syst}}$. Using these results, we are able to make predictions for the charged channel. 
These predictions are found to be in very good agreement with the BES-III measurements of this channel. 
Moreover, we were able to extract some information on the $I=1$ $\pi\eta$ phase shift at low energy. 

The theoretical framework developed here should be suitable for precision analyses of future experimental data. 


\section*{Acknowledgements}

The authors acknowledge discussions with Patrik Adlarson, Feng-Kun Guo, Zhi-Hui Guo and Stefan Leupold.
The work of S.G-S is supported in part by the CAS President's International Fellowship Initiative for Young International Scientists (Grant No.\,2017PM0031), by the Sino-German Collaborative Research Center \textquotedblleft Symmetries and the Emergence of Structure in QCD\textquotedblright\,(NSFC Grant No.\,11621131001, DFG Grant No.\,TRR110), and by NSFC (Grant No.\,11747601).
S.G-S also acknowledges Indiana University for hospitality.
The work of E.P. is supported in part by the U.S. Department of Energy (contract DE-AC05-06OR23177), Indiana University and a National Science Foundation Grant No. PHY-1714253.

\appendix

\section{Loop contributions to the decay amplitude}\label{loopcorrections}

\begin{eqnarray}
&&\nonumber\mathcal{M}^{\rm{Loop}}(s,t,u)=\frac{m_{\pi}^{2}}{9F_{\pi}^{2}}\left(\sqrt{2}{\rm{c}}^{2}\theta-{\rm{c}}\theta{\rm{s}}\theta-\sqrt{2}{\rm{s}}^{2}\theta\right)\times\Bigg[\frac{1}{3F_{\pi}^{2}}\left(m_{\pi}^{2}-4m_{K}^{2}\right)\times\\[1ex]
&&\nonumber\times\left(-2{\rm{c}}^{4}\theta+2\sqrt{2}{\rm{c}}^{3}\theta{\rm{s}}\theta+3{\rm{c}}^{2}{\rm{s}}^{2}\theta-2\sqrt{2}{\rm{c}}\theta{\rm{s}}^{3}\theta-2{\rm{s}}^{4}\theta\right)B_{0}(s,m_{\eta^{\prime}},m_{\eta})\\[1ex]
&&\nonumber+\frac{m_{\pi}^{2}}{F_{\pi}^{2}}\left({\rm{c}}\theta-\sqrt{2}{\rm{s}}\theta\right)^{2}B_{0}(t,m_{\eta},m_{\pi})+(t\leftrightarrow u)\\[1ex]
&&\nonumber+\frac{m_{\pi}^{2}}{F_{\pi}^{4}}\left(2{\rm{c}}^{2}\theta+2\sqrt{2}{\rm{c}}\theta{\rm{s}}\theta+{\rm{s}}^{2}\theta\right)B_{0}(t,m_{\eta^{\prime}},m_{\pi})+(t\leftrightarrow u)\Bigg]\\[1ex]
&&\nonumber+\frac{m_{\pi}^{2}}{54F_{\pi}^{4}}\left({\rm{c}}\theta-\sqrt{2}{\rm{s}}\theta\right)^{2}\Bigg[\sqrt{2}{\rm{c}}^{4}\theta\left(5m_{\pi}^{2}-8m_{K}^{2}\right)-{\rm{c}}^{3}\theta\left(8m_{K}^{2}+m_{\pi}^{2}\right){\rm{s}}\theta\\[1ex]
&&\nonumber+3\sqrt{2}{\rm{c}}^{2}\theta\left(4m_{K}^{2}-m_{\pi}^{2}\right){\rm{s}}^{2}\theta+4{\rm{c}}\theta\left(5m_{K}^{2}-2m_{\pi}^{2}\right){\rm{s}}^{3}\theta+4\sqrt{2}\left(m_{K}^{2}-m_{\pi}^{2}\right){\rm{s}}^{4}\theta\Bigg]B_{0}^{\rm{eq}}(s,m_{\eta})\nonumber\\[1ex]
&&\nonumber+\frac{m_{\pi}^{2}}{54F_{\pi}^{4}}\left(\sqrt{2}{\rm{c}}\theta+{\rm{s}}\theta\right)^{2}\Bigg[4\sqrt{2}{\rm{c}}^{4}\theta\left(m_{\pi}^{2}-m_{K}^{2}\right)+4{\rm{c}}^{3}\theta\left(5m_{K}^{2}-2m_{\pi}^{2}\right){\rm{s}}\theta\\[1ex]
&&\nonumber+3\sqrt{2}{\rm{c}}^{2}\theta\left(m_{\pi}^{2}-4m_{K}^{2}\right){\rm{s}}^{2}\theta-{\rm{c}}\theta\left(8m_{K}^{2}+m_{\pi}^{2}\right){\rm{s}}^{3}\theta+2\sqrt{2}\left(8m_{K}^{2}-5m_{\pi}^{2}\right){\rm{s}}^{4}\theta\Bigg]B_{0}^{\rm{eq}}(s,m_{\eta^{\prime}})\nonumber\\[1ex]
&&+\nonumber\frac{s}{24^{4}}\Bigg[2\sqrt{2}{\rm{c}}^{2}\theta\left(2m_{K}^{2}-m_{\pi}^{2}\right)+{\rm{c}}\theta\left(3m_{\eta}^{2}+2m_{\eta^{\prime}}^{2}+8m_{K}^{2}+2m_{\pi}^{2}-9s\right){\rm{s}}\theta\\[1ex]
&&\nonumber+2\sqrt{2}\left(m_{\pi}^{2}-2m_{K}^{2}\right){\rm{s}}^{2}\theta\Bigg]B_{0}^{\rm{eq}}(s,m_{K})+\frac{m_{\pi}^{2}\left(m_{\pi}^{2}-2s\right)}{6f^{4}}\left(\sqrt{2}{\rm{c}}^{2}\theta-{\rm{c}}\theta{\rm{s}}\theta-\sqrt{2}{\rm{s}}^{2}\theta\right)B_{0}^{\rm{eq}}(s,m_{\pi})\\[1ex]
&&\nonumber+\frac{1}{216F_{\pi}^{4}}\Bigg[-2\sqrt{2}{\rm{c}}^{2}\theta\left(2m_{K}^{2}+m_{\pi}^{2}\right)\left(3m_{\eta}^{2}+8m_{K}^{2}+m_{\pi}^{2}-9\left(m_{\eta}^{2}+m_{\eta^{\prime}}^{2}+2m_{\pi}^{2}-s-t\right)\right)\\[1ex]
&&\nonumber+{\rm{c}}\theta{\rm{s}}\theta\Big(32m_{K}^{4}-16m_{K}^{2}m_{\pi}^{2}-7m_{\pi}^{4}+3m_{\eta^{\prime}}^{2}\left(8m_{K}^{2}+m_{\pi}^{2}-9t\right)\\
&&+3m_{\eta}^{2}\left(3m_{\eta^{\prime}}^{2}+8m_{K}^{2}+m_{\pi}^{2}-9t\right)-144m_{K}^{2}t-18m_{\pi}^{2}t+81t^{2}\Big)\nonumber\\
&&+2\sqrt{2}\left(2m_{K}^{2}+m_{\pi}^{2}\right){\rm{s}}^{2}\theta\Bigg]B_{0}^{\rm{eq}}(t,m_{K})+(t\leftrightarrow u)\,,
\label{loops}
\end{eqnarray}
with $({\rm{c,s}})=(\cos,\sin)$ for abbreviation and where the loop functions are calculated in dimensional regulartization within the $\overline{MS}$-1 renormalization scheme defined by
\begin{eqnarray}
B_{0}(s,m_{a},m_{b})=\frac{1}{16\pi^{2}}\left(1-\log\frac{m^{2}_{b}}{\mu^{2}}+x_{+}\log\frac{x_{+}-1}{x_{+}}+x_{-}\log\frac{x_{-}-1}{x_{-}}\right)\,,
\label{loopfunction}
\end{eqnarray}
with
\begin{eqnarray}
x_{\pm}=\frac{s+m_{a}^{2}-m_{b}^{2}}{2s}\pm\frac{1}{-2s}\sqrt{-4s(m_{a}^{2}-i0^{+})+(s+m_{a}^{2}-m_{b}^{2})^{2}}\,.
\label{xplusminus}
\end{eqnarray}
for the case of different mesons masses, and
\begin{eqnarray}
B_{0}^{\rm{eq}}(s,m)=\frac{1}{16\pi^{2}}\left(1-\log\frac{m^{2}_{b}}{\mu^{2}}+\sigma(s)\log\frac{\sigma(s)-1}{\sigma(s)+1}\right)\,,
\label{loopfunctionidenticalmasses}
\end{eqnarray}
with
\begin{eqnarray}
\sigma(s)=\sqrt{1-\frac{4m^{2}}{s}}\,,
\end{eqnarray}
for the case of equal masses.
In Eq.\,(\ref{loops}), loops with identical particles have been multiplied by factor of $1/2$.

The tadpole contribution is given by
\begin{eqnarray}
\mathcal{M}^{\rm{Tadpole}}&=&\left(\sqrt{2}\cos^{2}\theta-\cos\theta\sin\theta-\sqrt{2}\sin^{2}\theta\right)\frac{2m_{\pi}^{2}}{3F_{\pi}^{4}}\mu_{\pi}\nonumber\\[1ex]
&&+\frac{1}{60F_{\pi}^{4}}\Bigg[-10\sqrt{2}m_{\pi}^{2}\cos^{4}\theta+8\sqrt{2}\left(m_{\pi}^{2}-m_{K}^{2}\right)+10m_{\pi}^{2}\cos^{3}\theta\sin\theta\nonumber\\[1ex]
&&+2\sqrt{2}\sin^{2}\theta\left(4m_{K}^{2}+m_{\pi}^{2}\left(5\sin^{2}\theta-4\right)\right)\nonumber\\[1ex]
&&+\cos\theta\sin\theta\Big(13m_{\eta}^{2}+13m_{\eta^{\prime}}^{2}+32m_{K}^{2}-18m_{\pi}^{2}\nonumber\\[1ex]
&&-25s+10m_{\pi}^{2}\sin^{2}\theta-25\left(m_{\eta}^{2}+m_{\eta^{\prime}}^{2}+2m_{\pi}^{2}-s-t\right)-25t\Big)\Bigg]\mu_{K}\,,
\end{eqnarray} 
where 
\begin{eqnarray}
\mu_{P}=-\frac{m_{P}^{2}}{16\pi^{2}}\log\frac{m_{P}^{2}}{\mu^{2}}\,.
\label{mutadpole}
\end{eqnarray}

\section{$\pi\pi$ scattering within one-loop $U(3)$ R$\chi$T}\label{pipiscattering}

The $S$-and $D$-wave of the $\pi\pi$ scattering, $t^{00}(s)^{(2)+\rm{Res+Loop}}$ and $t^{02}(s)^{(2)+\rm{Res+Loop}}$ entering Eq.\,(\ref{schannelunitarityamplitude}), are obtained through (omitting the superscripts associated to the perturbative expansion)
\begin{eqnarray}
t^{IJ}_{\pi\pi}(s)=\frac{1}{32\pi}\frac{1}{s-4m_{\pi}^{2}}\int ^{0}_{4m_{\pi}^{2}-s}dt P_{J}\left(1+\frac{2t}{s-4m_{\pi}^{2}}\right)T^{I}_{\pi\pi}(s,t,u)\,.
\end{eqnarray}
For the case that concerns us $I=0$ and so the corresponding isospin amplitude reads
\begin{eqnarray}
T^{0}_{\pi\pi}(s,t,u)=3A(s,t,u)+A(t,s,u)+A(u,t,s)\,,
\end{eqnarray}
where
\begin{eqnarray}
A(s,t,u)&=&\frac{s-m_{\pi}^{2}}{F_{\pi}^{2}}-\frac{2\left(2c_{m}m_{\pi}^{2}+c_{d}(s-2m_{\pi}^{2})\right)^{2}}{3F_{\pi}^{4}(s-m_{S_{8}}^{2})}-\frac{4\left(2\tilde{c}_{m}m_{\pi}^{2}+\tilde{c}_{d}(s-2m_{\pi}^{2})\right)^{2}}{3F_{\pi}^{4}(s-m_{S_{1}}^{2})}\nonumber\\
&&+\frac{1}{16\pi^{2}}\Bigg[\frac{1}{12F_{\pi}^{4}}\Big(32m_{\pi}^{4}+2m_{K}^{2}(4m_{\pi}^{2}-3s)-8m_{\pi}^{2}(4m_{\pi}^{2}-s)+2s(4m_{\pi}^{2}-s-t)\nonumber\\
&&+(4m_{\pi}^{2}-s-t)^{2}+2st+t^{2}\Big)+\frac{16\pi^{2}}{9F_{\pi}^{4}}\left(4m_{\pi}^{2}-4(4m_{\pi}^{2}-s)+5s\right)\mu_{\pi}\nonumber\\
&&+\frac{16\pi^{2}}{12F_{\pi}^{4}}\left(-8m_{\pi}^{2}+6s\right)\mu_{K}\nonumber\\
&&+\frac{m_{\pi}^{4}}{9F_{\pi}^{4}}\left(-\sqrt{2}\cos^{2}\theta+\cos\theta\sin\theta+\sqrt{2}\sin^{2}\theta\right)^{2}B_{0}(s,m_{\eta^{\prime}},m_{\eta})\nonumber\\
&&+\frac{m_{\pi}^{4}}{18F_{\pi}^{4}}\left(\cos^{2}\theta-2\sqrt{2}\cos\theta\sin\theta+2\sin^{2}\theta\right)^{2}B_{0}^{\rm{eq}}(s,m_{\eta})\nonumber\\
&&+\frac{m_{\pi}^{4}}{18F_{\pi}^{4}}\left(2\cos^{2}\theta+2\sqrt{2}\cos\theta\sin\theta+\sin^{2}\theta\right)^{2}B_{0}^{\rm{eq}}(s,m_{\eta^{\prime}})+\frac{s^{2}}{8F_{\pi}^{4}}B_{0}^{\rm{eq}}(s,m_{K})\nonumber\\
&&+\frac{\left(s^{2}-m_{\pi}^{4}\right)}{2F_{\pi}^{4}}B_{0}^{\rm{eq}}(s,m_{\pi})+\frac{\left(u-4m_{K}^{2}\right)\left(u+2s-4m_{\pi}^{2}\right)}{24F_{\pi}^{4}}B_{0}^{\rm{eq}}(u,m_{K})\nonumber\\
&&+\frac{1}{6F_{\pi}^{4}}\left(14m_{\pi}^{2}-2m_{\pi}^{2}(2s+5u)+(s+2u)u\right)B_{0}^{\rm{eq}}(u,m_{\pi})\nonumber\\
&&+\frac{\left(t-4m_{\pi}^{2}\right)\left(t+2s-4m_{\pi}^{2}\right)}{24F_{\pi}^{4}}B_{0}^{\rm{eq}}(t,m_{K})\nonumber\\
&&+\frac{1}{6F_{\pi}^{4}}\left(14m_{\pi}^{2}-2m_{\pi}^{2}(2s+5t)+(s+2t)t\right)B_{0}^{\rm{eq}}(t,m_{\pi})\Bigg]\,.
\end{eqnarray}

\newpage

\section{$\pi\eta$ scattering within one-loop $U(3)$ R$\chi$T}\label{pietascattering}

The corresponding $I=1$ $S$-wave of the $\pi\eta$-scattering $t^{10}_{\pi\eta}(t)^{(2)+\rm{Res+Loop}+\Lambda}$ entering Eq.\,(\ref{tchannelunitarityamplitude}) is given by (omitting the superscripts associated to the perturbative expansion)

\begin{eqnarray}
t^{10}_{\pi\eta}(s)=\frac{s}{16\pi}\frac{1}{\lambda(s,m_{\eta}^{2},m_{\pi}^{2})}\int ^{0}_{-\frac{\lambda(s,m_{\eta}^{2},m_{\pi}^{2})}{s}}ds P_{0}\left(1+\frac{2st}{\lambda(s,m_{\eta}^{2},m_{\pi}^{2})}\right)T^{1}_{\pi\eta}(s,t,u)\,,
\end{eqnarray}
with the $\pi\eta$ scattering amplitude given by
\begin{eqnarray}
T^{1}_{\pi\eta}(s,t,u)=T^{(2)}_{\pi\eta}+T^{\rm{Res}}_{\pi\eta}+T^{\rm{Loop}}_{\pi\eta}+T^{\Lambda}_{\pi\eta}+T^{\rm{mixing}}_{\pi\eta}\,,
\label{scatamppieta}
\end{eqnarray}
where
\begin{eqnarray}
T^{(2)}_{\pi\eta}=\frac{m_{\pi}^{2}}{F_{\pi}^{2}}\left(\cos^{2}\theta-2\sqrt{2}\cos\theta\sin\theta+2\sin^{2}\theta\right)\,,
\end{eqnarray}
\begin{eqnarray}
\label{pietascatt}
&&T^{\rm{Res}}_{\pi\eta}=\left(\cos^{4}\theta+2\sin^{2}\theta+\cos^{2}\theta(1-6\sin^{2}\theta)+2\sqrt{2}\cos\theta\sin\theta(-1+2\sin^{2}\theta)\right)\times\nonumber\\
&&\times\frac{16c_{d}c_{m}m_{\pi}^{2}(m_{\pi}^{2}-m_{K}^{2})}{9F_{\pi}^{4}M_{S_{8}}^{2}}\nonumber
+\frac{8}{3F_{\pi}^{4}}\left(\cos^{2}\theta-2\sqrt{2}\cos\theta\sin\theta+2\sqrt{2}\sin^{2}\theta\right)\times\Bigg\lbrace\nonumber\\[1ex]
&&+\frac{8c_{m}^{2}m_{\pi}^{4}+8c_{d}c_{m}m_{\pi}^{2}(s-m_{\eta}^{2}-m_{\pi}^{2})+2c_{d}^{2}(s-m_{\eta}^{2}-m_{\pi}^{2})^{2}}{M_{a}^{2}-s}\nonumber\\[1ex]
&&+\frac{8c_{m}^{2}m_{\pi}^{4}+8c_{d}c_{m}m_{\pi}^{2}(m_{\eta}^{2}+m_{\pi}^{2}-s-t)+2c_{d}^{2}(s+t-m_{\eta}^{2}-m_{\pi}^{2})^{2}}{M_{a}^{2}-u}\Bigg\rbrace\nonumber\\
&&+\frac{2}{9F_{\pi}^{4}(M_{S_{8}}^{2}-t)}\Big[2c_{m}\Big(\cos^{2}\theta(8m_{K}^{2}-5m_{\pi}^{2})+2\sqrt{2}\cos\theta\sin\theta(4m_{K}^{2}-m_{\pi}^{2})\nonumber\\
&&+4\sin^{2}\theta(m_{K}^{2}-m_{\pi}^{2})\Big)-3c_{d}\cos\theta(\cos\theta+2\sqrt{2}\sin\theta)(2m_{\eta}^{2}-t)\Big]\times\nonumber\\
&&\times\Big[2c_{m}m_{\pi}^{2}+c_{d}(t-2m_{\pi}^{2})\Big]+\frac{4}{3F_{\pi}^{4}(M_{S_{1}}^{2}-t)}\Big[2\tilde{c}_{m}\Big(\cos^{2}\theta(4m_{K}^{2}-m_{\pi}^{2})\nonumber\\
&&+4\sqrt{2}\cos\theta\sin\theta(m_{K}^{2}-m_{\pi}^{2})+\sin^{2}\theta(2m_{K}^{2}+m_{\pi}^{2})\Big)-2\tilde{c}_{d}(\cos^{2}\theta+\sin^{2}\theta)(2m_{\eta}^{2}-t)\Big]\times\nonumber\\
&&\times\Big[2\tilde{c}_{m}m_{\pi}^{2}+\tilde{c}_{d}(t-2m_{\pi}^{2})\Big]\,,
\end{eqnarray}
and
\begin{eqnarray}
T^{\Lambda}_{\pi\eta}=\frac{2m_{\pi}^{2}\Lambda_{2}}{3F_{\pi}^{2}}\sin\theta\left(2\sin\theta-\sqrt{2}\cos\theta\right)\,.
\end{eqnarray}
The $\pi\eta$ scattering loop contribution, $T^{\rm{Loop}}_{\pi\eta}$ in Eq.\,(\ref{scatamppieta}), is small and its expression it is not shown due to its length but rather can be provided by the authors upon request. 
The $T^{\rm{mixing}}_{\pi\eta}$ contribution is also tiny and we therefore refrain to show it.

\end{document}